\documentclass[manuscript]{acmart}
\AtBeginDocument{%
  }

\setcopyright{acmlicensed}
\copyrightyear{2025}
\acmYear{2025}
\acmDOI{XXXXXXX.XXXXXXX}

\usepackage{graphicx}
\usepackage{amsmath}
\usepackage{booktabs}
\usepackage[table]{xcolor}
\usepackage{colortbl}
\usepackage{longtable}
\usepackage{tabularx}
\usepackage{array}
\usepackage{caption}
\usepackage{etoolbox}
\usepackage{multirow}
\usepackage{tikz}
\usepackage{textcomp}
\usepackage[shortcuts]{extdash}
\usepackage{soul}
\usepackage{listings}
\usepackage{adjustbox}
\usepackage{pdflscape}
\usepackage{bigstrut}
\usepackage[most]{tcolorbox}
\usepackage{ltablex}

\usetikzlibrary{shapes, arrows.meta, positioning, fit, backgrounds, calc}

\keepXColumns

\definecolor{llgray}{gray}{0.95}

\lstdefinestyle{prompt}{
  backgroundcolor=\color{llgray},
  basicstyle=\normalfont\small,
  frame=single,
  columns=fullflexible,
  breaklines=true,
  aboveskip=0pt,
  belowskip=0pt,
  lineskip=-1.5pt,
}

\begin{document}
\title[MORPHEUS: A Multidimensional Framework of Human Factors in Cybersecurity]{MORPHEUS: A Multidimensional Framework for Modeling, Measuring, and Mitigating Human Factors in Cybersecurity}

\author{Giuseppe Desolda} 
\author{Francesco Greco}
\author{Rosa Lanzilotti}
\author{Cesare Tucci}
\affiliation{%
  \institution{Department of Computer Science, University of Bari Aldo Moro}
  \city{Bari}
  \state{BA}
  \country{Italy}
}

\ccsdesc[500]{Security and privacy~Human and societal aspects of security and privacy}
\ccsdesc[500]{Security and privacy~Usability in security and privacy}
\ccsdesc[500]{Security and privacy~Social engineering attacks}
\ccsdesc[500]{Human-centered computing~HCI theory, concepts and models}
\ccsdesc[300]{Human-centered computing~Empirical studies in HCI}
\ccsdesc[300]{General and reference~Empirical surveys}

\renewcommand{\shortauthors}{Desolda et al.}
\acmArticleType{Research}
\keywords{Human Factors, Cybersecurity, Social Engineering, Framework, Measurement, Psychological Traits, Phishing, Risk Assessment, CAB Model}

\begin{abstract}
Despite technical advancements, the human factor remains cybersecurity's most exploited vulnerability. Current research acknowledges this but remains fragmented, treating vulnerabilities as isolated, static traits. To address this, we introduce MORPHEUS, a holistic framework conceptualizing human-centric security as a dynamic, interconnected system. Grounded in the Cognition–Affect–Behavior (CAB) model and Attribution Theory, MORPHEUS consolidates 50 human factors influencing susceptibility to major cyberthreats (e.g., phishing, malware, password management, and misconfigurations). 

Beyond mere identification, the framework introduces a hierarchical Causal Pathway Architecture. Systematically mapping 302 empirical interactions (82.8\% architecture-compliant), we reveal how cognitive, affective, and behavioral processes jointly shape security outcomes, distilling them into 12 recurring interaction mechanisms. MORPHEUS further links theory to practice through an inventory of 99 validated psychometric instruments for empirical assessment. We illustrate its applicability through in-depth operational scenarios for risk diagnosis and targeted interventions. Overall, MORPHEUS provides a comprehensive theoretical foundation for advancing human-centered cybersecurity.
\end{abstract}
  
\maketitle

\section{Introduction}
Research increasingly demonstrates that cybersecurity is strongly shaped by human factors, with empirical findings indicating that almost 88\% of breaches stem from human error \cite{sjouwerman2021stanford}. Consequently, cybercriminals have refined their tactics to target individuals rather than focusing solely on technical vulnerabilities. Social engineering tactics, notably phishing and its variants (e.g., SMishing and spear-phishing), exploit psychological mechanisms such as trust, urgency, and fear to induce users to disclose sensitive information or execute malicious actions \cite{arevalo2023human}. Besides social engineering, other human-related causes of security failures, such as poor password management habits, misconfigured systems, and accidental malware downloads, highlight the extent to which human choices influence cybersecurity posture. Despite technological advancements in security control measures, human factors remain one of the most persistent and exploitable weak points in defending against cyberthreats and an under-evaluated dimension of cybersecurity for individuals and organizations.

To identify and mitigate human-generated cyber risk, it is essential to develop a systematic process to identify human-related aspects that underpin security breaches. The latter have been studied in the context of cognitive biases and affective responses in cybersecurity~\cite{fagan2017investigation}, as well as behavioral patterns and organizational situations that make such lapses more likely~\cite{alqarni2016toward, whitty2015individual, marin2023influence, renaud2021shame}. Other studies have attempted to classify human factors through frameworks such as the NIST Cybersecurity Framework~\cite{shen2014nist}, analyses of insider threat and time-pressure variables, and models that synthesize these constructs~\cite{uebelacker2014social, shen2014nist}. These frameworks may be valuable, but they are narrow or excessively constraining, partitioning cognition, emotion, behavior, and social influences against each other instead of capturing the dynamic relationships that occur in real-world environments. In turn, these models provide an incomplete picture of the potential weaknesses and are insufficient for developing an overall plan to address human-based risks.

A further obstacle to effective mitigation is the lack of standardized methodologies for measuring and integrating human factors across various cyberthreats. The variability of modern attacks and the increasing complexity of digital ecosystems require a unified paradigm that can categorize human weaknesses comprehensively and methodically. This kind of framework would serve as a platform for developing specific interventions and policies to minimize human-related cyber risks.

\subsection{Contributions}
To mitigate the current limitations of existing cybersecurity frameworks, we propose \textit{MORPHEUS} (Multidimensional framework for Organizational Resilience and Psychology-based Human factors for Effective Understanding of Security). It constitutes a comprehensive synthesis and multidimensional model of human factors within cybersecurity, thereby serving as a systematic lens for proactive strategies that improve overall organizational resilience. In this way, MORPHEUS aims at contributing to the existing state-of-the-art by the following means: 

\begin{enumerate}

\item Identification, through a systematic scoping review, of \textbf{50 human factors} and their explicit mapping to \textbf{six primary cyberthreats} (phishing, SMishing, spear-phishing, malware downloads, password management, and system misconfigurations). By anchoring human vulnerabilities to the most critical risks identified in recent landscape reports~\cite{Verizon2025, Clusit2025, ENISA2025, Microsoft2025, OWASP2021}, this contribution provides the empirical foundation for prioritizing mitigation strategies against real-world attack vectors.

\item Development of the \textbf{MORPHEUS Causal Pathway Architecture}. By integrating the CAB model and Attribution Theory, the framework organizes the 50 identified factors into a hierarchical causal system. It distinguishes between distal \textit{Modulators} (e.g., Personality, Demographics, Social/Organizational context) and proximal \textit{Direct Factors} (e.g., Cognitive, Affective, and Behavioral states), shifting the paradigm from a flat taxonomy to a dynamic, structural model of human vulnerability.

\item Identification and systematic mapping of \textbf{302 empirical interactions} among human factors. Crucially, we use this dataset to quantitatively validate the Causal Pathway Architecture, demonstrating an \textbf{82.8\% structural compliance} with the theoretical top-down flow. We further distill this complex network into \textbf{12 key interaction mechanisms} that explain the recurring failure pathways and feedback loops governing user insecurity.

\item Identification of \textbf{99 measurement solutions} for assessing human factors. This is critical in reframing insights into action and enabling organizations to track and assess the progress of their mitigation plan over time.

\item Definition of \textbf{in-depth operational scenarios} (supported by an extended appendix catalog), which can guide researchers and practitioners in practically using the framework to perform human-centric risk diagnosis, targeted interventions, risk prioritization, and integration into existing security practices.

\end{enumerate}

\begin{figure}[ht!]
    \centering
    \includegraphics[width=\textwidth]{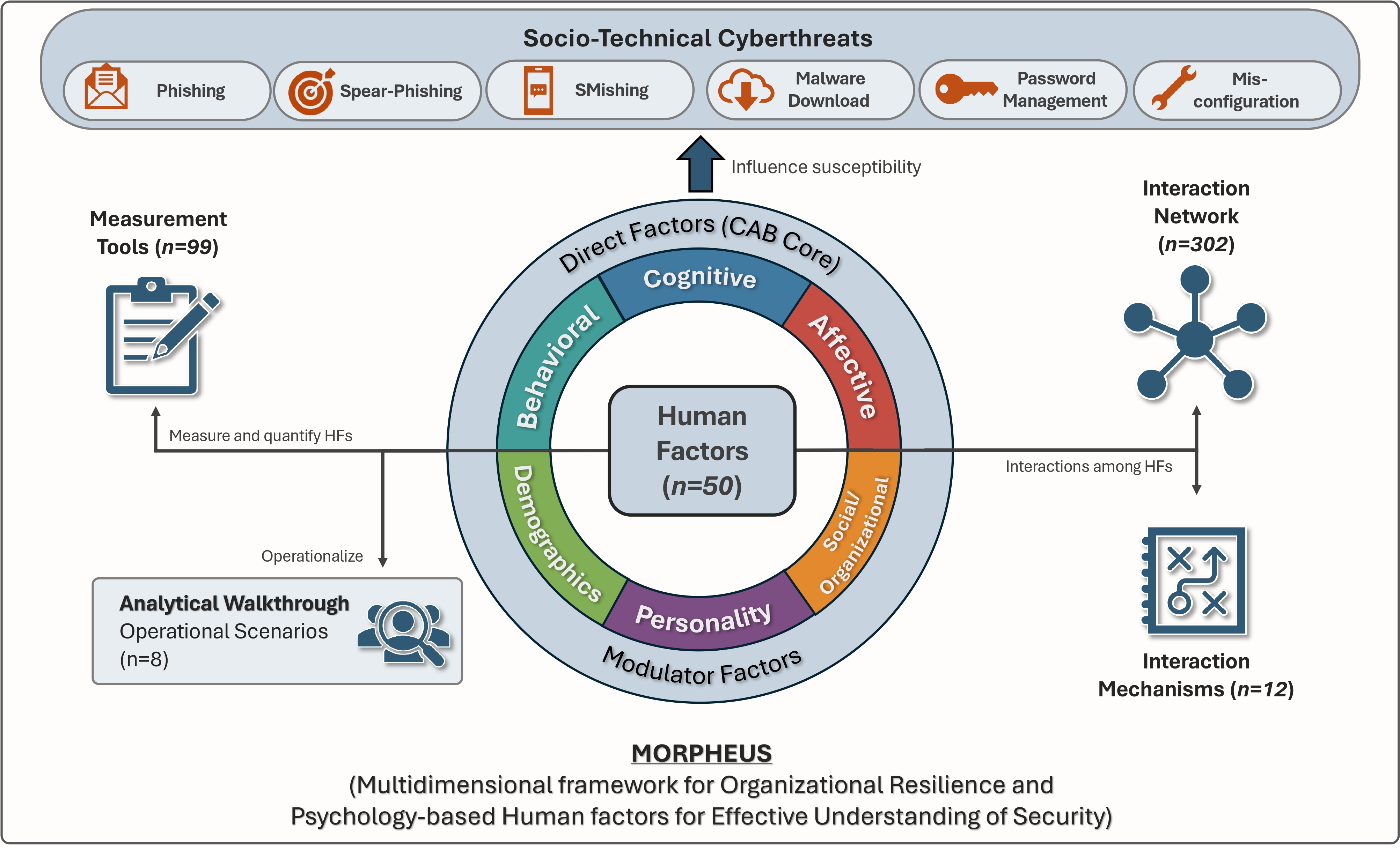} 
    \caption{\textbf{High-level overview of the MORPHEUS Framework.} 
    The central hub organizes $n=50$ human factors across six dimensions, distinguishing between proximal \emph{Direct Factors} aligned with the CAB Core (Cognition, Affect, Behavior) and distal \emph{Modulators} (Personality, Demographic, Social/Organizational). 
    The 50 human factors influence the susceptibility to the six cyberthreats treated in MORPHEUS. 
    The framework comprises an inventory of $n=99$ validated measurement solutions (Left) and a network of $n=302$ documented interactions among factors, together with $n=12$ Interaction Mechanisms (Right). The framework also comprises an Analytical Walkthrough (Bottom-Left) to operationalize the framework, also presenting $n=8$ operational scenarios for applying the framework in risk diagnosis and intervention.}
    \label{fig:morpheus_overview}
    \Description{A comprehensive diagram illustrating the MORPHEUS framework architecture. At the center is a circular hub labeled "HUMAN FACTORS (n=50)" containing a triangle divided into "Cognition," "Affect," and "Behavior," representing the CAB model's proximal processing structure. Surrounding this core is a segmented ring representing the six dimensions of human factors: Cognitive (blue), Affective (red), Behavioral (orange), Personality (purple), Demographic (green), and Social/Organizational (teal). These dimensions are categorized as either "Direct factors (proximal)" or "Modulators (distal antecedents)." Four arrows radiate outward from this central hub to four rectangular panels representing the framework's operational layers: Top Panel (cyberthreats): Labeled "cyberthreats exploited by / mapped to," displaying icons for Phishing, Spear-phishing, SMishing, Malware download, Password misuse, and Misconfiguration. Left Panel (Measurement): Labeled "Measurement Layer (n=99 measurement solutions)," showing a grid of icons representing diverse assessment tools like surveys, eye-tracking, and checklists. Right Panel (Interactions): Labeled "Interactions among factors (n=302 interactions)," depicting a complex network graph of interconnected nodes representing the relationships between factors. Bottom Panel (Operational Use Cases): Labeled "Operational Use Cases / Scenarios," listing eight workflows: Human-Centric Risk Diagnosis, Guiding Targeted Interventions, Risk Prioritization via Threat Mapping, Monitoring with Validated Tools, Psychological Red Teaming, Blameless Post-Incident Forensics, Breaking the Habitual Loop in Access Control, and Insider Threat Pre-emption.}
\end{figure}

To operationalize these contributions, MORPHEUS is structured as a holistic multidimensional framework, as visualized in Figure \ref{fig:morpheus_overview}. It is envisioned as a concentric architecture anchored by a central hub of $n=50$ Human Factors. Here, the CAB model (Cognition, Affect, Behavior) provides the proximal processing structure, surrounded by six dimensions where factors function as either Direct influences or distal Modulators. 
This core dictates susceptibility to specific mapped cyberthreats (top) and is quantified through a comprehensive Measurement Layer comprising $n=99$ validated solutions (left). 
Furthermore, the framework captures the complexity of behavioral dynamics by mapping a dense Network of Interactions $n=302$ relationships (right), channeling these insights into applied Operational Scenarios (bottom), such as risk diagnosis and targeted interventions. 
By integrating these layers, MORPHEUS addresses current gaps in the literature, providing actionable guidance for constructing resilient, human-centric cybersecurity defenses.

\section{Background and related work}
\label{sec:relwork}

\subsection{Frameworks of human factors in cybersecurity}

Understanding human factors in cybersecurity helps address vulnerabilities arising from technical flaws, human errors, and behaviors that may expose organizations to threats. Previous research in HCI and cybersecurity has proposed various frameworks to systematically analyze the impact of human factors on cybersecurity outcomes.

The \textit{Drivers of Insider Threats Framework} \cite{green2023understanding} categorizes the factors that lead to insider threats into themes such as technological awareness, leadership influence, and individual selfishness. While highly effective for analyzing internal malicious actors, its specific focus naturally limits the exploration of exhaustive interactions between factors and the broader, indirect role of organizational culture across different types of threats.

The \textit{Time Pressure Cybersecurity Behavior Framework} \cite{chowdhury2020time} revolves around time pressure as a driver of insecure behaviors, identifying scenarios, psychological mechanisms, and moderating factors, including task complexity, personal characteristics, and work environments. Given its specific focus on temporal constraints, this model is highly specialized; however, expanding its application requires integrating it with other resource constraints, such as inadequate tools or expertise, which also significantly impact security behaviors.

In contrast, the \textit{NIST Cybersecurity Framework} \cite{shen2014nist} integrates human factors into a broader framework by identifying twenty critical human behaviors, including password management and incident reporting. This framework provides an excellent high-level organizational standard. However, because it is designed for broad compliance rather than psychological modeling, it tends to categorize behaviors statically, which makes it challenging to capture the fluid and situation-specific occurrences of intentional deviations from standard practices. 


Al-Darwish and Choe \cite{aldarwish2019framework} presented an integrative information security framework explicitly addressing human factors alongside technical, organizational, and economic considerations. Their approach extends beyond traditional security methods and encompasses both direct human factors (e.g., errors, awareness, stress, negligence, training, and usability) and indirect factors (e.g., organizational culture, management support, communication, budgeting, and policy enforcement). The model successfully emphasizes human behavior as the primary factor in security breaches within organizational and economic contexts, though its macro-level perspective intentionally leaves highly individualized factors, such as specific personality traits, outside its scope.

According to the \textit{Cybersecurity Culture} (CSC) model \cite{mwim2023conceptual}, which relies on the \textit{Human Factor Diamond} framework, organizational and individual factors are divided into areas such as management, preparedness, and responsibility. The CSC model offers a robust perspective on structural preparedness. Because it is tailored to evaluate formal arrangements (policies and leadership), it naturally places less emphasis on the informal, affective, or personality-based factors that drive individual, moment-to-moment cybersecurity behaviors.

Choong \cite{choong2014cognitive} designed a cognitive-behavioral model centered on the password management lifecycle (i.e., generation, maintenance, and authentication phases). This model provides a highly valuable and deep analysis of its specific domain; however, its targeted design means that translating its insights to other cybersecurity-related settings requires a broader abstraction.

More recently, Khadka \& Ullah \cite{khadka2025human} presented a comprehensive interdisciplinary review and proposed a \textit{Human-Centric Cybersecurity Framework} composed of four pillars: (i) Psychological Resilience, which aims to strengthen individuals' ability to deal with stress, fear, fatigue, and affective manipulation; (ii) Adaptive Learning, which promotes dynamic and context-sensitive security education; (iii) Emotional Intelligence, which focuses on enhancing users' capacity to detect and interpret social engineering tactics; and (iv) Socio-Technical Integration, which aims to ensure that technical controls, organizational processes, and human-centered practices are aligned. This framework provides a solid conceptual foundation for interdisciplinary integration. However, as its goal is to establish macro-level pillars, it leaves room for future work to map the granular influence of individual human factors on specific cyberthreats and their interrelationships.

Despite the presence of several models and frameworks on human factors in cybersecurity, there is an evident lack of a holistic framework that provides a broader perspective on the human factors involved in the most common cyberthreats, that investigates the interaction among these factors, and that drives practitioners in measuring them. All of this is necessary to deliver actionable insights that can be used to construct a more robust cybersecurity defense.

\subsection{Existing Surveys or literature reviews on human factors in cybersecurity}
\label{sec:existing_surveys}

One of the primary contributions of our study is a literature review aimed at identifying human factors in cybersecurity. This activity is motivated by the opportunity to complement the deep, domain-specific insights of available reviews with a broader, integrated approach. For instance, Rahman et al. \cite{rahman2021human} investigated the human factors that significantly influence cybersecurity outcomes. The thorough examination of the concept of social influence is a notable feature of this work, as the authors emphasize the importance of users adopting secure practices in case people with similar habits support them. The study identifies a lack of established theoretical frameworks that could provide further insights into the nature of user behavior and improve its prediction.

A comprehensive and systematic review by Desolda et al. explores the interplay between human factors and phishing attacks \cite{desolda2021human}. The authors identified important issues affecting vulnerability to such attacks in groups, including a lack of security knowledge and training that keeps users uninformed about existing threats and how to protect themselves effectively. This work is supported by the thorough classification of essential factors, including the absence of security knowledge and overconfidence. Given its specific focus on phishing, this review provides profound insights into social engineering, paving the way for extending similar analyses to other cyberthreats.

In the overview of phishing training efficiency, Jampen et al. \cite{jampen2020don} identify various human and situational factors that contribute to phishing vulnerability. However, despite the thorough analysis offered by the authors, the findings on aspects such as age, gender, and user technical knowledge are somewhat contradictory in some cases, indicating that the issue requires further research to clarify these points.

A systematic review about the impact of human factors on malware and ransomware attacks has also been reported in~\cite{levesque2018technological}. According to this study, human traits play a significant role in the success of malware attacks. The actions of users, including browsing the internet, downloading files, and configuring security, directly influence malware vulnerability. Surprisingly, people with advanced computer knowledge proved to be more vulnerable, which is explained by their tendency to overestimate their knowledge and take more risks, or by being exposed to threats more frequently. It is also reported that peer-to-peer activity and high network usage can make the system more vulnerable, as well as specific behaviors, like failing to update the system or interacting with harmful websites. 

Similarly, the systematic review by Nifakos and colleagues \cite{nifakos2021influence} examines the growing intersection between human behavior and cybersecurity in healthcare environments, acknowledging that digital transformation has significantly increased the sector's vulnerability to cyberthreats. Phishing and other social engineering attacks have proven to be among the most prevalent and damaging types of cyberthreats in the healthcare sector. 
One of the key findings of the review is the recognition that healthcare professionals often unwittingly serve as entry points for cybercriminals. This can partly be explained by the fact that the healthcare setting is a high-stress environment, and employees are often overburdened, distracted, or uninformed about the specific methods attackers can use to target them. The review notes that healthcare workers are particularly vulnerable to such schemes, not necessarily because they are careless, but because they often lack adequate cybersecurity knowledge and its associated training. 
 
The review by Allendoerfer \cite{allendoerfer2005human} highlights various human factors that significantly impact the effectiveness of password security systems. Although human memory is strong, it is not resistant to lengthy and complex passwords, and a user must remember several distinct credentials to access various systems. It is also possible that cognitive constraints, including confusion among systems and difficulty remembering passwords that are not frequently used, may compromise password security. 

Finally, Tornblad et al. \cite{tornblad2021characteristics} reviewed prior studies to identify 32 human factors predicting phishing susceptibility, grouped into seven categories: personality traits, demographics, education, cybersecurity experience and beliefs, platform experience, email behaviors, and work commitment style. The review found no single strong predictor, but combinations of traits, such as impulsivity, trust, low awareness, and habitual platform use, tend to increase vulnerability.

In addition to threat-centric reviews, recent studies have produced highly valuable surveys focusing on specific micro-level dimensions of human factors. For instance, Burda et al. \cite{burda2024cognition} provided a comprehensive systematic review of cognitive factors in social engineering, mapping the information-processing stages involved in susceptibility. Similarly, Chen et al. \cite{chen2025deterrence} systematically reviewed the role of autonomous motivation, finding 17 factors that drive compliance with security policies, spontaneous secure behavior, or policy violation. Moreover, von Preuschen et al. \cite{vonpreuschen2024beyond} explored the holistic temporal lifecycle of emotions in cybersecurity, moving beyond the effects of in-the-moment fear and frustration to consider consequences and spill-over effects across behavioral, cognitive, and social dimensions. Finally, Borgert et al. \cite{Borgert2024} rigorously assessed the methodological approaches to assess cybersecurity self-efficacy by conducting a systematic literature review; the authors identified 276 variables as possible causes or outcomes of self-efficacy, along with 13 intervention designs. 
These dimension-specific surveys offer profound deep dives into individual psychological mechanisms; however, their scope remains bounded to their respective dimensions.

While the literature on human factors in cybersecurity is extensive---spanning both threat-specific and dimension-specific reviews---no comprehensive work exists that integrates these micro-level insights across all dimensions (cognitive, affective, behavioral, personality, and social) into a unified, theoretical framework covering multiple cyberthreats. Thus, with this work, we aim to fill this gap by identifying the most prevalent threats that exploit human errors and the human factors that may contribute to such errors, synthesizing them within the MORPHEUS framework.
We will detail the methodology in the following section.

\section{Methodology}
To develop the MORPHEUS framework, we employed a multi-methodological approach tailored to address three research objectives: identifying the core human factors that play a crucial role in the most significant cyberthreats, mapping their complex interactions, and retrieving validated instruments to measure these factors.

To identify and define the foundational taxonomy of human factors and align them with existing theoretical models, we adopted a \textit{Systematic Scoping Review}, ensuring qualitative depth and conceptual consistency (Section~\ref{sec:scoping_review}). Subsequently, to address the scalability challenge of mapping the extensive network of potential interactions among these identified factors and retrieving specific psychometric tools, we implemented an \textit{AI-assisted Systematic Screening} \cite{galli2025large, van2023artificial}. This second phase integrated AI-driven semantic search with strict Human-in-the-Loop (HITL) validation to manage the high-dimensional search space without compromising rigor (Section~\ref{sec:hybrid_protocol}). The following subsections detail the procedural steps, inclusion criteria, and validation mechanisms for each strategy. The complete dual-phase process, including the document selection at each screening step, is visually summarized in Figure~\ref{fig:methodology_flow}. All the datasets, protocols, and validation logs related to these activities are reported in the Additional Material.

\begin{figure*}[ht]
\centering
\resizebox{0.98\textwidth}{!}{%
\begin{tikzpicture}[
    node distance=0.8cm and 1.2cm,
    >=Latex,
    font=\small\sffamily,
    main_p1/.style={rectangle, draw=blue!80!black, fill=blue!5, thick, rounded corners=4pt, minimum width=4.2cm, minimum height=1cm, text width=4cm, align=center},
    main_p2/.style={rectangle, draw=purple!80!black, fill=purple!5, thick, rounded corners=4pt, minimum width=4.2cm, minimum height=1cm, text width=4cm, align=center},
    excl_p1/.style={rectangle, draw=black!50, fill=white, dashed, rounded corners=2pt, minimum width=3.2cm, minimum height=0.8cm, text width=3cm, align=center, font=\scriptsize},
    excl_p2/.style={rectangle, draw=black!50, fill=white, dashed, rounded corners=2pt, minimum width=3.2cm, minimum height=0.8cm, text width=3cm, align=center, font=\scriptsize},
    result/.style={rectangle, draw=red!80!black, fill=red!5, thick, rounded corners=4pt, minimum width=4.2cm, minimum height=1.2cm, text width=4cm, align=center},
    arrow/.style={->, thick, draw=black!70}
]

\node[main_p1, font=\bfseries] (p1_start) {Phase 1: Systematic Scoping Review};
\node[main_p1, below=0.6cm of p1_start] (p1_s0) {Initial Retrieval\\(Google Scholar)\\$n=6,000$ candidates};

\node[main_p1, below=1.2cm of p1_s0] (p1_s1) {Title \& Abstract Screening\\$n=1,476$ retained};
\draw[arrow] (p1_s0) -- (p1_s1) coordinate[midway] (m1_1);
\node[excl_p1, left=0.8cm of m1_1] (e1_1) {Excluded (Off-topic)\\$n=4,524$};
\draw[arrow] (m1_1) -- (e1_1.east);

\node[main_p1, below=1.2cm of p1_s1] (p1_s2) {Cross-reference with Prior Reviews\\$n=1,035$ retained};
\draw[arrow] (p1_s1) -- (p1_s2) coordinate[midway] (m1_2);
\node[excl_p1, left=0.8cm of m1_2] (e1_2) {Excluded (Already in existing reviews)\\$n=441$};
\draw[arrow] (m1_2) -- (e1_2.east);

\node[main_p1, below=1.2cm of p1_s2] (p1_s3) {Full-Text Eligibility\\$n=55$ eligible};
\draw[arrow] (p1_s2) -- (p1_s3) coordinate[midway] (m1_3);
\node[excl_p1, left=0.8cm of m1_3] (e1_3) {Excluded (Failed IN1-IN4 criteria)\\$n=980$};
\draw[arrow] (m1_3) -- (e1_3.east);

\node[main_p1, below=0.8cm of p1_s3] (p1_s4) {Snowballing Augmentation\\36 identified; $+33$ unique additions};
\draw[arrow] (p1_s3) -- (p1_s4);

\node[result, below=0.8cm of p1_s4] (p1_res) {\textbf{Final Corpus (Phase 1)}\\88 Unique Publications $\rightarrow$ \textbf{50 Factors}};
\draw[arrow] (p1_s4) -- (p1_res);

\node[main_p2, font=\bfseries, right=3.5cm of p1_start] (p2_start) {Phase 2: AI-Assisted Screening};
\node[main_p2, below=0.6cm of p2_start] (p2_s1) {LLM Retrieval (Prompts)\\$n=338$ candidates};

\node[main_p2, below=1.2cm of p2_s1] (p2_s2) {HITL Hallucination Check\\$n=307$ verified};
\draw[arrow] (p2_s1) -- (p2_s2) coordinate[midway] (m2_1);
\node[excl_p2, right=0.8cm of m2_1] (e2_1) {Excluded (Hallucinated)\\$n=31$};
\draw[arrow] (m2_1) -- (e2_1.west);

\node[main_p2, below=1.2cm of p2_s2] (p2_s3) {Quality Filter (Q3+/Core C+)\\$n=289$ eligible};
\draw[arrow] (p2_s2) -- (p2_s3) coordinate[midway] (m2_2);
\node[excl_p2, right=0.8cm of m2_2] (e2_2) {Excluded (Failed quality)\\$n=18$};
\draw[arrow] (m2_2) -- (e2_2.west);

\node[main_p2, below=1.2cm of p2_s3] (p2_s4) {Relevance Assurance (Full-Text)\\$n=178$ relevant};
\draw[arrow] (p2_s3) -- (p2_s4) coordinate[midway] (m2_3);
\node[excl_p2, right=0.8cm of m2_3] (e2_3) {Excluded (Misattribution)\\$n=111$};
\draw[arrow] (m2_3) -- (e2_3.west);

\node[main_p2, below=0.8cm of p2_s4] (p2_s5) {Snowballing Augmentation\\$+93$ relevant papers};
\draw[arrow] (p2_s4) -- (p2_s5);

\node[result, below=0.8cm of p2_s5] (p2_res) {\textbf{Final Corpus (Phase 2)}\\271 Publications $\rightarrow$ \textbf{302 Interactions} \& \textbf{99 Tools}};
\draw[arrow] (p2_s5) -- (p2_res);

\draw[->, dashed, line width=1.5pt, red!70, rounded corners=5pt] (p1_res.east) -- ++(1.0,0) |- (p2_start.west) node[pos=0.25, right, font=\scriptsize\bfseries, text=red!80!black, align=left, xshift=1mm] {Identified factors\\drive the LLM\\search prompts};

\end{tikzpicture}
}
\caption{PRISMA-style flowchart of the dual-phase methodology. Phase 1 (left) details the Systematic Scoping Review to identify the 50 human factors. Phase 2 (right) depicts the AI-Assisted Systematic Screening, explicitly showing the attrition during the Human-in-the-Loop (HITL) validation protocol to map interactions and retrieve measurement tools.}
\label{fig:methodology_flow}
\Description{This diagram visually maps both Phase 1 (Systematic Scoping Review) and Phase 2 (AI-Assisted Systematic Screening) side-by-side. It explicitly details the flow of documents, the attrition numbers at each exclusion step, and crucially, the Human-in-the-Loop (HITL) verification checkpoints used to validate the AI-retrieved data.}
\end{figure*}

\subsection{Identifying key human factors in cybersecurity: a systematic scoping review}
\label{sec:scoping_review}
A \textit{systematic scoping review} was carried out to answer the following research question: \textbf{``What human factors play a critical role in cyber-attacks?''}. The scoping review approach was selected to map concepts across multidisciplinary fields and update earlier reviews \cite{tricco2018prisma, petersen2015guidelines, munn2018systematic}. This methodology allowed us to efficiently expand upon existing systematic reviews (e.g., \cite{desolda2021human, tornblad2021characteristics, jampen2020don, nifakos2021influence, levesque2018technological}) while simultaneously exploring under-researched areas like misconfiguration.

This review focuses on the most significant attacks in which human behavior and decision-making are considered particularly impactful: \textit{phishing, SMishing, spear phishing, malware download, password management, and misconfiguration}. The selection of these six threats is grounded in their consistent classification as the most widespread and critical risks in the modern threat landscape, as explicitly demonstrated in authoritative industry analyses---such as the Verizon Data Breach Investigations Report (DBIR)~\cite{Verizon2025} and the Clusit Report~\cite{Clusit2025}. We strategically selected these six because they map to the primary ways human vulnerabilities are exploited across an attack's lifecycle: initial access (phishing, spear-phishing, SMishing), execution (malware downloads), credential compromise (password management), and environmental vulnerability (misconfigurations). In fact, these vectors remain the primary drivers of successful security breaches globally compared to purely technical vulnerabilities, as highlighted by the ENISA Threat Landscape, the Microsoft Digital Defense Report, and OWASP~\cite{ENISA2025, Microsoft2025, OWASP2021}.

\subsubsection{Data source and search strategy}

We relied on \textit{Google Scholar} as the primary search engine thanks to its broad, cross-disciplinary coverage—spanning cybersecurity, psychology, HCI, and behavioral science. Moreover, its choice was motivated by its ability to index publications across journals, conferences, workshops, and technical venues. This choice aligns with the goals of a scoping review to capture diverse literature and complements prior systematic reviews \cite{desolda2021human, tornblad2021characteristics, jampen2020don, nifakos2021influence, levesque2018technological}, which had already screened the major disciplinary databases (e.g., ACM DL, IEEE Xplore, Scopus). To mitigate the limitations of relying on a single search engine and ensure robustness, we additionally performed systematic backward and forward snowballing on all included papers.

The search was conducted between March and July 2025. No lower bound was imposed on the publication year, as our goal was to capture all potentially relevant studies that connect human factors to the selected cyberthreats.

The formulation of the search string was an iterative process conducted by the research team. Following consensus meetings to identify the most relevant terminology, the search terms were derived from the terminology and citation trails reported in previous reviews. These existing reviews served as a starting point to determine proper keywords, citation trails, and under-researched subtopics. Specifically, our strategy prioritized cross-disciplinary coverage by systematically extracting the most frequently occurring keywords from these foundational papers. These terms were then deliberately selected and grouped to capture three intersecting core concepts: the human dimension, the security event, and the targeted threat vector. The resulting search string was:

\begin{tcolorbox}[colback=llgray,title=Search String ]
\begin{center}
  (``human factors'' \textit{OR} ``psychological traits'' \textit{OR} ``human aspects'') \textit{AND} (``cyberattack'' \textit{OR} ``attack'' \textit{OR} ``incident'' \textit{OR} ``threat'') AND $attack\_type$ 
  
\end{center}
\end{tcolorbox}

where $attack\_type$ was systematically replaced with each of the specific cyberthreats investigated in this review: ``phishing'', ``SMishing'', ``spear phishing'', ``malware download'', ``password management'', and ``misconfiguration''. Consequently, we executed the search strategy six independent times---substituting the $attack\_type$ variable with the respective threat keyword in each iteration---obtaining six distinct lists of results. We deliberately used these established base terms rather than exhaustively listing all morphological variations and synonyms (e.g., ``misconfigured'', ``configuration errors''). This decision leveraged Google Scholar's native semantic matching capabilities, while the subsequent systematic snowballing phase acted as a structural net to capture relevant studies employing alternative terminologies.

Furthermore, it is worth noting that we intentionally omitted umbrella terms such as ``cyber security'' or ``cybersecurity'' from the core search string. This methodological decision was driven by the interdisciplinary nature of the relevant literature. Behavioral science, psychology, and HCI publications often examine specific vectors (e.g., referring directly to ``phishing attacks'' or ``password management threats'') without explicitly indexing the broader term ``cybersecurity'' in their titles or abstracts. Forcing the inclusion of ``cybersecurity'' via an \textit{AND} operator would have artificially constrained the search space, potentially excluding highly relevant human-centric studies. Instead, the combination of threat-related keywords (``attack'', ``incident'', ``threat'') with the specific $attack\_type$ successfully anchored the retrieval process to the correct domain without inadvertently limiting its scope.

\subsubsection{Inclusion Criteria}

The selection procedure to include an article in our review was based on these inclusion criteria: 
\begin{itemize}
    \item [IN1] The publication must discuss the impact of at least one human factor, not in a generic sense, but related to one or more cyberthreats. 
    \item [IN2] The publication must be peer-reviewed (we exclude gray literature, such as unpublished manuscripts, master's or doctoral theses, white papers, technical reports, and non-peer-reviewed preprints).
    \item [IN3] The publication must be a full paper (not a short paper, demo, poster, or workshop paper) published in recognized venues. Given the interdisciplinary nature of this research, we did not restrict the selection to a predefined list of disciplines or computer science venues. Instead, publications from any relevant field (e.g., Cybersecurity, HCI, Psychology, Behavioral Sciences) were included, provided the venue met a universal quality baseline: specifically, journals ranked at least Q3 in Scimago, or conferences ranked as at least class C in the Core Conference Portal.
    \item [IN4] The publication must be written in English. 
\end{itemize}

\subsubsection{Screening and selection process}

The search was conducted by running six queries on Google Scholar using the \textit{Publish or Perish} tool. For each of the six cyberthreats considered in this review, we adopted the following selection procedure:
\begin{enumerate}
\item We examined the retrieved publications (n = 6000, 1000 per query) by reading their titles and abstracts until the list of results no longer yielded relevant papers; from this phase, we obtained a total of \textbf{1476} results (phishing = 391, spear-phishing = 207, SMishing = 242, malware download = 228, password management = 197, misconfiguration = 211).
\item We excluded all articles already reported in existing reviews and surveys \cite{tornblad2021characteristics, desolda2021human, levesque2018technological, jampen2020don, nifakos2021influence, allendoerfer2005human}. To execute this systematically, we compiled a master bibliography of all primary studies synthesized in these foundational reviews and cross-referenced our newly retrieved results against this list, filtering out direct matches. From this phase, we obtained a total of \textbf{1035} results (phishing = 349, spear-phishing = 96, SMishing = 170, malware download = 115, password management = 135, misconfiguration = 170). This exclusion step does not imply disregarding the contribution of previous primary studies. Instead, we systematically extracted all the human factors reported in those prior reviews and integrated them into our analysis. The primary concern with re-including these previously reviewed primary studies is methodological redundancy: evaluating them again would constitute ``double-counting'' of evidence already rigorously synthesized by the community, thereby skewing the emphasis toward older findings. The current review, therefore, focuses primarily on studies published after the temporal coverage of the previous reviews, or not considered in these reviews, while still incorporating the human-factor evidence identified by them. 
\item The remaining candidate publications were assessed by reviewing their abstracts, introductions, and conclusions against the inclusion criteria. When necessary, the full text was read to determine eligibility. From this phase, we obtained a total of \textbf{55} articles, categorized as follows: phishing (17), spear-phishing (6), SMishing (9), malware download (10), password management (3), and misconfiguration (10).
\item A backward and forward snowballing phase was conducted, identifying \textbf{36} relevant publications. After deduplication against the manually screened corpus, \textbf{33} of these publications were retained as additional unique publications, bringing the Phase 1 corpus from 55 to \textbf{88} unique publications. It should be noted that during this specific snowballing phase, to streamline the process, we did not manually cross-reference all newly surfaced citations against the master list of previously reviewed papers. Consequently, a few papers overlapped with the manually screened corpus; however, as our framework maps qualitative characteristics rather than quantitative incidence, this overlapping does not skew the resulting taxonomy of 50 factors.
\end{enumerate}

The final set consists of \textbf{88} unique publications.
It is worth noting that the final number refers to unique publications after deduplication. A single paper in this final set may address multiple cyberthreats (e.g., phishing and SMishing), and a single paper can discuss multiple human factors. 

All the human factors that emerged in this process were then combined with the human factors already presented in the existing reviews and surveys \cite{desolda2021human, levesque2018technological, tornblad2021characteristics, jampen2020don, nifakos2021influence, allendoerfer2005human}, resulting in a total of \textbf{50 human factors}. To prevent inconsistent evidence granularity between newly extracted factors and those inherited from prior reviews, all identified factors underwent a rigorous process of conceptual harmonization. Rather than passively adopting legacy classifications, every factor—regardless of its origin—was uniformly standardized, redefined, and scoped according to our unifying theoretical architecture (the CAB model~\cite{breckler1984empirical} and Attribution Theory~\cite{heider2013psychology} dimensions). This ensured that the final taxonomy of 50 factors operates at a single, consistent level of conceptual granularity.

The next section presents all the resulting human factors, detailing how they are organized into broader dimensions, mapped to the 6 cyberthreats of this study, and how these elements informed the construction of the proposed framework of human factors in cybersecurity. 

\paragraph{Screening Reliability.}
Two researchers independently screened titles, abstracts, and, when necessary, full texts. 
Disagreements were resolved through discussion, and when consensus could not be reached, a third senior researcher acted as arbiter. 
Inter-rater agreement on inclusion/exclusion decisions was substantial, with Cohen’s $\kappa = .70$, which is consistent with similar reviews in HCI and usable security.

\subsubsection{Methodological considerations and limitations}

The scoping review method supported the findings of relevant studies; however, some limitations should be considered. \textit{Google Scholar} may have resulted in overlooking some relevant studies indexed in other databases or published in more specialized venues (although this is quite rare, it can still occur). 

Furthermore, the foundational selection of our three core search terms (``human factors,'' ``psychological traits,'' and ``human aspects'') warrants critical reflection. These specific macro-level terms were strategically chosen because they represent the established umbrella taxonomies across our target interdisciplinary domains: ``human factors'' is the standard IEEE/ACM/NIST classification, ``psychological traits'' captures the behavioral science literature, and ``human aspects'' is a standardized construct in usable security (e.g., the widely adopted HAIS-Q instrument~\cite{parsons2014haisq}). Starting with these broad umbrella terms, rather than a granular list of specific traits (e.g., ``stress'' OR ``neuroticism''), prevented us from introducing an \textit{a priori} bias where we would only retrieve the specific factors we already knew to look for.

Moreover, although our search string explicitly included these three macro-terms, we acknowledge that relevant studies may refer to individual-level constructs using alternative terminology (e.g., ``individual differences,'' ``user characteristics'', or by naming a specific trait directly in the title without the umbrella term). To mitigate these risks and prevent foundational bias, backward and forward snowballing were systematically applied to every included paper. Crucially, our baseline also integrated all human factors previously identified in the foundational systematic reviews \cite{desolda2021human, tornblad2021characteristics, jampen2020don, nifakos2021influence, levesque2018technological}. This tri-layered approach (umbrella keyword search + prior review integration + systematic snowballing) acted as a structural countermeasure, ensuring the inclusion of studies whose terminology differed from our initial search terms and guaranteeing the rigorous completeness of the 50 identified factors.

However, despite the focus of MORPHEUS on 6 specific cyberthreats that are mostly driven by human factors, the deliberate choice of excluding general terms (e.g., ``cybersecurity'', ``threat'', ``attack'', etc.) and other attacks (e.g., ``social engineering'', ``vishing'', etc.) may have led us to not consider human factors that affect susceptibility in either more generic or specific contexts. Future work can focus on extending the framework to include more specific cyberthreats, possibly also considering studies that provide evidence on the effects of human factors on generic cyber victimization, without being bound to a specific attack.

As a final limitation, focusing on recent literature and peer-reviewed sources, although they guarantee academic rigor, may have led to the exclusion of useful gray literature. Lastly, some variation in terminology and emphasis across papers was inevitable.

\subsection{AI-assisted Systematic Screening}
\label{sec:hybrid_protocol}
While the identification of the 50 human factors (Section \ref{sec:scoping_review}) relied on a manual review, the subsequent tasks — mapping pairwise interactions among factors and identifying measurement instruments — presented prohibitive scalability challenges. We estimate that manually executing these tasks would have required more than 2,500 researcher-hours. Specifically:
\begin{itemize}
    \item For \textbf{interactions}, the challenge was \textit{combinatorial volume}: managing over 2,500 interaction pairs would have required screening an estimated pool of 50,000+ titles/abstracts to determine relevance ($\approx$1,600 hours at 2 min/paper).
    \item For \textbf{measurement instruments}, the challenge was \textit{semantic depth}: distinguishing validated scales from theoretical discussions requires a deep methodological analysis. Processing just the top $\sim$55 relevant results for each of the 50 factors would have entailed examining $\approx$2,700 full texts ($\approx$900 hours at 20 min/paper).
\end{itemize}
To address these constraints in a manageable timeline while shifting human effort towards high-value validation, we employed an \textit{AI-assisted Systematic Screening} \cite{galli2025large, van2023artificial}. This approach integrates the semantic retrieval capabilities of Large Language Models (LLMs) with a strict Human-in-the-Loop (HITL) validation mechanism \cite{lieberum2025large, shah2024efficacy}.

We selected \textit{Perplexity.ai} and \textit{OpenAI Deep Research} as retrieval engines. This choice was driven by their capacity to minimize hallucinations compared to standard LLMs and their proven efficacy in complex scientific domains (e.g., health sciences and biology), where they have demonstrated superior accuracy in citation retrieval compared to keyword-based tools, while still needing heavy human supervision~\cite{liu2023evaluating, sawhney2023prometheus}. The search was conducted between March and July 2025.

The protocol follows a standardized workflow that comprises AI-assisted retrieval, a rigorous three-step validation process, and a snowballing phase, all of which are documented via a comprehensive audit trail.

\subsubsection{Phase 1: AI-assisted Retrieval}
Before starting the search, we conducted a \textit{pilot phase} to refine prompt engineering. Initially, generic prompts yielded broad but shallow results; we iteratively refined the structure to enforce \textit{evidence-based constraints}, requiring the model to extract ``direct quotes'' and ``methodological context'' alongside the citation. This constraint significantly reduced the retrieval of generic associations.

The final retrieval strategy was adapted to the specific topology of each research question:
\begin{itemize}
    \item \textit{Mapping pairwise interactions among factors (Combinatorial Discovery):} To cover the dense search space of over 2,500 potential combinations without overwhelming the AI's context window, we avoided asking the model to process all 50 factors simultaneously. Instead, we broke the task down into smaller, focused queries (a ``decomposed redundancy strategy''), designing six distinct prompts—one for each of the human factor dimensions. This forced the AI to conduct deep, exhaustive searches within one specific behavioral or cognitive category at a time.
    \item \textit{Identifying measurement instruments for the human factors (Targeted Retrieval):} Identifying psychometric tools required high precision to distinguish formally validated scales from generic ad-hoc surveys. To address the \textit{semantic filtering burden} and decrease the number of irrelevant papers merely mentioning the word ``survey'' coming from keyword searches, we employed a standardized prompt template. This single prompt structure was explicitly instructed to semantically extract validation metrics (e.g., ``validity'', ``reliability'', ``psychometric properties'') for the specific identified factors.
\end{itemize}

In both cases, each prompt has been executed iteratively five times to mitigate the non-deterministic nature of LLMs. We observed a complementary behavior between the two engines, driven by their underlying architectures: while \textit{Perplexity.ai} (employing Retrieval-Augmented Generation to ground responses in real-time web indexes) excelled in retrieving recent open-access literature via direct web crawling, \textit{Deep Research} (utilizing an agentic workflow capable of iterative searching and reading) demonstrated higher precision in parsing semantic relationships within older or paywalled abstracts, with a partial overlap in results that confirmed the necessity of a multi-engine approach. A total of 338 candidate publications (interactions = 251, measurements = 87) were retrieved. Prompts are reported in \hyperref[sec:appendix_4]{Appendix 4}.

\subsubsection{Phase 2: HITL Validation and Audit Trail}
Given the known risks of hallucination in generative models, \emph{no AI-retrieved publication was accepted without passing a rigorous manual validation}. To ensure robustness, we adopted a \textbf{double-blind independent review protocol}. Two researchers independently executed each filtering step, and discrepancies were resolved through iterative consensus meetings at the conclusion of each activity before proceeding to the next stage. This prevented error propagation and ensured that only fully validated entries advanced through the pipeline.

To ensure transparency and replicability, the entire process was thoroughly documented in the Additional Material. Starting from the initial pool of raw candidate publications retrieved, we applied the following three-step filtering process:

\begin{enumerate}
    \item \textbf{Hallucination Check (Verification):} Each of the 338 candidate studies was cross-referenced independently by both researchers against authoritative indexing services (e.g., Google Scholar, ACM Digital Library, IEEE Xplore). Entries that could not be verified or appeared to be ``hallucinated'' (i.e., having both non-existent titles and wrong authors, rendering the paper completely untraceable) were immediately discarded. This step identified 31 hallucinated entries (interactions = 19, measurements = 12), resulting in an observed Hallucination Rate of 9.2\% (interactions = 7.6\%, measurements = 13.8\%). This phase specifically targeted hallucinations manifested as \textit{Fabricated References}. Crucially, if a publication was real and verifiable (e.g., the title was correct and searchable), it was not discarded even if the AI generated completely wrong authors; instead, we retained the paper and manually corrected the metadata. For the surviving valid publications, this step also served to correct any other metadata inaccuracies.

    \item \textbf{Exclusion Criteria Filtering (Quality):} The remaining 307 verified papers (interactions = 232, measurements = 75) were subsequently assessed for bibliographic quality. We thus excluded 18 papers (interactions = 17, measurements = 1) that did not meet the publication standards (e.g., non-peer-reviewed, short papers, posters) or ranking thresholds (Scimago Q3+ or CORE Class C+).

    \item \textbf{Relevance Assurance (Scope):} The full texts of the surviving 289 publications (interactions = 215, measurements = 74) were independently examined by the two researchers to verify semantic consistency and identify \textit{Content Misattribution}. This qualitative screening ensured that (i) the paper actually investigated the specific interaction or measurement instrument, and (ii) the finding was relevant to the context. Following the consensus discussion for this final stage, a total of 111 publications were excluded (interactions
= 102, measurements = 9).
\end{enumerate}

\paragraph{Reliability}
While rigorous consensus checks were performed at the end of each factual verification step (Steps 1 and 2) to resolve objective discrepancies, the inter-rater agreement metrics reported here specifically refer to the \textbf{qualitative eligibility decisions} performed in Step 3, where subjective interpretation of relevance was required. The agreement was substantial (Cohen's $\kappa = .76$ for interactions, Cohen's $\kappa = .85$ for measurement tools). After resolving disagreements, the remaining valid studies (N=178; interactions = 113, measurements = 65) were processed for the final corpus.

\subsubsection{Phase 3: Snowballing Augmentation}
To mitigate potential algorithmic biases (e.g., \textit{popularity bias} toward highly cited papers, omission of older seminal works prior to digitization, or \textit{accessibility bias} toward open-access repositories), the AI-generated corpus was supplemented by a backward and forward snowballing phase. For measurement solutions, a further \textbf{34} publications were included, while, for interactions, a total of \textbf{59} new publications were added.
This unified protocol resulted in a final corpus of \textbf{271} publications (interactions = 172, measurements = 99), substantiating the \textbf{302} interactions (Section~\ref{sec:interplay}) and \textbf{99} measurement tools (Section~\ref{sec:hfmeasure}).

Crucially, we systematically cross-checked the 93 new publications discovered during these augmentation phases against our Phase 1 taxonomy. We confirmed that all human-centric variables discussed in this extended corpus successfully mapped onto our existing taxonomy of 50 factors (or their established sub-facets). While they provided much deeper evidence for interaction mechanisms and measurement tools, they did not introduce entirely new conceptual categories that necessitated expanding the baseline of 50 factors.

\subsubsection{Reflections on LLM-Assisted Reviews}

Given the growing adoption of LLM-assisted systematic reviews in the human-centered computing and cybersecurity communities, we reflect on our methodological experience to provide practical guidance for future researchers. The primary advantage of integrating LLMs (specifically Perplexity.ai and OpenAI Deep Research) was the ability to navigate a prohibitively large combinatorial search space and extract semantic context that traditional Boolean keyword searches often miss. We observed that the two engines played complementary roles: one excelled at real-time web crawling of recent open-access literature, while the other demonstrated superior semantic parsing of complex abstracts. This AI-driven triage drastically reduced the manual burden of initial identification.

However, this efficiency comes with significant caveats. First, the non-deterministic nature of LLMs means that identical prompts can yield different results; we mitigated this by running each prompt iteratively five times to sample a wider distribution and stabilize recall, effectively reducing the number of missed items. Second, the observed hallucination rate of 9.2\% highlights a persistent risk of fabricated references or misattributed findings. Consequently, our core recommendation for the community is that LLMs should never be viewed as autonomous screening agents. A strict Human-in-the-Loop (HITL) protocol remains absolutely non-negotiable to ensure \textit{precision} by eliminating false positives.

Finally, we must emphasize a critical limitation regarding \textit{recall}. While iterative prompting and HITL protocols validate the retrieved items, they cannot definitively rule out items that the AI simply misses (false negatives). To mitigate algorithmic omissions and popularity biases, researchers cannot rely solely on AI output; traditional backward and forward snowballing (as executed in our Phase 3) remains a mandatory human-driven step to recapture relevant literature. Acknowledging this intrinsic limitation of AI retrieval, we present our framework as a highly robust and representative mapping, while cautioning against claims of absolute exhaustion.

\section{MORPHEUS: a comprehensive framework of human factors in cybersecurity}
\label{sec:morpheus}

The results of the literature review led to the identification of 50 human factors contributing to cyberthreat susceptibility. Detailed definitions and theoretical backgrounds for each factor are reported in \hyperref[sec:appendix_1]{Appendix 1}.

To enhance clarity and organization, we classified these factors into six dimensions, i.e., Demographics, Personality Traits, Cognitive Factors, Affective Factors, Behavioral Factors, and Social/Organizational Factors (See Section~\ref{sec:classification}). Table~\ref{tab:hfs_vertical_larger} summarizes all the identified factors and their associations with the cyber threats, according to the results of the Systematic Scoping review (see \ref{sec:scoping_review}). More details are available in Table~\ref{tab:hfs} reported in the \hyperref[sec:appendix_1]{Appendix}, which details, for each human factor:
\begin{itemize}
    \item Their classification into one of the six dimensions;
    \item Their association with one or more cyberthreats;
    \item The publications supporting the connection(s) between that human factor and its associated cyberthreat(s);
    \item Their type (either \textit{internal} or \textit{external});
    \item Their functional role (either \textit{direct} or \textit{modulator}).
\end{itemize}
 The rationale and details of this classification are reported in the following sub-section.

 \begin{table}[htbp]
    \centering
    \footnotesize 
    
    \renewcommand{\arraystretch}{0.95} 
    
    \setlength{\tabcolsep}{2.5pt} 
    
    \caption{Compact Mapping of Human Factors to Cyberthreats. ($\bullet$) indicates documented evidence. ("Phish." = Phishing, "Spear." = Spear-phishing, "SMish." = SMishing, "Malw." = Malware Download, "Passw." = Password management, "Misco." = Misconfiguration). Full references in \hyperref[sec:appendix_1]{Appendix 1} in Table~\ref{tab:hfs}.}
    \label{tab:hfs_vertical_larger}
    
    \begin{tabular}{c l c c c c c c} 
        \toprule
        \textbf{Dim.} & \textbf{Human Factor} & \textbf{Phish.} & \textbf{Spear.} & \textbf{SMish.} & \textbf{Malw.} & \textbf{Passw.} & \textbf{Misco.} \\
        \midrule
        
        \parbox[t]{2mm}{\multirow{3}{*}{\rotatebox{90}{\textbf{Demog.}}}} 
        & Age & $\bullet$ & $\bullet$ & $\bullet$ & $\bullet$ & $\bullet$ & \\
        & Education & $\bullet$ & $\bullet$ & $\bullet$ & & & $\bullet$ \\
        & Gender & $\bullet$ & $\bullet$ & $\bullet$ & & & \\
        \midrule
        
        \parbox[t]{2mm}{\multirow{7}{*}{\rotatebox{90}{\textbf{Personality}}}} 
        & Agreeableness & $\bullet$ & $\bullet$ & & $\bullet$ & & \\
        & Conscientiousness & $\bullet$ & $\bullet$ & $\bullet$ & & & \\
        & Extraversion & $\bullet$ & $\bullet$ & $\bullet$ & & & \\
        & Greed & & & $\bullet$ & $\bullet$ & & \\
        & Openness & $\bullet$ & $\bullet$ & $\bullet$ & $\bullet$ & & \\
        & Narcissism & $\bullet$ & & & & & \\
        & Neuroticism & $\bullet$ & $\bullet$ & $\bullet$ & & & \\
        \midrule
        
        \parbox[t]{2mm}{\multirow{15}{*}{\rotatebox{90}{\textbf{Cognitive}}}} 
        & Bias & $\bullet$ & $\bullet$ & & & & $\bullet$ \\
        & Cognitive fatigue & $\bullet$ & & & & $\bullet$ & $\bullet$ \\
        & Cognitive reflectiveness & $\bullet$ & $\bullet$ & $\bullet$ & & & \\
        & Cyber risk beliefs & $\bullet$ & $\bullet$ & $\bullet$ & $\bullet$ & & \\
        & Cybersec. self-monitoring & & $\bullet$ & & & & \\
        & Decision fatigue & $\bullet$ & & & & & \\
        & Distraction & $\bullet$ & & & & & \\
        & Lack of awareness & $\bullet$ & $\bullet$ & $\bullet$ & $\bullet$ & $\bullet$ & $\bullet$ \\
        & Lack of knowledge & $\bullet$ & $\bullet$ & $\bullet$ & $\bullet$ & $\bullet$ & $\bullet$ \\
        & Misperception & & $\bullet$ & & & & $\bullet$ \\
        & Overconfidence & $\bullet$ & & $\bullet$ & $\bullet$ & & \\
        & Risk attitude & $\bullet$ & & & & & \\
        & Security self-efficacy & $\bullet$ & $\bullet$ & $\bullet$ & $\bullet$ & $\bullet$ & $\bullet$ \\
        & Uncertainty & $\bullet$ & $\bullet$ & & & & \\
        & Vigilance & $\bullet$ & $\bullet$ & & & & \\
        \midrule
        
        \parbox[t]{2mm}{\multirow{6}{*}{\rotatebox{90}{\textbf{Affective}}}}
        & Anxiousness & $\bullet$ & $\bullet$ & & & & \\
        & Digital anxiety & $\bullet$ & $\bullet$ & & & & \\
        & Fear & $\bullet$ & $\bullet$ & $\bullet$ & $\bullet$ & & \\
        & Frustration & & & & & $\bullet$ & $\bullet$ \\
        & Shame & & $\bullet$ & & & & $\bullet$ \\
        & Stress & $\bullet$ & & & & & \\
        \midrule
        
        \parbox[t]{2mm}{\multirow{8}{*}{\rotatebox{90}{\textbf{Behavioral}}}} 
        & Complacency & $\bullet$ & & & & $\bullet$ & \\
        & Compulsive behavior & $\bullet$ & & & & & \\
        & Impulsive behavior & $\bullet$ & $\bullet$ & & $\bullet$ & $\bullet$ & \\
        & Internet addiction & & & & & $\bullet$ & \\
        & Internet usage & $\bullet$ & $\bullet$ & $\bullet$ & $\bullet$ & & \\
        & Laziness & & & & & $\bullet$ & \\
        & Recurrence & $\bullet$ & $\bullet$ & & & & \\
        & Risk-taking & $\bullet$ & & & & & \\
        \midrule
        
        \parbox[t]{2mm}{\multirow{11}{*}{\rotatebox{90}{\textbf{Social \& Organiz.}}}} 
        & Attitude towards policies & $\bullet$ & $\bullet$ & & & $\bullet$ & \\
        & Lack of communication & $\bullet$ & $\bullet$ & & $\bullet$ & $\bullet$ & $\bullet$ \\
        & Lack of resources & $\bullet$ & $\bullet$ & $\bullet$ & $\bullet$ & $\bullet$ & $\bullet$ \\
        & Lack of trust & $\bullet$ & $\bullet$ & $\bullet$ & & $\bullet$ & $\bullet$ \\
        & Norms & $\bullet$ & $\bullet$ & & & $\bullet$ & \\
        & Online exposure & $\bullet$ & $\bullet$ & $\bullet$ & & $\bullet$ & \\
        & Organizational tenure & $\bullet$ & $\bullet$ & $\bullet$ & & & \\
        & Security posture & & & & $\bullet$ & & $\bullet$ \\
        & Social influence & $\bullet$ & $\bullet$ & $\bullet$ & & & \\
        & Social proof & & $\bullet$ & & & $\bullet$ & \\
        & Type of organization & & $\bullet$ & & $\bullet$ & $\bullet$ & \\
        \bottomrule
    \end{tabular}
\end{table}

\subsection{A classification of resulting human factors}
\label{sec:classification}
To systematically explain how and why online behavior turns out to be secure or unsafe, the MORPHEUS framework integrates two foundational structural models: the Cognitive-Affective-Behavioral (CAB) model~\cite{breckler1984empirical} and Heider's Attribution Theory~\cite{heider2013psychology}.
While the cybersecurity literature already features robust behavioral theories (such as the Protection Motivation Theory~\cite{Herath2009Protection, boer1996protection} or the General Deterrence Theory~\cite{chen2025deterrence}), these domain-specific models typically focus on particular mechanisms like motivation or compliance. Consequently, they are often too narrow to serve as a universal taxonomy capable of organizing all 50 diverse human factors identified in our review, which span from cognitive fatigue to organizational resources. Instead of replacing these existing theories, MORPHEUS utilizes CAB and Attribution Theory as a macro-level architectural scaffolding that accommodates them. Specifically, we strictly delineate the roles of these two models: the CAB model is used exclusively to define the proximal \textit{Direct Factors} (deliberately unpacking the broad concept of ``attitude'' into distinct cognitive, affective, and behavioral components to avoid oversimplification), whereas Attribution Theory is applied exclusively to categorize the distal \textit{Modulators} based on their internal (dispositional) or external (situational) origins.

To ensure these models are accessible to an interdisciplinary cybersecurity audience, it is important to briefly define their core premises. The \textbf{CAB model}~\cite{breckler1984empirical} posits that human attitudes and subsequent actions are not monolithic, but are formed through three distinct, interacting components: \textit{Cognition} (what a person knows, believes, or perceives), \textit{Affect} (how a person feels emotionally), and \textit{Behavior} (how a person acts or intends to act). On the other hand, \textbf{Attribution Theory}~\cite{heider2013psychology} explains how individuals and observers interpret the underlying causes of behaviors and events. As successfully adapted in previous HCI research \cite{Depping2017}, this theory fundamentally distinguishes between \textit{internal} (dispositional) attributions, where an outcome is driven by a person's inherent traits (e.g., personality), and \textit{external} (situational) attributions, where the outcome is forced by environmental or contextual constraints (e.g., social factors).

\subsubsection{Direct Factors (The CAB Core)}

The CAB model provides the proximal processing engine of MORPHEUS, moving beyond a static list of traits to capture the dynamic psychological state of a user at the moment of a security decision. Grounded in the tripartite model of attitude structure, this model allows the framework to operationalize how cognitive aspects, affective states, and behavioral intentions converge to drive security outcomes. We specifically rely on the seminal work of Breckler \cite{breckler1984empirical}, which empirically validated the distinctness and interaction among these three components, providing a rigorous structural basis for our ``Direct Factors'' layer. 
Direct factors can manifest as cognitive failures (e.g., incomplete threat knowledge), affective responses (e.g., fear, stress), or behavioral tendencies (i.e., dispositional or habitual patterns of action, such as automatic or impulsive clicking on an email link)~\cite{Budimir2021CybersecurityEmotions, hadlington2018media}. The CAB model is particularly relevant in cybersecurity as it can explain, for example, why users with high knowledge (cognitive) may still fail in avoiding a cyberthreat due to acute stress (affective) or habitual patterns (behavioral).

Regarding the \textbf{cognitive} dimension, we align our findings with the comprehensive survey by Burda et al. \cite{burda2024cognition}. Following their information-processing perspective, the 15 cognitive factors mapped in MORPHEUS (see Table~\ref{tab:hfs}) can be functionally structured into the three core stages of social engineering cognition:
\begin{itemize}
    \item \textbf{Perception} (translating stimuli based on prior knowledge): encompassing \textit{Lack of awareness, Lack of knowledge, Misperception,} and \textit{Uncertainty}.
    \item \textbf{Attention} (modulating access to conscious processing and managing cognitive load): encompassing \textit{Distraction}, \textit{Vigilance}, \textit{Cognitive fatigue}, and \textit{Decision fatigue}.
    \item \textbf{Elaboration} (reasoning and decision-making, balancing heuristics and systematic processing): encompassing \textit{Bias}, \textit{Cognitive reflectiveness}, \textit{Cybersecurity self-monitoring}, \textit{Cyber risk beliefs}, \textit{Overconfidence}, \textit{Risk attitude}, and \textit{Security self-efficacy}.
\end{itemize}
This structured view provides a nuanced theoretical lens, clarifying how different cognitive vulnerabilities disrupt distinct phases of a user's security decision-making process. It is worth noting that some factors might belong to more than one dimension; for example, \textit{Bias} belongs to the \textit{Elaboration} dimension as it directly affects decision making, but might also be considered a factor related to \textit{Perception}, as an individual's cognitive biases are also rooted in their past knowledge and prejudices.

Regarding \textbf{affective} responses, it is important to note that the factors currently mapped in MORPHEUS are predominantly negatively valenced (e.g., fear, stress, frustration, anxiousness). However, positively valenced emotions (e.g., enjoyment, pride, satisfaction, interest) can enhance security aspects such as perceptions of company policies and a sense of responsibility towards colleagues~\cite{vonpreuschen2024beyond}, as well as improve phishing reporting and training effectiveness~\cite{chen2024motivates}. Moreover, ``positive'' emotions such as joy or surprise can be weaponized by cybercriminals to bypass analytical thinking and deceive their victims, e.g., in social engineering attacks (see Section~\ref{sec:adversarial_triggers} for more details).
Therefore, while the emotions mapped in MORPHEUS are skewed towards the negative end of the affective spectrum, this does not imply that positively valenced emotions are irrelevant in cybersecurity. Rather, this skew is dictated by the focus of our systematic review on factors that specifically affect susceptibilities to the six considered cyberthreats, excluding factors that generically influence cybersecurity behavior, effectiveness of training programs, or adversarial triggers employed by cybercriminals. 


Finally, regarding the \textbf{behavioral} dimension, it is important to clarify that it maps what are technically defined as ``behavioral tendencies''---long-term dispositional patterns or habitual traits---rather than isolated, accidental behavioral events. These tendencies represent the primary human-centric source of systematic vulnerability that attackers actively target.

It is worth noting that all direct factors (i.e., those in the CAB Core) have an immediate impact on behavior. Consequently, interventions that focus on these factors are often situational and short-term, such as awareness campaigns, nudges, UI redesign, and stress buffer features in the interface~\cite{handri2024examining, li2022effects}.

\subsubsection{Modulators (Distal Antecedents)}
To organize the distal antecedents that set the stage for these proximal processes, we integrate Heider's Attribution Theory \cite{heider2013psychology}, which provides a logic for distinguishing the root causes of insecure behavior. By applying the distinction between dispositional (internal) and situational (external) attributions, MORPHEUS categorizes factors into ``Internal'' and ``External'' Modulators. This theoretical choice is critical for the framework's actionability: it allows security analysts to determine whether a vulnerability is an inherent trait of the user population that requires targeted monitoring or a product of the organizational environment that can be mitigated through systemic redesign.

\begin{itemize}
    \item \textbf{Internal Modulators (Personality Traits, Demographics):} These factors influence how an individual intrinsically regulates responses to a threat. For example, emotional reactivity and self-regulation are influenced by personality traits like neuroticism~\cite{uebelacker2014social, McCormac2017, rogers1975protection}. Similarly, demographic modulators (e.g., age, education) impact digital literacy, social anticipations, and inclinations towards trust~\cite{sheng2010falls, Blythe2011, mittal2019demographic}.
    \item \textbf{External Modulators (Social \& Organizational):} These factors define the environment where the behavior occurs. External conditions, such as security culture, norms, workload, social pressure, and policy clarity, shape risk perception and compliance behavior~\cite{Herath2009Protection, handri2024examining, allendoerfer2005human}.
\end{itemize}

Interventions that mitigate modulator factors are systemic and longitudinal in nature, including role-based training, policy change, personality-based segmentation, or organizational workflow redesign.

It is worth noting that some factors (e.g., \textit{Risk attitude} or \textit{Stress}) may overlap more than one dimension (e.g., Cognitive and Personality, or Affective and Cognitive); in these cases, we consider the most relevant dimension as representative for each factor. Furthermore, examples such as workload, security culture, or policy clarity can be understood as conceptual antecedents of the social and organizational factors explicitly represented in our framework (e.g., lack of resources, security posture, norms). We therefore use these examples illustratively, while grounding the final taxonomy strictly in the 50 human factors identified through the literature review.

\subsection{The MORPHEUS Causal Pathway Architecture}
\label{sec:causal_pathway}

To operationalize the theoretical distinctions introduced above and avoid viewing the 50 human factors as an unstructured list, MORPHEUS is structured as a hierarchical causal system: the \textit{Causal Pathway Architecture}. 

As illustrated in Figure~\ref{fig:causal_pathway}, the framework establishes a theoretical top-down directional flow—visually mapped from left to right—to represent the transition from distal antecedents to proximal states. Notably, while this architecture is grounded in the psychological integration of the CAB model and Attribution Theory, the figure deliberately anticipates its empirical validation by overlaying the quantitative data extracted from our comprehensive interaction network (the full analysis of which is detailed in Section~\ref{sec:interplay}). This immediate integration of theory and data is provided to demonstrate that our architectural layers are not merely conceptual groupings, but are directly corroborated by the statistical topology of the 302 pairwise interactions between human factors identified in the literature, thus confirming the structural validity of the framework's core flow:

\begin{itemize}
    \item \textbf{Layer 1 (Modulators):} These serve as the distal antecedents, categorized by their attributional origin. Internal characteristics, such as \textit{Personality Traits} (e.g., Neuroticism), \textit{Demographic traits} (e.g., Education), and external contextual factors (e.g., \textit{Organizational Norms}), set the baseline conditions for the user.
    \item \textbf{Layer 2 (Direct Factors):} This layer represents the proximal processing engine, aligned with the CAB model. It captures the user's transient state—comprising Cognitive, Affective, and Behavioral factors—which is the immediate driver of the security action.
    \item \textbf{Layer 3 (Outcome):} This represents the final susceptibility to specific threats (e.g., phishing, misconfiguration). Crucially, this outcome is precipitated by the specific \textit{Behavioral Factors} enacted in Layer 2, which result from the interplay of modulators, cognitive-affective states, and external \textit{Adversarial Triggers} (e.g., time pressure).
\end{itemize}

This architectural view clarifies that distal factors rarely cause incidents in isolation; rather, they modulate the intensity of the cognitive and emotional processes that lead to the behavioral outcome.

Within this causal architecture, it is also important to explicitly position \textit{Motivation}---a core construct in behavioral models such as the Protection Motivation Theory (PMT)~\cite{boer1996protection}, Fogg Behavior Model~\cite{Fogg2009Behavior}, and recent systematic reviews on autonomous security motivation~\cite{chen2025deterrence}. In MORPHEUS, motivation is not treated as a single, flat factor among the 50. Instead, it is conceptualized as an emergent, higher-order property that drives the transition from the proximal processing in Layer 2 to the final outcome in Layer 3. Specifically, motivation arises from the dynamic interplay within the CAB core: Cognitive factors (such as \textit{Cyber risk beliefs} and \textit{Security self-efficacy}) combine with Affective factors (such as \textit{Fear} or \textit{Anxiousness}) and are shaped by Social Modulators (such as \textit{Attitude towards policies}) to formulate a behavioral intention. This structural approach operationalizes how different motivational drivers are assembled from granular human factors before manifesting as secure or insecure actions.

It is worth noting that demographic characteristics (e.g., age, gender, education) are classified as \textit{internal modulators} in MORPHEUS, aligning with the dispositional perspective of attribution theory. However, functionally they differ from other internal traits: while they represent inherent attributes of the user, they cannot be modified through security training or behavioral correction. Instead, they serve as fixed constraints that systematically modulate susceptibility, requiring interventions focused on interface adaptation rather than user modification.

Such a layered framework will enable practitioners to develop sophisticated cybersecurity interventions, which may include tailored training programs and environmental or policy adjustments. However, some methodological issues might be associated with this approach. First, the direct and modulator types often overlap and require advanced analytic procedures, considering interactions~\cite{perrotin2022hos}. Second, cyberthreat environments are dynamic and demand adaptive processes that combine different data streams and contextualize them to inform policy decision-making. These challenges require the systematic study of empirical data to extrapolate from known models in the psychological domain to the particular area of cybersecurity~\cite{moore2019multi}.

\begin{figure*}[htbp]
\centering
\resizebox{\textwidth}{!}{%
\begin{tikzpicture}[
    node distance=1.6cm and 4.8cm, 
    >=Stealth, 
    font=\Large\sffamily,
    basebox/.style={rectangle, rounded corners=4pt, minimum width=4.5cm, minimum height=1.6cm, text width=5cm, align=center, draw, line width=1pt},
    modulator_int/.style={basebox, draw=purple!80!black, fill=purple!5},
    modulator_ext/.style={basebox, draw=orange!80!black, fill=orange!5},
    direct/.style={basebox, draw=cyan!80!black, fill=cyan!5, line width=1.2pt},
    outcome/.style={basebox, draw=red!70!black, fill=red!5, minimum height=4cm, text width=4.8cm, line width=1.5pt},
    trigger/.style={rectangle, rounded corners=4pt, draw=black!60, dashed, fill=gray!5, minimum width=5.5cm, minimum height=1.2cm, text width=5cm, align=center},
    group/.style={draw=black!20, line width=1pt, inner sep=22pt, rounded corners=8pt, fill=black!2},
    main_arrow/.style={->, line width=3.5pt, draw=cyan!70!black},
    loop_arrow/.style={->, line width=1.5pt, draw=blue!50!black, rounded corners=4pt},
    reverse_arrow/.style={->, dashed, line width=1.5pt, draw=orange!80!black},
    data_label/.style={font=\normalsize\bfseries, fill=white, inner sep=4pt}
]

\node[direct] (affective) {\textbf{Affective Factors}\\ \textit{\normalsize (e.g., Stress, Fear)}};
\node[direct, above=of affective] (cognitive) {\textbf{Cognitive Factors}\\ \textit{\normalsize (Perception, Attention, Elaboration)}};
\node[direct, below=of affective] (behavioral) {\textbf{Behavioral Factors}\\ \textit{\normalsize (e.g., Habits, Impulsivity)}};

\node[modulator_int, left=of cognitive] (personality) {\textbf{Personality}\\ \textit{\normalsize (Internal)}};
\node[modulator_int, left=of affective] (demo) {\textbf{Demographics}\\ \textit{\normalsize (Internal)}};
\node[modulator_ext, left=of behavioral] (social) {\textbf{Social/Org. Context}\\ \textit{\normalsize (External)}};

\node[outcome, right=of affective, xshift=1cm] (susceptibility) {\textbf{SUSCEPTIBILITY}\\ \medskip \normalsize Phishing, SMishing, \\ Malware, Passwords, \\ Misconfigurations};

\begin{scope}[on background layer]
    \node[group, fit=(cognitive) (behavioral), label={[font=\LARGE\bfseries, text=cyan!80!black, yshift=6pt]above:LAYER 2: CAB CORE}] (L2) {};
    \node[group, fit=(personality) (social) (demo), label={[font=\LARGE\bfseries, text=purple!80!black, yshift=6pt]above:LAYER 1: MODULATORS}] (L1) {};
    \node[group, fit=(susceptibility), label={[font=\LARGE\bfseries, text=red!70!black, yshift=6pt]above:LAYER 3: OUTCOME}] (L3) {};
\end{scope}


\node[trigger, below=1.8cm of L2.south] (trigger) {\textbf{Adversarial Triggers}\\ \textit{\normalsize (Time Pressure, Deceptive UIs, Persuasion, Emotional Drivers)}};

\node[draw=black!30, fill=white, line width=1pt, rounded corners=4pt, align=left, font=\normalsize, inner sep=8pt, below=1.8cm of L1.south] (legend) {
    \textbf{Empirical Interaction Network Data (n=302)}\\
    \medskip
    \textcolor{green!60!black}{\textbf{82.8\%}} \textit{Architecture-Compliant}\\
    \textcolor{orange!80!black}{\textbf{17.2\%}} \textit{Recursive Feedback}
};


\draw[main_arrow] (L1.east) -- node[midway, align=center, data_label, yshift=20pt] {Top-down\\Cascades (32.1\%)} (L2.west);

\draw[reverse_arrow] (L2.west) to[out=210, in=330, looseness=1.3] node[midway, align=center, data_label, below=4pt] {Recursive\\ Feedback (17.2\%)} (L1.east);

\draw[loop_arrow] (L1.north west) -- ++(0, 1.5cm) coordinate (t1) -- node[midway, data_label] {Intra-Modulator (22.2\%)} (t1 -| L1.north east) -- (L1.north east);

\draw[loop_arrow] (L2.north west) -- ++(0, 1.5cm) coordinate (t2) -- node[midway, data_label] {Internal CAB Loops (28.5\%)} (t2 -| L2.north east) -- (L2.north east);

\draw[->, line width=1.5pt, dashed, draw=black!70] (trigger.north) -- node[right, font=\normalsize\bfseries, text=black!70] {Adversarial Impact} (L2.south);

\draw[->, line width=3pt, draw=red!70!black] (L2.east) -- node[above, font=\large\bfseries, text=red!80!black] {Motivation} (L3.west);

\end{tikzpicture}
}
\caption{The Causal Pathway Architecture of MORPHEUS comprising three layers: \textbf{Layer 1 – Modulators} (Personality, Demographics, and Socio-Organizational factors), which affect \textbf{Layer 2 – CAB Core} (Cognitive, Affective, and Behavioral factors), which in turn motivates \textbf{Layer 3 – Outcome} (Susceptibility to the six cyberthreats).
The arrows and the numbers between Layer 1 and Layer 2 represent interaction effects between the two layers. Internal feedback loops are also represented by recursive arrows in both layers.
Adversarial Triggers are positioned as external stimuli acting upon the CAB Core.
The empirical data coming from 302 unique interactions (see Section~\ref{sec:interplay}) validates the architecture flow, with only a minority of the interactions (17.2\%) indicating Recursive Feedback from Layer 2 to Layer 1.}
\Description{A flowchart illustrating the three-layered causal pathway of the MORPHEUS framework. On the left is Layer 1: Modulators (Distal Antecedents), divided into three categories: "Personality Traits (e.g., Neuroticism)" and "Demographics" (e.g., Age, Education), which are internal modulators, and Social and Organizational (e.g., Norms, Lack of resources), which are external attribution factors. Layer 1 is linked to Layer 2 with an arrow representing the direct interactions (L1 to L2, with 32.1\% of cases from our data); a dashed arrow from Layer 2 to Layer 1 represents recursive feedback interaction (with 17.2\% of cases from our data). In the middle is Layer 2: CAB Core, containing three boxes, i.e., "Cognitive Factors (e.g., Perception, Attention, Elaboration)", "Affective Factors (e.g., Stress, Fear)", and "Behavioral Factors (e.g., Impulsivity, Habits)." Below Layer 2 is a box labeled "Adversarial Triggers (Stimuli)," listing "Persuasion Principles," "Time pressure," "Deceptive UIs," and "Emotional Drivers". This is linked to Layer 2, indicating the Adversarial Impact of these triggers on the CAB factors. On the right, there is Layer 3: Outcome, containing a single box labeled "Susceptibility (e.g., Phishing, Misconfiguration)", linked to Layer 2 via an arrow indicating that Motivation stems from the CAB factors. Layer 1 and Layer 2 have a recursive arrow indicating Intra-Modulator interactions (22.2\%) and Internal CAB loops (28.5\%), respectively. These numbers represent interactions coming from the Empirical Network Data of 302 interactions.}
\label{fig:causal_pathway}
\end{figure*}

\subsection{Distinction between Human Factors and Adversarial Triggers}
\label{sec:adversarial_triggers}
A fundamental theoretical contribution of MORPHEUS lies in the clear demarcation between \textit{human factors} and \textit{adversarial triggers}, which are external stimuli or tactics deployed to exploit these characteristics. While often conflated in cybersecurity literature, distinguishing the \textit{stimulus} from the \textit{response mechanism} is critical for accurate risk diagnosis. 

Crucially, this boundary also clarifies why systemic conditions—such as Social and Organizational elements (e.g., lack of resources)—are treated as human factors rather than external triggers. In the context of MORPHEUS, the term ``human factor'' does not equate to ``user fault.'' Rather, it encompasses the entire ecosystem of variables that structurally shape human decision-making. Adversarial triggers are \textit{active, malicious stimuli} injected by an attacker (e.g., a deceptive UI or an engineered time constraint), whereas organizational factors constitute the \textit{passive, systemic environment} in which the user operates. By explicitly categorizing these systemic issues as \textit{External Modulators} (via Attribution Theory, see Section~\ref{sec:classification}), MORPHEUS demonstrates that when a vulnerability is driven, e.g., by a lack of resources or poor security culture, the resulting breach is a symptom of an external systemic failure, not an internal individual psychological deficit.

We identify four primary categories of adversarial triggers that function as amplifiers of human vulnerability: \textit{persuasion principles}, \textit{emotional drivers}, \textit{time pressure}, and \textit{deceptive user interfaces}.

Drawing from social psychology \cite{cialdini2009influence}, \textit{persuasion principles} such as \textit{Authority}, \textit{Scarcity}, and \textit{Reciprocity} are not human traits but manipulation tactics. Within our framework, these principles act as keys to unlock specific human factors. For instance, an attack leveraging \textit{Authority} (tactic) targets the human factor of \textit{Norms} (obedience) and \textit{Trust}. Research~\cite{zhuo2023sok, ferreira2015principles} has confirmed that these principles—along with other adversarial triggers, such as Gragg's psychological triggers \cite{Gragg2003Social} or Stajano's principles \cite{10.1145/1897852.1897872}—are strategically embedded in phishing campaigns to bypass analytical processing (System 2) and force a reliance on heuristics.

Similarly, \textit{emotional drivers} are typically used by attackers to manipulate victims and make them act irrationally~\cite{wang2021Social, Phelps2014Emotion}. In general, individuals under the influence of ``visceral influences'' seek immediate satisfaction of their visceral desires, even if that entails riskier actions~\cite{Wang2012Phishing, langenderfer2001}. For example, in the case of phishing attacks, attackers can exploit emotions such as curiosity, fear, greed, anger, joy, surprise, and compassion~\cite{Cybsafe2023Ultimate, Hadnagy2015Phishing}. It is worth noting that, even if these affective factors can be exploited by attackers, they are not all included as MORPHEUS factors because they act as external adversarial triggers, which fall outside the scope of the framework.

\textit{Time pressure} is an environmental constraint imposed by the attacker (e.g., a phishing deadline) or the context (e.g., incident response), rather than a dispositional trait of the user \cite{chowdhury2020time}. In the MORPHEUS framework, time pressure functions as a cognitive inhibitor, restricting the resources available for \textit{Vigilance}, \textit{Cognitive reflectiveness}, and systematic processing (System 2 – slow, deliberate, and logical thinking~\cite{kahneman2011thinking}), thereby forcing a reliance on \textit{Biases} and heuristics (System 1 - fast, automatic, and intuitive thinking~\cite{kahneman2011thinking}) \cite{Butavicius2022People}.

In threats such as malware downloads and system misconfigurations, attackers utilize \textit{deceptive user interfaces} (also known as dark patterns) or contextual mimicry (such as spear-phishing pretexting). These are external design choices intended to maximize \textit{Misperception} and exacerbate \textit{Cognitive Fatigue} or \textit{Frustration}. For example, a confusing security alert (trigger) exploits a user's \textit{Lack of Knowledge} or \textit{Distraction}, leading to an insecure default choice \cite{manfredi2022empirical}.

Analysts should consider these external triggers to move beyond generic ``awareness'' solutions toward targeted interventions that disrupt the specific interaction between tactic and trait.

\section{Human factors affecting the most prevalent threats}
\label{sec:cybthreats}

This section applies the multidimensional MORPHEUS framework to six of the most pervasive cyberthreats—phishing (and its spear-phishing and SMishing variants), malware downloads, password management errors, and system misconfigurations—to highlight how specific human factors drive vulnerability in each context. The analysis is grounded in the reviewed literature and organized around the 50 human factors identified across demographic, personality, cognitive, affective, behavioral, and organizational dimensions. The next sub-sections report the details of how each threat is influenced by the associated human factors.

\subsection{Phishing and variants}

Phishing remains one of the most prevalent social-engineering threats, exploiting psychological biases and habitual decision processes through deceptive communication channels. Its targeted (spear-phishing) and mobile (SMishing) variants further amplify these mechanisms by leveraging personalization and immediacy cues. The empirical evidence from various studies consistently reveals that human susceptibility is systematically related to individual and contextual characteristics.

\paragraph{Demographic factors}

Findings on demographic factors on phishing susceptibility are mixed and sometimes contradictory. Some researchers argue that these are rather indirectly related to phishing victimization, helping to capture specific characteristics related to user knowledge and behavior at most~\cite{zhuo2023sok}. However, other studies in the literature report significant effects of demographics on susceptibility. 
\textit{Age} effects appear non-linear, with some evidence suggesting that older users remain more vulnerable to phishing and variants compared to younger individuals~\cite{lin2019susceptibility, tabassum2024drives, Rahman2023Users, Redmiles2018Examining}, while others indicate that younger users are instead more susceptible~\cite{ge2021personal, greitzer2021experimental, Huseynov2024Using, eftimie2022spear, sheng2010falls}.

For \textit{education}, in general, higher educational attainment (not only in cybersecurity or IT) can lead to individuals being more susceptible to phishing attacks~\cite{abroshan2021covid, Butavicius2022People}. However, higher levels of user education in general have proven to help people detect phishing techniques (e.g., domain spoofing) from known and unknown senders~\cite{Huseynov2024Using, moody2017phish}. 

\textit{Gender} differences are similarly controversial: several studies reported greater female susceptibility to all variants of phishing~\cite{Abdelhamid2020concerns, iuga2016baiting, sheng2010falls, Huseynov2024Using, abroshan2021phishing, ge2021personal, lin2019susceptibility, alhaddad2023study}, whereas some studies found men to be more at risk~\cite{yeboah2014phishing, greitzer2021experimental, Timko2025Understanding}.

\paragraph{Personality traits}

Several studies have investigated the effects of personality traits on phishing susceptibility, yielding contradictory results~\cite{zhuo2023sok}. Some results regarding the study of the Big 5 in phishing contexts suggest that there is an indirect effect on email processing, rather than a direct effect on susceptibility~\cite{ge2021personal}. In fact, we must consider that personality is a very complex aspect of individuals, and the Big 5 encompasses several subtraits (facets)~\cite{Fleenor2023BigFive}, which are difficult to directly relate to phishing vulnerability.

\textit{Agreeableness}, associated with its sub-facets dispositional trust and cooperativeness, can increase compliance with deceptive requests for all phishing variants~\cite{Workman2008Wisecrackers, eftimie2022spear, LopezAguilar2025Phishing, Abdelhamid2020concerns}. On the contrary, low dispositional trust (i.e., being suspicious by nature) is a protective factor for phishing~\cite{Wright2010Influence}.
However, for SMishing attacks, the \textit{morality} facet of \textit{agreeableness} was found to lead to increased resistance~\cite{Huseynov2024Using}.

\textit{Conscientiousness} is generally a protective factor for phishing and its variants~\cite{Frauenstein2020Susceptibility, Vishwanath2015Examining, Huseynov2024Using, LopezAguilar2025Phishing, marin2023influence, eftimie2022spear, ge2021personal}. However, there is evidence that a high lack of focus, which stems from low levels of conscientiousness, can be protective in spear-phishing scenarios~\cite{moody2017phish}. 

\textit{Extraversion} correlates with impulsivity, excitement seeking, and higher interaction with social content, and generally leads to elevated click-through rates~\cite{eftimie2022spear, Huseynov2024Using, LopezAguilar2025Phishing}; however, extraversion can be associated with positive cybersecurity behaviors when they are meant to help others (e.g., reporting phishing emails)~\cite{marin2023influence}. 

\textit{Greed} can be an influencing factor that increases SMishing vulnerability, with greedy users being more likely to fall for SMSs that promise rewards~\cite{Rahman2023Users}.

\textit{Openness} to experience tends to mitigate risk, as it also involves higher values of Intellectualism, fostering more analytical evaluation and intuitive thinking~\cite{ge2021personal, eftimie2022spear, Huseynov2024Using, Buckley2023Indicators}. On the other hand, openness also encompasses Curiosity, which is correlated with a higher susceptibility to all variants of phishing attacks \cite{moody2017phish, benenson2017unpacking, yeng2022investigation}.

The trait of \textit{Narcissism} was identified as a personality trait that increases phishing vulnerability, as it may lead to more impulsivity and overconfidence~\cite{Curtis2018DarkTriad, Hart2025Phishing}. 

Finally, high values of \textit{Neuroticism} (i.e., low emotional stability) can amplify vulnerability to phishing and variants  \cite{LopezAguilar2025Phishing, ge2021personal, eftimie2022spear}. On the other hand, low neuroticism has been shown to correlate with resistance to manipulation and better adherence to verification routines~\cite{Vishwanath2015Examining, greitzer2021experimental}. This dualism of neuroticism for SMishing might be explained by its subtraits, with the Vulnerability facet leading to increased susceptibility, but the Self-consciousness facet decreasing susceptibility~\cite{Huseynov2024Using}.

\paragraph{Cognitive factors}

Cognitive \textit{Biases} can increase susceptibility to phishing and spear-phishing by altering how phishing threats are processed by victims. Heuristic processing (i.e., using the faster, biased System 1 of thinking) can increase susceptibility, as suspicious elements are not properly analyzed \cite{Frauenstein2020Susceptibility, Vishwanath2015Examining, Chou2021Mindless, Williams2018Exploring, Waqas2023Enhancing}. Other specific biases can negatively affect phishing detection performance, such as the anchoring effect, which influences subsequent decisions based on the first piece of information received~\cite{iuga2016baiting}. 

If, on the one hand, heuristic processing (i.e., relying on biases) increases phishing susceptibility, on the other hand, \textit{Cognitive reflectiveness} and systematic, deeper processing (e.g., checking links in emails) positively impacts user performance in distinguishing between phishing attempts and genuine content~\cite{Waqas2023Enhancing, Butavicius2022People, Gallo2024HumanFactor, Buckley2023Indicators, dawn2024who, Vishwanath2015Examining, Vishwanath2018Suspicion, greitzer2021experimental, ge2021personal, Wang2017Coping}. However, there is evidence that, despite requiring more cognitive effort to analyze a phishing email, it can still trick users if it is complex enough or well-targeted~\cite{Wang2012Phishing, iuga2016baiting}. For SMishing, evidence suggests that cognitive reflectiveness serves as a protective factor, but only when combined with security awareness messages or training~\cite{Kamar2023Moderating}. 

Levels of individuals' \textit{Cognitive fatigue} and \textit{Decision fatigue} are positively associated with the likelihood of employees clicking on a phishing link~\cite{Jalali2020Employees, vishwanath2011people, nifakos2021influence}. However, a study by Musuva et al.~\cite{Musuva2019Cognitive} found a contradictory, yet intuitive result: participants under a higher volume of emails were less susceptible to phishing, as they might be less likely to promptly answer unimportant emails.

More negative \textit{Cyber risk beliefs}, such as thinking about potential consequences and severity of falling victim to phishing links, can increase motivation to comply with good security practices, ultimately decreasing susceptibility to all variants of phishing~\cite{greitzer2021experimental, Martens2019Investigating, Timko2025Understanding, Musuva2019Cognitive, boer1996protection, aleroud2020examination, jampen2020don} and increasing intention to report spear-phishing emails \cite{Kwak2020Spear}. However, for risk-prone users, the perceived risk can translate into a higher likelihood of checking phishing links~\cite{moody2017phish}.

\textit{Cybersecurity self-monitoring} tendencies in individuals can reduce victimization to spear-phishing~\cite{Kwak2020Spear}. On the other hand, 
\textit{Distraction} negatively affects users' performance, including the correctness of decision-making when dealing with phishing emails~\cite{zhuo2023sok}.

\textit{Lack of awareness} and \textit{Lack of knowledge} about security and scam tactics generally increase susceptibility to phishing \cite{Huseynov2024Using, ge2021personal, Jaeger2021Eyes, dawn2024who, desolda2021human, zhuo2023sok, Wash2020Experts, Ribeiro2024Factors, tornblad2021characteristics, nifakos2021influence, jampen2020don}, SMishing \cite{Timko2025Understanding, Huseynov2024Using}, and spear-phishing~\cite{Wang2012Phishing}. 
Conceptual knowledge should, however, be combined with procedural knowledge for more phishing protection~\cite{Arachchilage2014Security}.
However, there is empirical evidence that the more proficient users are, the more they are susceptible to phishing~\cite{Greco2025Enhancing} and spear-phishing~\cite{Ion2015Noone}. 

Another factor that hinders users' protection against spear-phishing is the \textit{Misperception} that a malicious email may be relevant, for example, when its context fits their expectations~\cite{benenson2017unpacking, Distler2023Influence}.

Similarly, the general trait of \textit{Risk attitude} is reflected in online contexts can increase vulnerability to phishing~\cite{moody2017phish}; nevertheless, it was found that propensity to financial risk predicts lower susceptibility~\cite{moody2017phish}.

\textit{Security self-efficacy} has an important role in decreasing phishing~\cite{Buckley2023Indicators, Martens2019Investigating, Arachchilage2014Security, Wright2010Influence, Jansen2018Persuading, House2020Phishing} and SMishing~\cite{Verkijika2019SelfEfficacy} susceptibility, and it has a core component in the motivation to avoid threats and the actual behavior, together with perceived threat~\cite{boer1996protection}. A high self-efficacy (and perceived high utility) can even foster individuals to report phishing and spear-phishing attempts~\cite{marin2023influence, Kwak2020Spear, chen2024motivates}.
However, high values of security self-efficacy have also been shown to lead to higher phishing and SMishing victimization~\cite{Ribeiro2024Factors, Lee2023Thwarting}. In fact, too much self-efficacy can translate to \textit{Overconfidence}, a factor that leads users to let their guard down and be more susceptible to phishing \cite{wang2016overconfidence, Canfield2019Metacognition, jampen2020don} and SMishing attacks \cite{yeng2022investigation}.

\textit{Uncertainty} about the legitimacy of a communication can prompt users to investigate the cause of receiving a phishing email, thereby increasing their risk~\cite{benenson2017unpacking}. Uncertainty can also affect employees who are unsure about their organization's technical security measures against phishing~\cite{Williams2018Exploring} or how to report a (spear-)phishing email~\cite{Distler2023Influence}.

Finally, \textit{Vigilance} is a protective factor for both phishing and spear-phishing, as giving attention to visceral triggers and phishing indicators in an email can reduce susceptibility~\cite{greitzer2021experimental, Wang2012Phishing, jampen2020don}.

\paragraph{Affective factors}
Emotions can affect susceptibility to phishing and variants in different ways. 

\textit{Anxiousness} has been shown to increase compliance with malicious requests~\cite{alhaddad2023study}, especially if users are led to feel concerned about their health~\cite{abroshan2021covid, Abdelhamid2020concerns}.
On the other hand, \textit{Digital anxiety} can deter users from interacting with suspicious content, thereby increasing their spear-phishing protection~ \cite{alhaddad2023study, moody2017phish}.

Among affective drivers, \textit{Fear} is surely one of the most impactful factors for falling victim to phishing~\cite{abroshan2021covid}, SMishing~\cite{Rahman2023Users}, and spear-phishing attacks~\cite{benenson2017unpacking}; e.g., a user might feel pressured to comply with deceiving requests, fearing that a stranger may have compromising pictures of them~\cite{benenson2017unpacking}.
Fear can also prevent reporting phishing emails~\cite{Martens2019Investigating}; e.g., if a malicious email comes from a colleague's address, the fear of getting that colleague into trouble can prevent reporting~\cite{Distler2023Influence}.
However, fear can also have the opposite effect and constitute a protective factor if an individual is affected by ``anticipated regret'', i.e., the anticipation of regret pre-behaviorally to avoid experiencing this unpleasant emotion~\cite{Sandberg2008AnticipatedRegret}; anticipated regret can in fact have a positive influence on protection from phishing and variants~\cite{Verkijika2019SelfEfficacy, Williams2018Exploring, Jansen2018Persuading, House2020Phishing}.

\textit{Shame} is an important emotion in the scenario of reporting phishing attacks; in fact, an employee who has fallen victim to a phishing email might not report it to the organization's IT department to avoid feeling ashamed~\cite{Distler2023Influence}.

Finally, \textit{Stress} is a critical factor in increasing susceptibility to phishing attacks, as it can impair productivity and rational decision-making when, e.g., dealing with emails~\cite{abroshan2021covid, zhuo2023sok, chen2024motivates}, especially in high-pressure environments such as public health facilities~\cite{nifakos2021influence}.

\paragraph{Behavioral factors}

\textit{Complacency} is a susceptibility factor in phishing, as users might not assume proper security behaviors if they do not feel they have anything worthwhile to make themselves a target for attackers~\cite{desolda2021human}.

Email-related \textit{Compulsive behaviors}, such as promptly checking and responding to e-mails quickly, can increase phishing vulnerability~\cite{Vishwanath2015Examining}; compulsive behaviors can be exacerbated by emergency scenarios and can, e.g., include compulsively checking about health issues during a pandemic~\cite{abroshan2021covid}.

\textit{Impulsive behaviors} also predict more victimization to phishing and spear-phishing attacks, as impulsivity is linked to faster and less thoughtful reactions to phishing stimuli~\cite{greitzer2021experimental, dawn2024who, Butavicius2015HumanFirewall, benenson2017unpacking, tornblad2021characteristics, jampen2020don}.

Higher \textit{Internet Usage} is generally a risk factor when it comes to phishing and variants~\cite{Greco2025Enhancing, Vishwanath2015Habitual, Huseynov2024Using, Ribeiro2024Factors, moody2017phish, alhaddad2023study, Reyns2015Routine, Ngo2020Victimization}. This might have different explanations, such as the fact that regular internet users are more exposed to threats, or feel more confident online, thereby assuming a more reckless behavior.
However, there is also evidence that increased Internet usage and experience are associated with a better understanding of phishing attacks and detection strategies, thereby reducing phishing susceptibility~\cite{iuga2016baiting, Redmiles2018Examining, ge2021personal}.

\textit{Recurrence} of insecure Internet- and email-related behaviors greatly increases phishing susceptibility of users~\cite{vishwanath2011people, Williams2018Exploring, tornblad2021characteristics}, while having good cybersecurity habits increases protection and the likelihood to report attacks~\cite{marin2023influence}. There is evidence that detrimental behaviors can persist after victimization and can even predict future misbehavior by the same people~\cite{greitzer2021experimental}.

Finally, \textit{Risk-taking} behavior patterns also predict more susceptibility to phishing~\cite{Abdelhamid2020concerns, abroshan2021covid, abroshan2021phishing}.

\paragraph{Organizational and social factors}

Social and organizational contexts play a decisive role in shaping phishing-related decisions. 
First of all, a correct \textit{Attitude toward policies} of the organization regarding cybersecurity and precautionary behavior is critical to guarantee compliance and reduce phishing susceptibility~\cite{Lee2022Phishing, Petric2022Impact, Martens2019Investigating}.

A \textit{Lack of communication} can lead to isolation and increase attacks' impact, as when an employee falls for a phishing scam, they often choose not to tell anyone or report it to the IT department~\cite{Distler2023Influence}, even though peers may have been targeted by the same attack. Communication can, therefore, increase protection, as support from peers can help make informed decisions about suspicious spear-phishing emails~\cite{Williams2018Exploring}. It is also important for an organization to communicate to its employees how reported phishing incidents are managed~\cite{chen2024motivates}.

A \textit{Lack of resources}, including time, organizational and technical support, and training material, can discourage employees from following correct security practices \cite{desolda2021human, chen2024motivates, nifakos2021influence}. Security training and the availability of educational content are crucial in improving users' protection against all phishing variants~\cite{tabassum2024drives, sheng2010falls, iuga2016baiting, burns2019spear, Chen2024GroupDiscussion}; specifically, this can occur either with sparse security awareness messages or via SMS-based training exercises~\cite{Kamar2023Moderating}.
Adequate technical resources can also improve protection against phishing for individuals, including automated filters, security software that warns users about phishing attempts, and simplified reporting options~\cite{zhuo2023sok, Greco2025Enhancing, Petelka2019Warning, rastenis2025credulitySpear}.
Moreover, lack of time and time pressure were found to be critical factors in increasing victimization to all phishing variants in urgency situations, as being very busy can lead users to not dedicate appropriate time to email processing, overlook phishing cues, or even click links by accident~\cite{yeng2022investigation, Butavicius2022People, Williams2018Exploring}. 

A \textit{Lack of trust} in the source of a (spear-)phishing message is a highly protective factor, as users are less likely to click links in emails coming from dubious sources~\cite{benenson2017unpacking}. Contrarily, a higher trust in the received attack material and in its sender leads to increased susceptibility to all phishing variants~\cite{aleroud2020examination, moody2017phish, yeng2022investigation, greitzer2021experimental, tornblad2021characteristics, jampen2020don}. Higher trust can be achieved by attackers by making phishing links more familiar and understandable \cite{moody2017phish} or, in the case of spear-phishing attacks, by sending malicious emails from senders with whom the victim is familiar~\cite{Williams2018Exploring}.

Socially accepted \textit{Norms} in the working environment (e.g., how others deal with emails and talk about security) change how employees interact with phishing emails, potentially increasing the spread of bad habits and thus phishing susceptibility, but also promote better security behavior~\cite{Petric2022Impact, desolda2021human, Distler2023Influence, marin2023influence}. 
Internal work norms can also influence how individuals perceive emails, as employees accustomed to receiving emails from outside the organization are less likely to consider an external email as suspicious~\cite{Williams2018Exploring}.

This is also linked to the degree of \textit{Online exposure} of an individual, as employees who regularly receive external emails are more vulnerable to spear-phishing~\cite{Williams2018Exploring}. Moreover, more reckless sharing of private information online (e.g., posting phone numbers on social media) can make individuals more susceptible to phishing~\cite{Lee2022Phishing, Reyns2015Routine, Ngo2020Victimization} and spear-phishing attacks~\cite{nifakos2021influence}.

\textit{Organizational tenure} is an intricate factor that can dictate phishing susceptibility. There is evidence that employees who have less working experience in an organization are more susceptible to phishing and spear-phishing, while long-time workers have been found to be less vulnerable~\cite{bullee2017spear, Taib2019SocialEngineering}. Low job-experienced employees are also less likely to report phishing attacks~\cite{Taib2019SocialEngineering}. The type of job can also impact phishing susceptibility, as full-time workers resulted in being less susceptible than adjuncts~\cite{greitzer2021experimental}. There is also evidence that employees with a very long tenure at a company (12 or more years) are more susceptible to phishing than those with shorter job experience (3-11 years)~\cite{Gallo2024HumanFactor}. Finally, an interesting result is that people with even minimal exposure to the work environment can quickly develop poor security habits and become more susceptible to SMishing than, for example, university students~\cite{Huseynov2024Using}.

\textit{Social influence} plays a major role in phishing and variants. Specifically, phishers leverage social influence techniques to create persuasive attacks and greatly increase victimization, with the most effective messages including urgency/scarcity, authority, and consensus cues~\cite{Distler2023Influence, Butavicius2015HumanFirewall, lin2019susceptibility, DeBona2020RealWorld, vishwanath2011people, Chou2021Mindless, Wright2014Influence, Taib2019SocialEngineering, Rahman2023Users, Gallo2024HumanFactor, Workman2008Wisecrackers}.

\textit{Social proof} from peers is important for defining how users make decisions when dealing with phishing emails, for example, by following shared tips~\cite{Williams2018Exploring}.

Finally, different \textit{Types of organizations} can exhibit varying levels of vulnerability to spear-phishing attacks and differences in how such incidents are managed~\cite{rastenis2025credulitySpear}.

\subsection{Malware download}

Malicious software, also known as malware, typically infiltrates systems by convincing users to open infected files or visit insecure websites. Human behavior underpins this vulnerability.

\paragraph{Demographic factors.}
There is limited evidence in the literature regarding the influence of demographic factors on malware downloads. However, a study by Lévesque and colleagues found that middle-aged users (ages 25–40) were more susceptible than younger individuals~\cite{levesque2018technological}.

\paragraph{Personality factors.}
A study by Yilmaz and colleagues~\cite{Yilmaz2023Personality} specifically investigated whether personality traits correlated with user susceptibility in the context of ransomware victimization, but found no significant effect. However, a specific subtrait of \textit{Openness}, namely curiosity, can be a driving factor in downloading insecure software~\cite{Mott2024Ransomware}. In another study, conversely, it emerged that higher levels of \textit{Agreeableness} — particularly facets related to trustfulness and compliance — may indirectly increase exposure to malicious downloads when users are inclined to comply with enticing or socially framed requests in online and mobile environments~\cite{Huseynov2024Using}. However, such effects appear less consistent than those observed in phishing contexts.
Moreover, a \textit{greedy} personality can lead to increased malware victimization, as users may be more prone to making insecure downloads in order to obtain, e.g., something for free~\cite{Mott2024Ransomware}.

\paragraph{Cognitive factors.}
Several cognitive variables heighten malware susceptibility. 
A critical factor in following best practices and avoiding risky downloads is user motivation to comply with good security practices~\cite{Abraham2010Overview}. This implies, according to the Protection Motivation Theory, that an individual should both perceive a situation as threatening (\textit{cyber risk beliefs}) and perceive their defending skills as adequate (\textit{self-efficacy})~\cite{boer1996protection, vance2012motivating}. Specifically, the perception of cyberthreats significantly influences user behavior when dealing with risky downloads, with higher risks generally perceived as leading to more secure behaviors~\cite{Choi2024Enhancing, Abraham2010Overview, Onarlioglu2012Insights}. This perceived severity can also be increased by exposing an employee to the potential consequences of malware, thereby ultimately decreasing their susceptibility \cite{Yilmaz2023Personality}. 

The \textit{lack of awareness} about protective measures and security policies has been identified as a major cause of malware victimization~\cite{Abraham2010Overview, Choi2024Enhancing}. 

A \textit{lack of knowledge} on malware can lead to a higher victimization~\cite{Mott2024Ransomware}. Computer expertise alone is not enough to protect against malware; for example, even figures such as software developers have been shown to be more at risk compared to other categories of users~\cite{ovelgonne2017understanding}.
In particular, \textit{overconfidence} in one's security skills (which might be caused by a higher general computer knowledge) has been shown to lead to more risky behavior in malware downloads \cite{levesque2018technological}.

\paragraph{Affective factors}
Regarding the affective dimension, fear is a major driver for malware downloads; this is the case of scareware, whose intent is to frighten users with fake pop-up warnings or loud sounds~\cite{Abraham2010Overview}.

\paragraph{Behavioral factors.}
Among behavioral factors, \textit{impulsive behavior} can decrease brain activity in users when dealing with malware warnings, possibly increasing their susceptibility to dangerous software~\cite{Neupane2016Neural}. Moreover, a higher amount of time spent online on social networks or banking applications (\textit{internet usage}) can increase susceptibility to malware downloads~\cite{levesque2018technological, Reyns2015Routine, Ngo2020Victimization}.

\paragraph{Organizational and social factors.}

A \textit{lack of communication} between security specialists and employees is also a critical factor that can influence users' susceptibility to malware \cite{Yamagishi2025Collaborative}. A proper communication between previous victims of malware and other employees can also be a factor in decreasing susceptibility \cite{Mott2024Ransomware}.

A \textit{lack of resources}, such as proper user training, can translate to employees being less aware and protected against the download of malware~\cite{Mott2024Ransomware, Abraham2010Overview, Vaclavik2025Lessons, Neupane2016Neural}. This can also be aggravated by unclear security best practices in the organization~\cite{Yamagishi2025Collaborative, Choi2024Enhancing}. 

\textit{Online exposure} of an individual can also influence susceptibility to malware~\cite{Reyns2015Routine, Ngo2020Victimization}.

Organizations with poor \textit{security posture} (e.g., poor culture and weak enforcement of cybersecurity norms or technologies) generally record high infection rates \cite{Mott2024Ransomware, Vaclavik2025Lessons}. 

Finally, malware attacks are more severe on private-sector enterprises, especially small/medium ones, compared to those in the public sector, which is explained by the more stringent security rules in place and their enforcement \cite{yuryna2020empirical}. 

\subsection{Password management}

The challenge of managing multiple unique and complex passwords for various digital services remains ongoing. In line with this, the spread of weak, recycled, or inadequately secured credentials is a critical weakness in individual and business cybersecurity settings. Moreover, the adoption of facilities, like password managers, is still limited by human factors, such as a lack of trust.

\paragraph{Demographic and Cognitive factors.}
According to the results of Whitty et al. \cite{whitty2015individual}, younger generations are significantly more likely to share their credentials.

Empirical evidence suggests that cognitive factors play a significant role in password management failures. Users' \textit{cyber risk beliefs} can indeed lead them to underestimate the threats of identity theft, making them more likely to adopt poor password management behaviors~\cite{allendoerfer2005human}. This may also originate from a \textit{misperception} of individuals that "nobody would target [them]" or that "[an attacker] could not do much damage anyway" if they had their password stolen~\cite{allendoerfer2005human}.

\textit{Cognitive fatigue}, mainly caused by the difficulty of maintaining multiple complex passwords, frequently prompts users to recycle credentials across different sites or to adopt predictable modification patterns \cite{das2014tangled, Komanduri2011, allendoerfer2005human}.

A \textit{lack of awareness} regarding password security often leads users to underestimate the dangers of reusing passwords, inadequate password storage, and weak password choices. A common misconception among users is that small password changes (e.g., mere replacements) are much more effective in improving protection, without realizing that attackers can easily crack foreseeable alteration patterns \cite{fagan2017investigation, Stanton2005Analysis, das2014tangled, allendoerfer2005human}. 

A \textit{lack of knowledge} on best practices or techniques for strong and effective password creation (e.g., using mnemonics) consistently emerges as a central human factor contributing to insecure, weak, or non-compliant passwords~\cite{allendoerfer2005human, ngandu2025strengthening}. Moreover, in \cite{inglesant2010true}, the authors demonstrated that unclear or technical password rules leave users uncertain and with scarce knowledge about what constitutes a secure password, prompting insecure coping behaviors such as writing passwords down or using predictable patterns. Despite the adoption of password managers can alleviate these issues, users tend to mistrust and avoid these tools mainly because they do not understand how password managers work and their underlying security mechanisms \cite{fagan2017investigation}. There is evidence that expertise can be beneficial and lead to avoiding misbehavior, such as writing passwords down or using two-factor authentication, but experts may change their passwords much more rarely compared to non-experts~\cite{Ion2015Noone}.

In the paper \cite{fagan2017investigation}, the authors highlight that \textit{security self-efficacy} has a decisive influence on the adoption of password management tools: users who perceive themselves as more competent in the autonomous management of their credentials are less inclined to delegate security to a password manager, while those who doubt their abilities rely more on automated solutions. This result underscores how perceived self-efficacy in security practices guides behavioral choices and trust in technical support tools.

\paragraph{Affective factors.}

\textit{Frustration} is a critical emotion that can influence secure password behavior~\cite{inglesant2010true, allendoerfer2005human}. Overly stringent and confusing password requirements, as well as delays in the authentication process, may indeed cause frustration in users. As a consequence, users try alternative solutions such as password reuse, simplification, or external recording, often compromising overall security compliance and increasing vulnerability.

\paragraph{Behavioral factors.}
In \cite{das2014tangled}, it was found that, due to the complexity of remembering passwords, users adopt simple and predictable rules for creating and changing passwords, mistakenly trusting that such changes are sufficient to guarantee security (\textit{complacency}).

\textit{Impulsive behaviour}, both attentional and motor, emerges as a positive predictor of password reuse and sharing, revealing that acting quickly and without reflection can lead to poor choices in password management \cite{hadlington2017human}. Moreover, a specific aspect of impulsivity, i.e., lack of perseverance, is associated with a greater likelihood of sharing passwords. This indicates a reduced ability to persevere with tedious and burdensome tasks, leading more easily to risky actions such as sharing credentials \cite{whitty2015individual}.

\textit{Internet addiction} has been identified as a behavioral factor contributing to insecure password management, as users with this disorder tend to take greater risks to maintain online engagement, showing reduced compliance with security protocols and poorer password hygiene as a consequence of habitual and impulsive digital behavior \cite{hadlington2017human}.

\textit{Laziness} can influence poor password management. In \cite{whitty2015individual}, it was found that participants share passwords primarily for convenience and to reduce the effort to comply with security rules. This result suggests that avoiding effortful or tedious security behaviors can lead to unsafe practices, such as password reuse and disclosure.

\paragraph{Organizational and social factors.}

\textit{Attitude toward policies} is associated with fewer risky behaviors. This implies that negative attitudes toward rules correlate with lax password practices (e.g., reuse, sharing) \cite{hadlington2017human, karlsson2021effect}. For example, when security policies are perceived as inconvenient or strict, users may develop negative attitudes toward them, such as using simplistic passwords, reusing passwords, writing passwords down, using predictable patterns, and sharing passwords ad hoc to get the job done \cite{Stanton2005Analysis, Komanduri2011, inglesant2010true, allendoerfer2005human}. This problem is amplified in cases where requirements are unclear, frequent changes are required for users, and the solution has a poor fit with workflows, undermining policy compliance \cite{ngandu2025strengthening}.

The adoption of alternative wrong strategies to create and manage passwords can also be affected by a \textit{Lack of communication}, which in this context can be characterized by inconsistent messaging about password rules, unclear reset procedures, and weak coordination between IT support and staff \cite{ngandu2025strengthening}. 

\textit{Lack of trust} in using password managers can be caused by fears of vendor access, cloud synchronization, and a “single point of failure” if the vault is breached \cite{fagan2017investigation}. These aspects emerged as a barrier to the adoption of password managers, while relying instead on less secure routines. On a social side, too much trust towards colleagues can lead employees to easily share their passwords~\cite{allendoerfer2005human}. 

The \textit{lack of resources} (technical, temporal, or cognitive) emerged as a critical cause of insecure password management. In \cite{inglesant2010true}, users reduce security efforts when faced with unmanageable demands and conflicting requirements, while \cite{ngandu2025strengthening} highlights how insufficient support and training degrade adherence to secure password practices. 

Compliance with \textit{norms}, typical of bureaucratic cultures, helps promote higher adherence with security policies, supporting correct behavior even in the creation and management of passwords \cite{karlsson2021effect}. On the other hand, informal work procedures and common bad practices may easily override formal password policies~\cite{allendoerfer2005human}.

In \cite{fagan2017investigation}, users expressed concerns about the adoption of password managers based on cloud synchronization, fearing that this can increase their \textit{online exposure} and the risk of large-scale credential compromise. As a result, this led them to reject these tools in favor of less secure yet perceived as safer local practices.

\textit{Social influence}, such as the pressure to be helpful to co-workers, constitutes a primary cause of password mismanagement~\cite{allendoerfer2005human}. In fact, employees might recklessly share their passwords in response to an explicit request, which may also be part of a social engineering attack.

Similarly, \textit{social proof} can play a major role in complying with password management practices, such as writing passwords down or not sharing them with colleagues~\cite{allendoerfer2005human}; in fact, observing bad security behaviors in peers or supervisors directly increases an employee's intention to violate security policies~\cite{Wang2023SecurityLocal}. 

The \textit{type of organization} might be a determining factor in understanding whether or not employees follow security rules, including those relating to passwords. For example, a high-stakes environment such as the Federal Aviation Administration differs significantly from a traditional office environment in terms of password policies~\cite{allendoerfer2005human}. Karlsson et al. highlight that internally oriented organizational cultures—such as bureaucratic and clan cultures—promote greater adherence to cybersecurity policies, including those related to password management, whereas more flexible cultures (adhocracy) tend to reduce compliance \cite{karlsson2021effect}.

\subsection{Misconfiguration}

Misconfigurations occur when systems or applications are configured, deployed, or provisioned incorrectly, thus exposing the affected systems to unauthorized access or data breaches. The problem is rooted in multiple human factors, which are discussed in the following.

\paragraph{Demographic factors.}
According to \cite{mushi2018human}, \textit{education} might undermine systems' security. Indeed, the authors found that "academic background was a notable factor associated with network instability". This is mainly because employees often lack the necessary qualifications and skills to properly configure the systems; for example, engineers commit fewer misconfigurations than administrators.

\paragraph{Cognitive factors.}

Cognitive \textit{biases} such as optimism and availability biases emerged as possible causes of misconfiguration \cite{chen2024}. They contribute to a false sense of adequacy, leading users to ignore critical options (e.g., alerts or spending limits) or misinterpret the implications of advanced configurations. Confirmation bias can also lead users to trust familiar names or services while overlooking malicious clues, which, when combined with alertness or fatigue, can result in incorrect configuration \cite{rahman2024towards}.

\textit{Cognitive fatigue} amplifies misconfiguration for several reasons. Indeed, when reports fail to consider explanatory context, for instance, by visualizing only summary messages or error codes, they place the burden on professionals to independently research and interpret the cause of the problem and the most appropriate mitigation. These activities increase cognitive load, leading to errors, misunderstandings, or partial application of corrective measures \cite{manfredi2021, renaud2021shame}. Similarly, \cite{rahman2024towards} demonstrated that implementing and maintaining security controls can lead to cognitive fatigue due to the high number of alerts, repetitive procedures, and complex technical decisions. These activities reduce user concentration and critical thinking. Moreover, they can lead users to automatic responses, omissions, and suboptimal choices that increase the likelihood of incorrect or incomplete configurations.  

Different studies indicate that misconfiguration is often facilitated by a \textit{lack of awareness}. For example, in the case of Non-Fungible Tokens (NFTs), lay users might not recognize relevant risks and configurations. At the same time, they tend to activate settings that they easily understand, leaving defaults unchanged, connect their wallet to marketplaces without preliminary checks, do not disconnect or log out, and ignore critical features such as 2FA \cite{chen2024}. This lack of awareness extends to backup and recovery, resulting in exposure to configuration errors and social engineering vectors. In the case of Border Gateway Protocol, administrators may not perceive the impact of their actions (e.g., incorrect filters or inconsistent policies), with errors persisting until they generate large-scale routing disruptions \cite{mahajan2002understanding}. Additionally, the configuration of the containers' orchestration system may be affected. Administrators of Kubernetes clusters often overlook the presence or severity of misconfigurations in manifests, as vulnerabilities are not immediately visible or reported by development tools \cite{Rahman2023Misconfigurations}. This leads to an underestimation of risk and a failure to take corrective action, revealing the need to design solutions that better highlight the signs of insecure configuration. Finally, in the cloud environment, the importance of awareness and staff education in preventing configuration errors is emphasized \cite{gupta2023vulnerability}. 

\textit{Lack of knowledge} in misconfiguration has been explored by different studies. For instance, overly descriptive reports force technicians to ‘fill in the information gaps’, indicating a lack of operational knowledge about the changes to be applied \cite{manfredi2021} . The previously mentioned study on Kubernetes configuration \cite{Rahman2023Misconfigurations} highlighted that practitioners often lack the necessary knowledge to mitigate misconfigurations. Finally, the study on BGP configuration associated several BGP incidents to operators' inadequate understanding of configuration semantics, pointing to a ‘lack of knowledge’ as the structural cause of misconfiguration at the network level \cite{mahajan2002understanding}.

Also \textit{misperception} emerged as a potential cause of misconfiguration. Specifically, in \cite{rahman2024towards} the authors stated that during the implementation of security controls, many requirements are formulated in an abstract or ambiguous manner, leaving room for interpretation. This ambiguity fosters misperceptions, which in this case is a distorted perception of the meaning or priority of controls. As a consequence, final configurations may diverge from the original security intent, resulting in systematic misconfigurations due to incorrect assumptions or a partial understanding of control requirements.

Regarding \textit{security self-efficacy}, the study reported in \cite{manfredi2021} proved that reports with actionable mitigations (explanations and operational snippets) increase accuracy and speed of correction, and reduce the need for external research. The authors also stated that such reports can “improve both the knowledge and skills” of administrators, thereby reducing the risk of systematic misconfigurations due to incorrect assumptions or errors in configuration tasks.

\paragraph{Affective factors.}
Frustration and shame emerged as the main affective factors affecting misconfiguration. The study by Rahmati et al. \cite{hasan2025} revealed that Wi-Fi configuration interfaces may cause significantly higher \textit{frustration} levels, due to repetitive errors and confusing configuration steps. These problems lead users to adopt insecure shortcuts or bypass essential security controls, thereby increasing the likelihood of misconfigurations. The study reported in \cite{renaud2021shame} revealed that \textit{shame} emerged as a factor influencing the willingness to disclose and address security errors, thereby indirectly contributing to the persistence of misconfigurations.

\paragraph{Social and Organizational factors.}

\textit{Lack of communication} can be a crucial factor for misconfiguration. For example, in \cite{renaud2021shame}, the authors found that shame is associated with misconfiguration errors because it can hinder internal communication, leading to delays or omissions in reporting, thereby turning an isolated error into a systemic vulnerability. The same study highlights that shame-based responses within organizations foster a \textit{lack of trust}: employees who feel monitored or publicly blamed after an incident perceive diminished credibility, while managers become more suspicious and controlling. This mutual erosion of trust limits an open dialogue and collaborative problem-solving, exacerbating the persistence and recurrence of configuration errors.

Several studies focused on the \textit{lack of resources}, highlighting the importance of this factor. For example, missing information, time, tools, and training increase the likelihood that configurations will remain incorrect or be corrected late \cite{manfredi2021}. In an empirical validation, it has been proven that explanations and executable snippets serve as a cognitive resource that reduces time and errors, while their absence hinders remediation \cite{manfredi2022empirical}. Also, limited budgets, inadequate tools, and insufficient human resources/training lead to flawed control implementations, resulting in misconfiguration and poor security hygiene \cite{rahman2024towards}. In the study on Kubernetes manifests, it also emerged that practitioners report knowledge gaps due to a lack of training material that should help prevent and correct misconfigurations \cite{Rahman2023Misconfigurations}. Finally, in the context of cloud analysis, the absence of periodic audits, continuous monitoring, and education paths leaves misconfigurations unresolved \cite{gupta2023vulnerability}.

Finally, \textit{security posture} can impact misconfiguration. Rahman et al. demonstrated that configuration errors weaken the overall security posture of the system, making privilege escalations and unauthorized access more likely \cite{Rahman2023Misconfigurations}. Similarly, during peak workloads and with limited resources, some administrators consciously accept violations of best practices (e.g., leaving default VLANs active) to “make the network work,” leaving it vulnerable to attacks; this represents a clear degradation of security posture due to misconfiguration and organizational choices \cite{mushi2018human}.

\section{Interactions among factors}
\label{sec:interplay}

A thorough comprehension of the human factors that affect security incidents is incomplete unless these factors are examined within an integrative framework that illustrates their interdependence. Cognitive biases, affective states, behavioral patterns, personality factors, demographic traits, and socio-organizational contexts do not operate in isolation; instead, they dynamically interact, simultaneously magnifying vulnerabilities and offering protective buffering processes.

Applying the \textit{AI-assisted Systematic Screening} detailed in Section~\ref{sec:hybrid_protocol}, we retrieved 172 peer-reviewed publications specifically investigating relationships among these variables. The analysis of these studies yielded \textbf{484} items of empirical evidence, which consolidate into \textbf{302 unique pairwise interactions} among human factors. It is critical to clarify the epistemological nature of these links. During our Human-in-the-Loop validation (Section~\ref{sec:hybrid_protocol}), mere theoretical ``co-mentions'' were strictly excluded; every mapped interaction represents an empirically tested relationship. However, because these relationships are derived from highly heterogeneous study designs (ranging from cross-sectional surveys to controlled lab experiments), they encompass a broad spectrum of statistical effects—including simple correlations, moderating effects, mediating pathways, and direct causal links. Consequently, these 302 interactions carry varying degrees of causal weight. As such, unless a specific causal link is explicitly established by the original experimental study, these interactions should be conservatively interpreted as strong probabilistic associations rather than deterministic rules.

Given the high dimensionality of this network, the exhaustive dimension-by-dimension analysis is provided in \hyperref[sec:appendix_2]{Appendix 2} and summarized in Tables \ref{tab:demographics_interactions}--\ref{tab:social_interactions}. In this section, we instead synthesize the results into high-level systemic mechanisms, identifying the most important patterns that drive vulnerability.

\subsection{Quantitative Empirical Validation of the Architecture}
\label{sec:validation}

Before detailing specific psychological and behavioral mechanisms, it is crucial to address a foundational question: is the integration of Attribution Theory (Layer 1) and the CAB model (Layer 2) a mere conceptual convenience, or does it accurately reflect the structural reality of human factors in cybersecurity?

To validate the theoretical assumptions of MORPHEUS—specifically, the distal-to-proximal cascading nature of the causal pathway—we analyzed the directional flow of all 302 extracted interactions. By classifying every mapped factor into its respective architectural layer, we calculated the structural compliance of the empirical literature against our model.

The topological flow provides robust quantitative support for our architectural choices (as visually summarized in Figure~\ref{fig:causal_pathway}):
\begin{itemize}
    \item \textbf{82.8\% (250 interactions) are ``Architecture-Compliant''}: The vast majority of empirical evidence flows exactly as predicted by MORPHEUS. Specifically, influence moves hierarchically from Modulators to Direct Factors (L1 $\rightarrow$ L2 Cascades: 32.1\%), circulates within the proximal CAB sub-systems (L2 $\rightarrow$ L2 Internal Loops: 28.5\%), or represents intra-modulator baseline dynamics (L1 $\rightarrow$ L1: 22.2\%).
    \item \textbf{17.2\% (52 interactions) represent ``Recursive Feedback'' (L2 $\rightarrow$ L1)}: A minority of interactions exhibit an inverse flow. A qualitative review of these edges reveals that they do not invalidate the conceptual model; rather, they capture longitudinal, ecological shifts within the Socio-Organizational context  (e.g., recurrent habits eventually altering external organizational norms, or affective states like shame and fear degrading communication among peers).
\end{itemize}

This high structural compliance demonstrates that the proposed Layer 1 $\rightarrow$ Layer 2 architecture is not an arbitrary taxonomy, but the statistically dominant causal topology underlying the human-factor cybersecurity literature.

\subsection{The Twelve Key Interaction Mechanisms}

Having quantitatively validated the framework's macro-topology, we distilled the dense network of 302 pairwise micro-interactions into \textbf{twelve} high-level \textbf{Key Interaction Mechanisms}. This distillation is the result of a rigorous qualitative thematic synthesis, deeply informed by the directional flow of the mapped interactions. To ensure methodological transparency, the process followed three structured phases: (1) \textit{Initial Coding:} The 302 micro-interactions were mapped as a network of nodes and edges, and assigned descriptive codes based on their functional dynamics; (2) \textit{Pathway Clustering:} We iteratively grouped these coded edges by analyzing their directional asymmetry (e.g., mapping how demographic traits act exclusively as systemic sources, while specific affective states form dense bidirectional sinks); (3) \textit{Architectural Alignment:} These functional clusters were evaluated against the causal flow of our framework to ensure they represent complete, end-to-end vulnerability pathways. All data supporting this analysis, including the list of interactions identified in each pattern, are available in the FigShare repository as additional material.

This data-informed process yielded twelve dominant thematic clusters, which functionally divide into two distinct operational categories: 
\begin{itemize}
    \item \textbf{Group A: Architectural Cascades}: Mechanisms that strictly follow the theoretical pathway from distal Modulators (Layer 1) to proximal Direct Factors (Layer 2).
    \item \textbf{Group B: Internal Feedback Loops}: Mechanisms representing horizontal, self-reinforcing spirals trapped entirely within the proximal CAB core (Layer 2), naturally emerging from bidirectional links.
\end{itemize}

These mechanisms, which are reported in summary in Table~\ref{tab:mechanisms_summary}, serve as representative archetypes strictly anchored to the densest patterns in the interaction matrix, illustrating how isolated factors systematically converge to drive vulnerability.

\begin{table*}[htbp]
    \centering
    \small
    \caption{Summary of the Twelve Key Interaction Mechanisms (derived from 302 unique empirical links; $n$ reports the total volume of supporting interaction instances). The table integrates human factors directly into the functional dynamic to highlight the architectural flow.}
    \label{tab:mechanisms_summary}
    \renewcommand{\arraystretch}{1.5}
    \begin{tabularx}{\textwidth}{l >{\hsize=0.8\hsize}X c >{\hsize=1.2\hsize}X}
        \toprule
        \textbf{\#} & \textbf{Key Interaction Mechanism} & \textbf{Support ($n$)} & \textbf{Architectural Flow \& Functional Dynamic} \\
        \midrule
        \rowcolor{gray!10} \multicolumn{4}{l}{\textbf{Group A: Architectural Cascades (Layer 1 $\rightarrow$ Layer 2)}} \\
        1 & Distal Modulation of Cognitive Resources & 152 & \textit{Personality} and \textit{Demographics} act as chronic regulators of affective thresholds and cognitive resource depletion. \\
        2 & The Dual Cognitive Process: Heuristic vs. Systematic & 23 & The resource-dependent shift between analytical reasoning (System 2) and heuristic shortcuts (System 1) driven by \textit{Decision Fatigue}. \\
        3 & Demographic Divergence in Risk Profiles & 65 & \textit{Age, Gender,} and \textit{Education} systematically bifurcate the causal flow, targeting distinct affective or cognitive CAB dimensions. \\
        4 & The ``Double-Edged Sword'' of Desirable Traits & 29 & \textit{Agreeableness} and \textit{Conscientiousness} facilitate compliance but simultaneously induce cognitive tunneling and social engineering vulnerability. \\
        5 & The Trust and Bias Overconfidence Trap & 14 & \textit{Lack of Trust} and \textit{Biases} converge to artificially inflate perceived competence, decoupling it from actual operational skill. \\
        6 & Social Amplification and Silence Loops & 59 & \textit{Social Influence} and \textit{Norms} generate cultural friction and \textit{Shame}, inhibiting incident reporting and organizational communication. \\
        7 & The Resource-Constraint Cascade & 16 & Systemic \textit{Lack of Resources} (time/support) triggers a top-down degradation of the user's \textit{Stress} and \textit{Self-Efficacy} levels. \\
        8 & The ``Dark Traits'' Risk Pathway & 14 & \textit{Narcissism} and \textit{Greed} bypass standard analytical filters, leading to \textit{Impulsivity} and willful non-compliance. \\
        
        \midrule
        \rowcolor{gray!10} \multicolumn{4}{l}{\textbf{Group B: Internal Feedback Loops (Intra-Layer 2 Spirals)}} \\
        9 & The Motivation Balance: Risk Beliefs and Self-Efficacy & 31 & Non-linear feedback between \textit{Risk Beliefs} and \textit{Self-Efficacy}; threat perception without coping capacity triggers defensive avoidance. \\
        10 & The Cognitive-Affective Bottleneck & 68 & A bidirectional loop where affective \textit{Stress} depletes executive functions, while \textit{Cognitive Fatigue} further impairs affective regulation. \\
        11 & The Habitual Autopilot Loop & 33 & \textit{Recurrence} establishes a behavioral-to-cognitive bypass, where automaticity suppresses \textit{Reflectiveness} and exacerbates \textit{Bias}. \\
        12 & The Compulsive Risk Spiral & 51 & \textit{Internet Addiction} actively overrides risk assessment through \textit{Impulsivity}, driving acute operational violations. \\
        \bottomrule
    \end{tabularx}
\end{table*}

\subsection*{Group A: Architectural Cascades}

\paragraph*{\textbf{1. Distal Modulation of Cognitive Resources.}}
\label{int_pat:distal}
Our empirical analysis quantitatively validates the top-down cascading architecture of MORPHEUS, demonstrating that the Internal Modulators of \textit{Layer 1} (i.e., Personality and Demographics) act as distal antecedents that actively dictate the operational baseline for the proximal CAB core in \textit{Layer 2 (Direct Factors)}. Within this cascade, \textit{Personality Traits} function as chronic regulators of the user's affective thresholds. The interaction data show that \textit{Neuroticism} acts as a systemic amplifier for the affective dimension, exhibiting strong positive associations with \textit{Stress}, \textit{Anxiety}, and \textit{Fear}~\cite{Budimir2021CybersecurityEmotions, Cuadrado2024Technostress, Maier2019Technostress, Leslie2025Gender, Regzedmaa2024Systematic, Rossier2004NEO16PF, Thorp1993Personality, Maricutoiu2014Meta, Dong2022Anxious}. This chronic hyper-arousal subsequently triggers a cross-dimensional chain reaction: it consumes executive functions, causing a premature depletion of cognitive resources and thereby increasing users' \textit{Cognitive Fatigue}~\cite{Botvinick2001Conflict, Mizrak2025DigitalDetox, Amin2024TechnostressEducators, Cuadrado2024Technostress, Tarafdar2019Technostress}. 

Similarly, \textit{Demographics} show complex structural trade-offs within the causal flow: while older \textit{Age} is directly associated with increased \textit{Cognitive Fatigue}~\cite{Bouche2025Mental}, our interaction matrix identifies a compensatory mechanism. Higher age is simultaneously coupled with higher \textit{Conscientiousness} and  \textit{Risk Aversion}~\cite{Nicholson2005Personality, Allemand2008Personality, Soto2012Development, Vecchione2012Gender}, acting as a robust psychological buffer that neutralizes impulsive errors despite the reduced cognitive stamina.

\paragraph*{\textbf{2. The Dual Cognitive Process: Heuristic vs. Systematic Processing.}}
\label{int_pat:dual_cognitive}
A critical interaction pattern emerges from the network topology, detailing how cognitive resources are dynamically allocated within Layer 2. This dynamic heavily references the Heuristic-Systematic Model (HSM)~\cite{Chaiken1980HSM}. The data reveals a constant, resource-dependent shift between two modes of thinking: \textit{systematic processing} (System 2: deep, reflective analysis) and \textit{heuristic processing} (System 1: mental shortcuts and biases)~\cite{kahneman2011thinking}.

\textit{Layer 1} dispositional traits act as the initial gatekeepers for these competing systems: \textit{Cognitive reflectiveness} is positively correlated with \textit{Openness} and \textit{Conscientiousness}~\cite{ge2021personal}, leading these individuals to resist heuristic shortcuts~\cite{Frauenstein2020Susceptibility}, while low self-regulation (associated with high \textit{Neuroticism}) drives users toward impulsive, heuristic reactions~\cite{Waqas2023Enhancing}.

However, the architectural flow demonstrates that cognitive depletion can easily disrupt these dispositional baselines. Specifically, \textit{Decision Fatigue}~\cite{Pignatiello2020Decision, stanton2016security} and external adversarial triggers like time pressure~\cite{williams2017individual} actively reallocate resources away from System 2, forcing even conscientious users to default to biased, heuristic processing. Moreover, while active \textit{Cognitive Reflectiveness} can reduce the reliance on heuristics~\cite{AlosFerrer2016CognitiveReflection, Chou2021Mindless} and heighten threat perception~\cite{Musuva2019Cognitive}, it can be bypassed by the behavioral inertia of dysfunctional \textit{Recurrence} patterns~\cite{Vishwanath2015Examining}.

To counteract this heuristic default, systemic interventions must artificially inject cognitive friction. Foundational technical knowledge~\cite{Wash2020Experts} and specific training~\cite{iuga2016baiting} provide the necessary schemas to engage System 2. Furthermore, organizations should deploy environmental constraints---such as security tools that embed in-the-moment awareness messages (e.g., warning banners for external phishing)---to interrupt System 1 and force an allocation of resources back to \textit{Cognitive Reflectiveness}~\cite{Kamar2023Moderating, Williams2018Exploring}.
Even subtle design interventions, such as intentionally introducing perceived \textit{Uncertainty} in an interface, can trigger the necessary analytical reflection to evaluate deceptive information~\cite{Vishwanath2015Examining}.

\paragraph{\textbf{3. Demographic Divergence in Risk Profiles.}}
\label{int_pat:demographic}
Our topological analysis highlights that \textit{Layer 1} demographic variables do not dictate a universal baseline of security; rather, demographic factors affect specific dimensions of the CAB core differently. 
Specifically, regarding \textit{Gender}, the male demographic profile systemically amplifies vulnerabilities within the cognitive and behavioral dimensions, driving higher \textit{Overconfidence} in their skills~\cite{aldarwish2019framework, anwar2017gender, bell2022exploring, Verkijika2019SelfEfficacy, Sun2016Mediating} and higher \textit{Risk-Taking} behavior~\cite{kennison2020risks, Pavlicek2021Personality, Nicholson2005Personality}. Conversely, the female demographic profile predominantly modulates the affective dimension, exhibiting higher levels of \textit{Digital Anxiety}~\cite{Yoon2016Computer}, as well as increased \textit{Neuroticism}~\cite{Weisberg2011Gender, Leslie2025Gender}, which induces hesitancy and acts as a psychological buffer by fostering higher risk aversion~\cite{fatokun2019Impact, kennison2020risks}.

A similar structural divergence occurs across \textit{Age}. Older users exhibit a more secure cognitive baseline, being more risk-averse~\cite{Nicholson2005Personality} and reflective~\cite{ge2021personal}; however, this is offset by self-reported lower ICT skills~\cite{bell2022exploring} and a heightened susceptibility to \textit{Social Proof}~\cite{fatokun2019Impact} as a compensatory heuristic. Conversely, younger users, despite being digital natives, exhibit pronounced vulnerabilities across the affective and behavioral dimensions, reporting higher \textit{Stress}~\cite{nwachukwu2020covid} and a greater systematic propensity to \textit{Internet Addiction}~\cite{LozanoBlasco2022Internet, Sechi2021Addictive}.

These divergent interactions suggest that a ``one-size-fits-all'' approach to human factors is architecturally flawed, as the Layer 2 mechanisms that shield one demographic profile may serve as the primary points of exploitation for another. Because demographic modulators represent immutable external constraints rather than trainable cognitive skills, these findings underscore the necessity of moving toward context-aware security architectures that dynamically adapt to the user's specific risk profile.

\paragraph{\textbf{4. The ``Double-Edged Sword'' of Desirable Traits.}}
\label{int_pat:double}
A counterintuitive architectural pattern emerges within the dispositional and demographic nodes of Layer 1: traits that are typically considered positive in a general workforce context create unintended vulnerability pathways in adversarial contexts. For instance, while \textit{Conscientiousness} typically increases systemic compliance with security policies~\cite{marin2023influence} and strengthens the effect of positive behavioral attitudes~\cite{Shropshire2015Personality}, our topological data reveals that it can simultaneously induce cognitive tunneling: highly conscientious users may indeed develop rigid heuristic loops---such as over-processing every inbound communication---paradoxically amplifying their susceptibility to phishing~\cite{Vishwanath2015Examining}. Furthermore, cross-dimensional demographic analysis indicates that within specific clusters, such as female users, high conscientiousness can actually predict riskier self-reported cybersecurity behaviors, functioning as a latent vulnerability vector rather than a shield~\cite{kennison2020risks}.

Similarly, \textit{Agreeableness} structurally supports organizational alignment and policy compliance~\cite{Shropshire2015Personality}, but it also serves as a vulnerability vector for social engineering. The interaction data explicitly shows that agreeableness suppresses critical distrust (\textit{Lack of Trust}), inherently lowering alert levels in users. This trait also increases their vulnerability to adversarial \textit{Social Influence}~\cite{Cusack2018Personality, Shropshire2015Personality, Palm2025Influence}, while also depleting the cognitive resources required to maintain high \textit{Vigilance}~\cite{Rossier2004NEO16PF}.

Finally, \textit{Education} presents a nonlinear mapping to the cognitive layer. Contrary to intuitive assumptions, formal educational attainment does not mitigate the \textit{Lack of Awareness} factor. Counterintuitively, evidence suggests that undergraduate students may possess a more robust cybersecurity awareness than post-graduate students~\cite{fatokun2019Impact}, and highly educated employees may exhibit deeper blind spots regarding password hygiene and organizational data protection~\cite{Sari2023Demographic}. This indicates that general cognitive skills acquired through academia do not automatically translate into cybersecurity resilience, meaning high-level education cannot serve as a proxy for operational threat readiness.

\paragraph{\textbf{5. The Trust and Bias Overconfidence Trap.}}
\label{int_pat:trust}
Beyond a simplistic knowledge deficit model commonly associated with the Dunning-Kruger effect ~\cite{Kruger1999UnskilledUnaware}, our analysis reveals that \textit{Overconfidence} functions as a cognitive sink node, driven by a convergence of dispositional, demographic, cognitive, and social factors. Specifically, the systemic reliance on heuristics (\textit{Bias}), coupled with high interpersonal trust baseline (inverse of \textit{Lack of Trust}), artificially inflates the user's self-assessed competence~\cite{wang2016overconfidence, Levine2014TDT, williams2017individual, aleroud2020examination}. 

Furthermore, the interaction matrix demonstrates that this cognitive distortion is heavily catalyzed by Layer 1 distal antecedents. Specifically, it exhibits strong positive modulation from the male \textit{Gender} demographic~\cite{bell2022exploring, aldarwish2019framework, anwar2017gender} and specific dispositional traits such as \textit{Extraversion}~\cite{Fawad2020Personality, Schaefer2004Overconfidence} and \textit{Narcissism}~\cite{Campbell2004Narcissism, Paulhus2002DarkTriad}. Ultimately, this architectural convergence forms a dangerous ``confidence trap'': users possessing these targeted dispositional profiles suffer a structural disconnect between their perceived and actual operational competence, neutralizing their threat perception and rendering standard educational warnings ineffective.

\paragraph*{\textbf{6. Social Amplification and Silence Loops.}}\label{int_pat:social_silence}
Social and organizational factors act as powerful multipliers of individual cognitive vulnerabilities. The interaction matrix isolates a critical, cross-dimensional ``Silence Loop'' driven by the mutual reinforcement between \textit{Lack of Communication} and \textit{Shame}. Topological data reveals that this affective node is heavily modulated by Layer 1 distal factors: demographics structurally predisposed to higher baseline shame---often younger individuals~\cite{Gambin2018Relations, Orth2010Shame} or female users~\cite{malinakova2020Psychometric} are significantly less likely to communicate or seek assistance. Crucially, this affective paralysis is worsened when paired with a cognitive \textit{Lack of Knowledge}~\cite{Wang2019Technology}. This resulting silence severs the organizational feedback loop, preventing the timely reporting of critical threat vectors such as spear-phishing~\cite{Distler2023Influence}. Architecturally, this implies that organizations must engineer psychologically safe reporting environments, avoiding punitive policies that inadvertently transform shame into a systemic barrier against threat visibility. 

In a parallel social dynamic, the \textit{Social Proof} node operates as a systemic double-edged sword. While peer influence can validate secure baseline behaviors, it frequently overrides individual analytical processing to dictate compliance, blindly shaping users' \textit{Attitude toward Policies}~\cite{Ifinedo2014Information, Cheng2013Understanding, Safa2015Information} and artificially skewing their perceived \textit{Knowledge}~\cite{Hong2023Influence}. This structural bypass demonstrates that reliance on collective validation can act as a viral vector, propagating systemic ignorance and noncompliance whenever the localized group norm is compromised.

\paragraph{\textbf{7. The Resource-Constraint Cascade.}}
\label{int_pat:resource}
Layer 1 organizational constraints exert a top-down, cascading pressure across the proximal CAB core. The interaction matrix isolates \textit{Lack of Resources} (e.g., time deficits and inadequate support) as a primary structural antecedent of \textit{Stress}~\cite{Amin2024TechnostressEducators} while demonstrating an inverse association with \textit{Security Self-Efficacy}~\cite{Chen2024GroupDiscussion}. Architecturally, this cross-dimensional cascade reframes human error not as an isolated behavioral failure, but as a systemic symptom of an under-resourced environment. Enforcing strict policy compliance without providing adequate operational resources inevitably leads to affective burnout and a collapse of the user's cognitive belief in their capability to perform security tasks effectively.

\paragraph{\textbf{8. The ``Dark Traits'' Risk Pathway.}}
\label{int_pat:dark}
Moving beyond accidental errors, the architectural framework isolates a distinct vulnerability pathway for intentional non-compliance driven by antagonistic Layer 1 dispositional traits. Specifically, \textit{Narcissism} and \textit{Greed} structurally bypass standard analytical filters within the CAB core. The interaction matrix reveals that \textit{Narcissism} acts as a systemic catalyst, driving positive modulations across multiple risk nodes: it amplifies \textit{Overconfidence}~\cite{Paulhus2002DarkTriad, Campbell2004Narcissism}, exacerbates \textit{Impulsive Behavior}~\cite{Jones2011Impulsivity}, increases adversarial susceptibility to \textit{Social Influence}~\cite{Hart2025Phishing} and baseline \textit{Online Exposure}~\cite{Leite2023Dark, Kircaburun2018Dark}, and fosters procrastination~\cite{Meng2024Procrastinators}. At the same time, it affects organizational alignment, worsening the user's \textit{Attitude toward Policies}~\cite{Maasberg2020DarkTriad}. 

On an architectural level, this defines an opportunistic insider profile that violates security norms not out of a cognitive knowledge deficit, but through a dispositional sense of entitlement or gain-seeking that actively overrides standard risk perception mechanisms. Consequently, this pathway represents a structural vulnerability that cannot be effectively mitigated through standard educational training, necessitating instead strict environmental constraints and active behavioral monitoring.

\subsection*{Group B: Internal Feedback Loops}

\paragraph{\textbf{9. The Motivation Balance: Risk Beliefs and Self-Efficacy.}}
\label{int_pat:motivation}
Within the cognitive dimension of the CAB core, the topological data empirically isolates a critical motivation balance between \textit{Cyber Risk Beliefs} and \textit{Security Self-Efficacy}, firmly anchoring the framework in Protection Motivation Theory (PMT)~\cite{boer1996protection}. The interaction matrix demonstrates that these two cognitive nodes must maintain a structural equilibrium to effectively drive secure baseline behaviors~\cite{Kwak2020Spear, Lee2023Thwarting, Verkijika2019SelfEfficacy} and foster a resilient\textit{ Attitude toward Policies}~\cite{DelsoVicente2025ComplianceReview, Li2022Investigation, Martens2019Investigating}.

An important architectural discovery is the feedback loop between threat perception and perceived self-competence. Initially, high \textit{Cyber Risk Beliefs} act as a cognitive catalyst, increasing the mental resources a user allocates to analytical reflection~\cite{vishwanath2011people, Kwak2020Spear} . However, if \textit{Cyber Risk Beliefs} spike without a compensatory high \textit{Security Self-Efficacy}, the topological flow exhibits a direct inverse modulation, actively suppressing the user's remaining self-efficacy~\cite{Wang2017Coping}. This cognitive collapse may then propagate into the affective dimension, driving severe spikes in \textit{Anxiousness}, \textit{Fear}, and \textit{Stress}~\cite{Wang2017Coping, Hengen2021Stress}. Consequently, this affective overload induces a defensive paralysis, replacing proactive protection with avoidance~\cite{Brennan2010Fear}. Operationally, this can manifest as a ``lie bias'', where fear corrupts analytical filters, causing users to misjudge benign operational tasks as threats, thereby hampering productivity~\cite{Wang2017Coping}. 

Therefore, when organizations deploy targeted affective stressors (e.g., fear appeals) to motivate compliance, they must precisely calibrate these interventions against the specific cognitive baselines and biases of their specific employees~\cite{DelsoVicente2025ComplianceReview}. When architecturally balanced, targeted fear appeals can indeed neutralize employees' \textit{Misperception} and realign the user's \textit{Cyber Risk Beliefs} with reality~\cite{stanton2016security, Jansen2019Design}.

\paragraph*{\textbf{10. The Cognitive-Affective Bottleneck.}}\label{int_pat:cognitive_emotional}
Within Layer 2, the topological interaction network highlights a critical structural bottleneck centered on the interplay between \textit{Stress} and \textit{Cognitive Fatigue}. These two factors establish a detrimental, self-reinforcing feedback loop. Specifically, affective-to-cognitive interference involves high \textit{Stress} and \textit{Anxiousness} substantially depleting executive functions and driving \textit{Cognitive Fatigue}~\cite{Amin2024TechnostressEducators, Botvinick2001Conflict, Cuadrado2024Technostress}, while cognitive-to-affective interference ensures that a fatigued cognitive state impairs affective regulation. This lowers the psychological vulnerability threshold, increasing the user's susceptibility to further \textit{Stress}, \textit{Frustration}, and \textit{Anxiousness}~\cite{Tarafdar2019Technostress, Mizrak2025DigitalDetox}. 

Operationally, this severe cognitive-affective bottleneck degrades users' security by exhausting their analytical resources. It forces users to bypass System 2 and default to heuristic processing~\cite{stanton2016security} while structurally exacerbating their baseline \textit{Risk Attitude}~\cite{Jia2022MentalFatigue}.

\paragraph{\textbf{11. The Habitual Autopilot Loop.}}
\label{int_pat:habitual}
Within the proximal Layer 2 CAB core, the interaction matrix isolates a robust behavioral inertia mechanism driven by the \textit{Recurrence} factor. Specifically, repetitive behavior establishes a behavioral-to-cognitive bypass, acting as a modulator that increases reliance on \textit{Biases} and subsequently suppresses \textit{Cognitive Reflectiveness}~\cite{vishwanath2011people, Vishwanath2015Examining}. This topological flow indicates that as operational tasks transition into automaticity, they systematically circumvent System 2 analytical processing. Architecturally, this mechanism clarifies the long-term decay of standard awareness campaigns: the sheer repetition of daily tasks induces cognitive tunneling, effectively desensitizing users to interface friction and critical warning signals~\cite{Anderson2015Polymorphic, Kalsher2006BehavioralCompliance}.

\paragraph{\textbf{12. The Compulsive Risk Spiral.}}
\label{int_pat:compulsive}

Architecturally distinct from the passive inertia of the ``Habitual Autopilot Loop'', the topological interactions surrounding the \textit{Internet Addiction} node define a pathological pathway to vulnerability. The interaction matrix isolates robust, mutually reinforcing positive feedback loops between this compulsive behavior and both \textit{Impulsive Behavior} and \textit{Risk-Taking}~\cite{Diotaiuti2022Internet, hadlington2017human, Marzilli2020Internet, Salehi2023Impulsivity}. Unlike habitual automaticity---which is characterized by an unconscious bypass of attentional resources---this compulsive spiral is driven by active behavioral dysregulation. The systemic drive for continuous online connectivity actively overrides the user's cognitive risk assessment mechanisms, driving acute operational violations, such as the intentional circumvention of established security controls.

\section{Measuring human factors}
\label{sec:hfmeasure}

To understand and mitigate cybersecurity risks caused by human behavior, it is necessary not only to identify relevant human factors but also to develop reliable methods for assessing their presence, intensity, and impact on security decisions.
Through the targeted retrieval and expert augmentation process described in Section~\ref{sec:hybrid_protocol}, we consolidated \textbf{99 measurement solutions} derived from 99 validated sources.
We did not aim at artificially restricting our selection to a single ``recommended'' instrument per factor, but we exhaustively included \textit{all} measurement tools retrieved by our search protocol that successfully passed our strict quality filters (as detailed in Section~\ref{sec:hybrid_protocol}). Specifically, an instrument was retained if its source met the bibliographic thresholds (peer-reviewed, Q3+ journal or Core Class C+ conference) and explicitly reported empirical evidence of psychometric validity and reliability. Consequently, ad-hoc or randomly adapted items lacking formal validation were systematically excluded.
These 99 tools enable researchers and practitioners to perform quantitative risk assessments, monitor changes over time, and evaluate the effectiveness of interventions. Table \ref{tab:all_tools_compact} reports a summary of the measurement instruments mapped to MORPHEUS factors, while Table~\ref{tab:measurement_tools} reported in the \hyperref[sec:appendix_3]{Appendix} provides a systematic mapping of each human factor to its respective measurement instrument(s).

\begin{table*}[ht]
\centering
\caption{Complete catalog of all 99 measurement instruments mapped to MORPHEUS factors. Tools are grouped by construct to allow for a compact visualization within the main text. Acronyms refer to the full list in \hyperref[sec:appendix_3]{Appendix 3}.}
\label{tab:all_tools_compact}
\footnotesize
\renewcommand{\arraystretch}{1.3}
\setlength{\tabcolsep}{3pt}
\begin{tabular}{|p{1.8cm}|p{3.2cm}|p{9.4cm}|}
\hline
\rowcolor[HTML]{EFEFEF} 
\textbf{Dimension} & \textbf{Target Factor(s)} & \textbf{Validated Instruments (Citations)} \\ \hline

\textbf{Demographics} & Age, Gender, Education & Standard Demographic Questionnaire (Self-report) \\ \hline

\multirow{3}{*}{\textbf{Personality}} 
 & General Traits & Big Five Inventory (BFI) \cite{john1991big}, BFI-2 \cite{soto2016bfi2}, HEXACO-24 \cite{DEVRIES2013871}, IPIP-FFM \cite{goldberb2006IPIP}, NEO-PI-R \cite{costa1992neo} \\ \cline{2-3} 
 & Narcissism & Five-Factor Narcissism Inv. (FFNI) \cite{glover2012five}, NARQ \cite{back2013narcissistic}, NPI \cite{raskin1979narcissistic} \\ \cline{2-3} 
 & Greed & Dispositional Greed Scale (DGS) \cite{zeelenberg2021dispositional}, MDGA \cite{Lambie31122022} \\ \hline

\multirow{6}{*}{\textbf{Cognitive}} 
 & Awareness \& Knowledge & HAIS-Q \cite{parsons2014haisq}, SeBIS \cite{egelman2015scaling}, CAS \cite{10.1108/OIR-01-2022-0023}, SART \cite{doi:10.1518/001872095779049499}, Attack Simulations, Focus Groups, Interviews, Video Games \cite{rahim2015systematic} \\ \cline{2-3} 
 & Fatigue \& Load & Cognitive Failures Quest. (CFQ) \cite{broadbent1982cognitive}, Fatigue Severity Scale (FSS) \cite{krupp1989fatigue}, FSMC \cite{Penner2009TheFS}, Pupil Dilation \cite{morad2000pupillography}, ECG \cite{schmitt2015monitoring} \\ \cline{2-3} 
 & Reflectiveness \& Bias & Cognitive Reflection Test (CRT) \cite{frederick2005cognitive}, A-DMC (Bias) \cite{bruine2007individual}, Decision Styles Scale (DSS) \cite{hamilton2016development}, TRDM \cite{paternoster2009rational}, Elaboration Scale \cite{vishwanath2011people}, Ad-hoc Misperception Quest \cite{walpole2020extending} \\ \cline{2-3} 
 & Risk Attitude & DOSPERT \cite{blais2006domain}, GRiPS \cite{zhang2019development}, Nicholson Risk Scale \cite{nicholson2005RTI}, RScB \cite{hadlington2017human}, Risk Perception Scale \cite{wilson2019developing} \\ \cline{2-3} 
 & Vigilance \& Distraction & Mackworth Clock Test \cite{mackworth1948mct}, PVT \cite{10.1093/sleep/34.5.581}, CPT \cite{homack2006conners}, ARCES \cite{cheyne2006absent}, MAAS (Mindfulness) \cite{brown2003mindful}, Eye-tracking \cite{zhang2006identification}, EEG \cite{othmani2023eeg} \\ \cline{2-3} 
 & Beliefs \& Efficacy & IUIPC (Privacy) \cite{gross2021validity}, SEIS (Self-Efficacy) \cite{rhee2009self}, Intolerance of Uncertainty (IUS) \cite{carleton2007fearing}, Short Overconfidence Scale \cite{Schaefer2004Overconfidence} \\ \hline

\multirow{4}{*}{\textbf{Affective}} 
 & Anxiety (General \& Tech) & STAI \cite{spielberger1999measuring}, GAD-7 \cite{spitzer2006gad7}, Beck Anxiety Inv. (BAI) \cite{beck1988beck}, ATAS (Tech) \cite{wilson2023development}, CARS \cite{heinssen1987assessing}, DAS \cite{wilson2023development} \\ \cline{2-3} 
 & Stress & Perceived Stress Scale (PSS) \cite{cohen1994perceived}, DASS \cite{lovibond1995DASS}, GSR \cite{williams2001arousal} \\ \cline{2-3} 
 & Fear & Fear Questionnaire (FQ) \cite{marks1979brief}, Cyber-Paranoia Scale \cite{mason2014ever}, EPPM \cite{witte1994fear}, PMT \cite{boer1996protection} \\ \cline{2-3} 
 & Frustration \& Shame & NASA-TLX \cite{hart1988development}, FDS \cite{harrington2005fds}, TOSCA-3 \cite{tangney2000tosca3}, ESS \cite{andrews2002predicting}, Facial Expression \cite{grafsgaard2013automatically} \\ \hline

\multirow{3}{*}{\textbf{Behavioral}} 
 & Impulsivity & Barratt Impulsiveness Scale (BIS-11) \cite{patton1995bis11}, ABIS \cite{coutlee2014abbreviated} \\ \cline{2-3} 
 & Compulsivity \& Addiction & Internet Addiction Test (IAT) \cite{young2009internet}, Internet Usage Scale (IUS) \cite{monetti2011factor}, OCI \cite{foa2002obsessive}, CIUS \cite{merkerer2009CIUS}, OCS \cite{davis2002OCS} \\ \cline{2-3} 
 & Habit \& Complacency & SRHI (Habit) \cite{verplanken2003reflections}, SRBAI \cite{gardner2012}, CPRS \cite{singh1993automation}, AICP-R \cite{merritt2019automation}, BART (Risk-taking) \cite{lejuez2002bart}, GPS (Laziness) \cite{lay1993trait}, Diaries/Logs \cite{inglesant2010true} \\ \hline

\multirow{3}{*}{\textbf{Social/Organiz.}} 
 & Policy \& Norms & Theory of Planned Behavior \cite{ajzen1991theory}, Attitude towards Security (ASR) \cite{toro2024not}, CSII \cite{netemeyer1992consumer}, StP-II \cite{modic2018stp2}, Ad-hoc Policy Quest. \cite{sommestad2019Theory}, Normative Conduct Items \cite{CIALDINI1991201}, Affective/Normative Commitment \cite{allen1990measurement} \\ \cline{2-3} 
 & Trust & Rotter's Trust Scale \cite{rotter1967new}, OTI \cite{cummings1996organizational}, McAllister's \cite{mcallister1995affect}, Trust in Automation \cite{jian2000foundations}, McKnight Measures \cite{McKnight2009TRUSTIT}, Trust Inventory \cite{mayer1995integrative} \\ \cline{2-3} 
 & Resources \& Posture & Perceived Org. Support (POS) \cite{eisenberger1986perceived}, DLOQ \cite{marsik2003DLOQ}, GoSafe \cite{al2022gosafe}, ATC-IB \cite{hadlington2017human}, Global Risk Survey \cite{MarshMicrosoft2019CyberRisk} \\ \hline

\end{tabular}
\end{table*}

\subsection{Organization of the assessment tools in three dimensions}
To better frame the role and scope of the retrieved tools, we identified three main dimensions that focus on the type of data collected by the tool: \textit{first-person}, \textit{second-person}, and \textit{third-person}. This classification is in line with well-established categorization often adopted in behavioral science and HCI, which distinguishes between these dimensions, offering insights into the cognitive processes and behavioral patterns relevant to cybersecurity.

\textit{First-person data: self-reported assessments}
These data include the direct reports of individuals about their attributes, beliefs, attitudes, and experiences, which are collected through questionnaires, scales, and interviews. Structured psychometric instruments, such as the \textit{Big Five Inventory} \cite{john1991big}, provide insight into an individual's personality traits. In contrast, Likert-scale self-reports of affective states, including fear, stress, and related factors, constitute a less structured yet equally standard tool \cite{cohen1994perceived}. Since such data reflect internal attitudes and perceived danger, first-person instruments provide valuable insights into long-term planning, organizational cultures, and behavioral tendencies. Their relative accessibility and scalability across heterogeneous populations make these tools easy to use for researchers who require a broad descriptive understanding. However, a clear downside of these methods is that they remain susceptible to \textit{self-report bias}, necessitating caution when interpreting quantitative results.

\textit{Second-person data: behavioral observation and simulation}
These data consist of first-hand observations of user activity in well-defined or semi-structured situations, such as experimental tasks, phishing simulations, decision-making games, and performance-based assessments. For example, a factor like \textit{overconfidence} can be assessed through simulated investment scenarios \cite{bruine2007individual}, while \textit{vigilance} can be evaluated with attentional tasks such as the Mackworth Clock Test \cite{mackworth1948mct}. Since they can measure observable behavior, these measures are more objective than self-reported data. However, they tend to be ecologically valid only in experimental conditions; the control of an environment can often be artificial, which limits external validity.

\textit{Third-person data: biometric and physiological indicators}. Thanks to more advanced sensor technologies, we can regularly collect third-person data, such as biometric and physiological measures in real-time. These instruments, such as electroencephalograms (EEG), eye-tracking gaze direction, heart-rate variability, and galvanic skin response, enable us to record involuntary and subconscious reactions associated with various human factors, including mental workload, stress, distraction, and fatigue \cite{othmani2023eeg, schmitt2015monitoring}. The resulting temporal resolution and objective evidence are appealing; however, these methods generally require specialized apparatus and theories to interpret the results.

Although each data source offers unique strengths and limitations, the combined use of first, second, and third-person methods can yield a more comprehensive understanding of the human factors that influence cybersecurity behavior. Integrating data from various sources is essential for developing adaptive systems that can dynamically respond to the state, weaknesses, and local needs of users. 

The following subsections present, for each dimension, the measurement strategies for each human factor related to that dimension, providing practical guidance for researchers and practitioners who aim to enhance cybersecurity through increased awareness of human risk. 
Table~\ref{tab:measurement_tools} systematically maps each human factor to the related measurement instruments.

\subsection{Demographic Factors}

Demographic factors such as \textbf{Age}, \textbf{Gender}, and \textbf{Education} are typically assessed through standard self-report demographic questionnaires. These instruments usually include a small set of items (e.g., year of birth, highest level of education attained, gender identity) and are administered alongside the main study measures. Although they do not require dedicated psychometric scales, demographic variables play a crucial role as control or moderating factors in cybersecurity research, as they systematically influence exposure, digital literacy, and preferences for protective strategies. Their simplicity and ubiquity make them easy to integrate into any empirical study that aims to contextualize human factors within broader population characteristics.

\subsection{Personality Traits}
Personality traits (\textbf{Agreeableness}, \textbf{Conscientiousness}, \textbf{Extraversion}, \textbf{Openness}, and \textbf{Neuroticism}/\textbf{Affective stability}) are commonly evaluated through validated questionnaires like the \textit{Big Five Inventory} (BFI) \cite{john1991big}, the \textit{International Personality Item Pool} (IPIP-FFM) \cite{goldberb2006IPIP}, both of which provide reliable assessments of personality dimensions. For example, the IPIP Big-Five scales correlate strongly with the commercial \textit{NEO-PI-R measure} \cite{costa1992neo}, a paid, licensed questionnaire. A more recent version of the BFI is the BFI-2 \cite{soto2016bfi2}, which offers several advantages over the original version, including a robust hierarchical structure and enhanced predictive power. The BFI-2 was developed in three versions: a full 60-item version, a compact 30-item version, and an abbreviated 15-item version. 
Another alternative is the \textit{HEXACO-24} \cite{DEVRIES2013871}, a short self-reported questionnaire designed to assess personality traits, with four items per trait. 

\textbf{Greed} can be measured through the \textit{Dispositional Greed Scale (DGS)} \cite{zeelenberg2021dispositional}. It is a seven–item self-report instrument that measures the tendency to desire “more” and to experience persistent dissatisfaction with current possessions or achievements. An alternative tool is the \textit{Multidimensional Dispositional Greed Assessment (MDGA)}, a questionnaire designed to measure adults’ dispositional greed, which encompasses the desire to acquire more than one currently has or to retain possessions at all costs, and the tendency to never be satisfied \cite{Lambie31122022}.

\textbf{Narcissism} can be assessed by using different questionnaires. The first is the \textit{Narcissistic Admiration and Rivalry Questionnaire (NARQ)} \cite{back2013narcissistic}. It conceives narcissism as a dual process encompassing self-promoting and antagonistic tendencies: \textit{admiration}, which reflects assertive self-enhancement and the pursuit of uniqueness, and \textit{rivalry}, which captures defensive self-protection through devaluation of others. The second questionnaire is the \textit{The Five-Factor Narcissism Inventory} \cite{glover2012five},  a self-report measure of the traits linked to grandiose and vulnerable narcissism, as well as narcissistic personality disorder. The last questionnaire is the \textit{Narcissistic Personality Inventory} (NPI), one of the first tools used to assess non-clinical narcissism as a stable personality trait \cite{raskin1979narcissistic}. It captures key dimensions such as grandiosity, entitlement, and exhibitionism, and is among the most widely used measures of narcissistic traits in the general population.

\subsection{Cognitive Factors}

Since cognitive \textbf{biases} are diverse, researchers often use different approaches. One useful tool is the \textit{Adult Decision-Making Competence} (A-DMC) battery \cite{bruine2007individual}, which includes seven tasks that users are required to perform to quantify their susceptibility to common biases (e.g., framing, overconfidence, resistance to sunk costs). Another quick gauge is the \textit{Cognitive Reflection Test} (CRT) \cite{branas2019cognitive}, a 3-item test that assesses reflective versus intuitive thinking; lower CRT scores indicate a higher reliance on heuristic bias and have been associated with greater phishing susceptibility \cite{ackerley2022errors}. Finally, \textit{electroencephalography} (EEG) has been proven effective in revealing heuristic biases in individuals \cite{gholami2018attentional}.

The Internet Users' Information Privacy Concerns (IUIPC) scale is widely employed to operationalize \textit{Cyber Risk Beliefs} when evaluating human factors in privacy-enhancing technologies (PETs) and investigating the privacy paradox. Specifically, it assesses users' subjective views of fairness regarding data handling through the three dimensions of Control, Awareness, and Collection
\cite{gross2021validity}.

Also, the \textbf{Cognitive fatigue} can be measured with questionnaires. The \textit{Cognitive Failures Questionnaire} (CFQ) \cite{broadbent1982cognitive} is a tool that captures frequent lapses in memory and attention due to fatigue or overload. The CFQ is a 25-item scale that predicts real-world absent-minded errors (e.g., forgetting to save work). A high CFQ score implies cognitive fatigue and attentional lapses. 
The \textit{Fatigue Severity Scale (FSS)} \cite{krupp1989fatigue} is a 9-item scale that assesses the impact of fatigue on daily functioning. In more specific settings, mental workload indexes like the \textit{NASA-TLX} \cite{hart1988development} (which includes a "mental fatigue" component) have also been used; the \textit{NASA-TLX} is widely validated for subjective workload, including cognitive strain. Another viable solution is the \textit{Fatigue Scale for Motor and Cognitive Functions (FSMC)} \cite{Penner2009TheFS}, a 20-item instrument (10 items each for cognitive and physical fatigue). Finally, physiological measurements such as EEG \cite{othmani2023eeg}, \textit{Heart Rate Variability} (HRV) \cite{schmitt2015monitoring}, and pupil dilatation \cite{morad2000pupillography} can be used to assess this factor.

\textbf{Cognitive reflectiveness} might be measured by using different questionnaires. An example is the \textit{Cognitive Reflection Test (CRT)} \cite{frederick2005cognitive}, which measures an individual’s propensity to override intuitive, yet incorrect, heuristic responses in favor of deliberate reasoning. Higher CRT scores indicate greater engagement in analytic processing and have been linked to improved phishing detection and reduced reliance on superficial cues. Another example is the \textit{Thoughtfully Reflective Decision Making} (TRDM) scale, which quantifies an individual's tendency to collect information, generate alternatives, and systematically deliberate before making a decision, as opposed to relying on impulsive or 'gut feeling' choices \cite{paternoster2009rational}. As an alternative, the \textit{Elaboration scale} evaluates the degree to which the user actively connects received stimuli (e.g., an email) with prior knowledge \cite{vishwanath2011people}. Finally, the \textit{Decision Styles Scale} (DSS) also measures an individual's propensity towards an analytical decision-making process, characterized by systematic information gathering and thorough evaluation of alternatives \cite{hamilton2016development}.

Regarding \textbf{Decision fatigue}, the \textit{Decision Fatigue Scale} (DFS) \cite{hickman2018decision} has been proposed as a tool to measure subjective decision fatigue. 

General attentional distractibility (i.e., \textbf{distraction}) can be assessed by the CFQ's distractibility sub-score \cite{broadbent1982cognitive} or by mindfulness scales. Alternatively, the \textit{Mindful Attention Awareness Scale} (MAAS) \cite{brown2003mindful} (15 items, free) measures the tendency to be attentive versus distracted; it's well-validated as an inverse measure of everyday distraction (higher mindfulness = less distraction). Other approaches include the \textit{Attention-Related Cognitive Errors Scale} (ARCES) \cite{cheyne2006absent}, which is similar to the CFQ and validated for attention lapses. Each of these has been used in studies to quantify how easily one's attention deviates, a factor linked to missing security cues. Eye-tracking \cite{le2020evaluating, zhang2006identification} and EEG \cite{almahasneh2014deep, ke2021monitoring} can also detect distraction and mind-wandering episodes.

Security-specific \textbf{lack of awareness} and \textbf{lack of knowledge} are often measured with questionnaires. An example is the \textit{Human Aspects of Information Security Questionnaire} (HAIS-Q) \cite{parsons2014haisq}, which assesses employees' knowledge of and awareness of security policies and procedures. The HAIS-Q covers domains such as email, passwords, and internet use, effectively serving as a security awareness test (available for free for research purposes). The \textit{Security Behavior Intentions Scale} (SeBIS) \cite{egelman2015scaling}, a 16-item questionnaire, and the Cybercrime Awareness Scale (CAS) \cite{10.1108/OIR-01-2022-0023} are two alternatives that measure the user's level of awareness in the cybersecurity domain. A revised version of this scale has also been proposed \cite{Timko2025Understanding}. In addition, the Situation Awareness Rating Technique (SART) \cite{doi:10.1518/001872095779049499} is a more general self-assessed tool for measuring situational awareness in human-machine environments, which can also be used in cybersecurity contexts \cite{brynielsson2016cyber}. Awareness and knowledge of cybersecurity incidents can be assessed through interviews, video games, focus groups, and attack simulations; a review of solutions on these is reported in \cite{rahim2015systematic}. 

\textbf{Misperception} is often evaluated through risk perception scales. For instance, Wilson et al. developed a questionnaire to measure risk perception \cite{wilson2019developing} and validated a 21-item form \cite{walpole2020extending} that includes four subscales: perceived affect, perceived exposure, perceived severity, and perceived susceptibility. 

\textbf{Overconfidence} is measured by comparing self-assessed ability to actual performance. In security, a typical measure is the \textit{short overconfidence scale} (some exist in psychology, e.g., a 6-item scale validated by Schaefer et al. \cite{Schaefer2004Overconfidence}), though these are less common. In all cases, evidence of validity comes from correlations between the overconfidence metric and actual errors (peer-reviewed studies consistently show that the most overconfident users are often the least skilled \cite{wang2016overconfidence}).

A well-known validated instrument for assessing risky tendencies (i.e., \textbf{risk attitude}) is the \textit{Domain-Specific Risk-Taking Scale (DOSPERT)} \cite{blais2006domain, blais2006DOSPERT}. DOSPERT is a questionnaire that assesses the likelihood of engaging in risky activities across various domains (ethical, financial, health/safety, recreational, and social). Higher DOSPERT scores have been linked to risky online behavior \cite{egelman2015scaling}. Another measure is the \textit{General Risk Propensity Scale} (GRiPS) \cite{zhang2019development}, a short unidimensional scale for overall risk-taking tendency. In studies of phishing, researchers have used \textit{Nicholson's risk propensity scale} \cite{Nicholson2005Personality} and found that higher risk-taking scores correlate with greater phishing susceptibility \cite{moody2017phish}. The \textit{Risky cybersecurity behaviours scale} (RScB) \cite{hadlington2017human} is a 20-item scale that measures users' risky cybersecurity behavior. Finally, research also proved that risk-taking behavior can be predicted through EEG analysis \cite{vance2014using}. 

\textbf{Security self-efficacy} can be assessed using the \textit{Self-Efficacy in Information Security Scale (SEIS)} \cite{rhee2009self}. It includes items that evaluate users' confidence in their ability to implement and manage protective actions, such as installing and updating security software, removing spyware, applying system patches, and configuring browser or privacy settings. This scale captures the cognitive belief that one can effectively protect personal or organizational digital assets, reflecting the motivational component that underpins proactive security engagement and resilience in the face of cyberthreats.
Furthermore, as highlighted by a recent systematic review of self-efficacy in security behavior \cite{Borgert2024}, the literature often suffers from a proliferation of randomly adapted items. To ensure rigorous assessment, it is critical to rely on formally validated scales like the SEIS or the specific self-efficacy sub-dimensions of comprehensive tools, avoiding ad-hoc formulations.

A standard instrument for \textbf{uncertainty} assessment is the \textit{Intolerance of Uncertainty Scale} (IUS) \cite{carleton2007fearing}. The IUS (27-item or the short 12-item version) is a well-validated measure of one's affective/cognitive reactions to ambiguity. It has excellent validity and reliability across cultures \cite{simos2023factor, article, 0368645f445544deb931a2afcdda11fd}. 

\textbf{Vigilance} is typically measured via performance tasks, for example, the \textit{Mackworth Clock Test} \cite{mackworth1948mct}, the \textit{Psychomotor Vigilance Task (PVT)} \cite{10.1093/sleep/34.5.581}, or the \textit{Continuous Performance Test} \cite{homack2006conners}, where lapses in detecting signals indicate reduced vigilance. Additionally, the CFQ mentioned above \cite{broadbent1982cognitive} includes items for absent-mindedness that reflect vigilance lapses. High CFQ scores or poor sustained-attention task performance would signal low vigilance, consistent with known effects of fatigue and stress on vigilance. Decreased vigilance can also be identified using eye-tracking \cite{bodala2017measuring, mcintire2014detection} (fixation duration, blink rate) and EEG alpha/theta wave patterns, which signal reduced cognitive engagement \cite{akin2008estimating, berka2007eeg}.

\subsection{Affective Factors}
\label{sec:emotional-factors}

A common measure for \textbf{anxiousness} is the \textit{State-Trait Anxiety Inventory} (STAI) \cite{spielberger1999measuring}, a 40-item questionnaire that distinguishes between baseline anxiety tendency and the current state. The STAI is a gold standard with strong validity, but it is a licensed instrument. As free alternatives, the \textit{Beck Anxiety Inventory (BAI)} \cite{beck1988beck} (for clinical anxiety) and the \textit{Generalized Anxiety Disorder 7 (GAD-7)} \cite{spitzer2006gad7} scale are often used. The GAD-7 is a brief 7-item scale validated for measuring anxiety severity in the general population, as it shows good reliability and construct validity \cite{lowe2008validation}. Moreover, affective reactions to cyberthreats (including anxiety-related negative affect) can be measured using the \textit{Positive and Negative Affect Schedule} (PANAS) and its extended form PANAS-X \cite{watson1994panas}. For instance, it has been adopted to quantify changes in positive and negative affect before and after exposure to phishing emails, privacy or security threats, and security-related tasks \cite{smith2020phishingAffect,nwadike2016affectStates,coopamootoo2016evidencebased,coopamootoo2017niftynine}.
Other solutions to measure anxiousness are based on electrodermal activity \cite{zangroniz2017electrodermal} and reduced EEG activity \cite{herman2021emotional}.

Concerning physiological assessment, anxiousness can also be detected through facial expression analysis \cite{perkins2012facial}. 

\textbf{Digital anxiety} can be measured using the \textit{Digitalisation Anxiety Scale (DAS)}, a questionnaire comprising 35 items that specifically measure anxiety related to digitalisation (as opposed to general or tech-specific anxiety) \cite{wilson2023development}.  It can be used to identify triggers of digital anxiety within organizations for targeted interventions. Another tool is the \textit{Computer Anxiety Rating Scale (CARS)}, which focuses on the behavioral, cognitive, and affective components of computer anxiety \cite{heinssen1987assessing}. During a computer interaction, greater computer anxiety was associated with lower expectations and poorer task performance, as well as with greater state anxiety, reported physiological arousal, and debilitative thoughts. Finally, the \textit{Abbreviated Technology Anxiety Scale} (ATAS) can be used to measure this factor \cite{wilson2023development}. 

A well-known instrument for assessing \textbf{fear} tendencies is the \textit{Fear Questionnaire (FQ)} \cite{marks1979brief}, which consists of a self-report scale to assess phobic fear and related avoidance, plus associated anxiety or depression. The \textit{Cyber-Paranoia and Fear Scale} \cite{mason2014ever} is a self-report questionnaire designed to measure individuals’ perceptions of threats related to information technology, encompassing both realistic and unrealistic fears, comprising 11 items. In the context of phishing or malware warnings, studies employ Likert items, such as "I am afraid I could lose important data," to gauge fear. These items are often drawn from the \textit{Extended Parallel Process Model} (EPPM) \cite{witte1994fear} or \textit{Protection Motivation Theory} \cite{boer1996protection} constructs (perceived severity and vulnerability). Another approach involves detecting physiological proxies (e.g., heart rate \cite{wu2019amusement, sartory1977investigation} and skin conductance \cite{williams2001arousal}) in laboratory studies. 

The NASA-TLX questionnaire \cite{hart1988development} is widely used to measure user \textbf{frustration} in addition to workload, thanks to its frustration component. 
Frustration can also be measured through the \textit{Frustration Discomfort Scale (FDS)}, a questionnaire developed and validated by Harrington \cite{harrington2005fds}.
Besides self-assessment tools, facial expressions can reveal individuals' frustration without the need for any survey \cite{grafsgaard2013automatically}.

\textbf{Shame} proneness can be measured by the \textit{Test of Self-Conscious Affect 3 (TOSCA-3)} \cite{tangney2000tosca3}. It presents scenarios and asks how the respondent would feel; higher shame scores indicate a tendency to respond with shame. Alternatively, the \textit{Experiential Shame Scale (ESS)} \cite{andrews2002predicting} is another validated self-report that directly asks about feelings of shame in various situations. Finally, shame may also be detected through facial expression analysis \cite{ho2004guilt}.

The \textit{Perceived Stress Scale} (PSS) \cite{cohen1994perceived, cohen1983pss10} is a classic 10-item scale for general \textbf{stress} appraisal. It measures how overwhelmed and stressed someone feels in the last month. A shorter 4-item version also exists. In cybersecurity research, PSS can be used to gauge chronic stress levels (e.g., job stress, which might correlate with errors). Another measure is the \textit{Depression Anxiety Stress Scales (DASS)} \cite{lovibond1995DASS}, which includes a stress subscale (available in both 42-item and 21-item forms). 
Literature has also proved the efficacy of stress and anxiety assessment through EEG \cite{hou2015eeg, seo2010stress, katmah2021review}, cortisol levels \cite{bozovic2013salivary, pollard1995use}, HRV \cite{article839}, Galvanic Skin Response (GSR) \cite{article728}, and electrocardiography (ECG) \cite{takada2022human}.

\subsection{Behavioral Factors}
\label{sec:behavioral-factors}

The \textit{Complacency-Potential Rating Scale (CPRS)} \cite{singh1993automation} is a specific tool developed to measure an individual's predisposition to automation \textbf{complacency}. It is a questionnaire that gauges attitudes like over-trust and low vigilance when using automated systems. Merrit et al. also proposed the \textit{Automation Induced Complacency Potential-revised scale} (AICP-R) \cite{merritt2019automation}, a 10-item questionnaire that focuses on attitudes about using and monitoring automation under high workload. In security contexts, high scores for automation can indicate that a user may become too complacent and overtrust the system to handle security tasks (e.g., dismissing warnings, failing to monitor for problems).

When referring to  \textbf{compulsive behavior}, the \textit{Obsessive-Compulsive Inventory} (OCI) \cite{foa2002obsessive} by Foa and colleagues can measure general compulsiveness. The \textit{Compulsive Internet Use Scale} (CIUS) \cite{merkerer2009CIUS} is another targeted instrument consisting of a 14-item scale that has demonstrated good psychometric properties for assessing compulsive, potentially pathological internet use. Similarly, the \textit{Online Cognition Scale} (OCS) \cite{davis2002OCS} measures problematic Internet use using a 36-item questionnaire. Finally, the \textit{Self-Report Habit Index} (SRHI) operationalizes the compulsive behavior by assessing the strength of habitual actions by measuring dimensions such as history of repetition, lack of control, and the psychological discomfort experienced when the behavior is inhibited \cite{verplanken2003reflections}.

Concerning \textbf{impulsive behaviour}, a common measure is the \textit{Barratt Impulsiveness Scale} (BIS-11) \cite{patton1995factor, patton1995bis11}, a validated 30-item questionnaire. Shorter scales like the \textit{Abbreviated Impulsiveness Scale} (ABIS) \cite{coutlee2014abbreviated} are also validated for quick assessment. Measures like the SeBIS \cite{egelman2015scaling}, which were introduced previously, also assess impulsivity and link it to risky cybersecurity behaviors. There is also evidence that impulsivity may be revealed by different EEG patterns of resting-state activity \cite{lee2017resting}.

The most widely used tool for measuring \textbf{internet addiction} is Young's \textit{Internet Addiction Test} (IAT) \cite{young2009internet}, a 20-item questionnaire that assesses symptoms of excessive internet use (salience, withdrawal, tolerance, etc.). Higher IAT scores correlate with negative outcomes.

\textbf{Internet usage} can be measured by adopting the \textit{Internet Usage Scale (IUS)}. It has been developed specifically for adolescent populations and includes 26 items that measure participants' beliefs about how their Internet usage impacts their behavior \cite{monetti2011factor}. 

There is no specific scale that measures \textbf{laziness} by that name, but this concept is inversely related to conscientiousness and proactive behavior \cite{bogg2013case}. Researchers often use procrastination scales as proxies. For example, Lay's \textit{General Procrastination Scale} (GPS) \cite{lay1993trait} is a validated 20-item measure of the tendency to delay tasks. A high procrastination score could indicate "laziness" in security upkeep (like postponing software updates).

\textbf{Recurrence} is often measured through \textit{longitudinal observation}. For instance, an organization might track how often the same individual repeats a security violation (clicks on phishing emails repeatedly over months). A high recurrence rate indicates this factor is present. Since ``recurrence'' is borrowed from clinical terminology (relapse in behavior \cite{beshai2011relapse}), its assessment is usually binary or frequency-based \cite{kelly2011predicting} (e.g., number of repeat incidents). However, a one-time approach is provided with the \textit{Self-Report Habit Index (SRHI)} \cite{verplanken2003reflections}, a 12-item scale that measures the intensity of a given habit. A shorter version, the \textit{Self-Report Behavioural Automaticity Index (SRBAI)} \cite{gardner2012}, comprises four items from the SRHI and was validated for capturing habitual behavior patterns. There is no standalone psychometric scale for recurrence, but its definition and measurement have been discussed in the literature on behavioral relapse \cite{kelly2011predicting, yamini2024psychometric}. If needed, diaries \cite{inglesant2010true} or logs \cite{jensen2017training} are valid tools (a consistent pattern over time confirms the behavior's recurrence).

Regarding \textbf{risk-taking} behaviour, the \textit{Balloon Analogue Risk Task} (BART) \cite{lejuez2002bart} is a widely used behavioral measure for assessing this aspect. In the BART, participants inflate a virtual balloon to earn rewards, balancing potential gains against the risk of the balloon bursting, which would result in losing the accumulated earnings for that balloon. This task simulates real-world risk-taking by forcing participants to weigh uncertain outcomes. The test can predict real-world risk behaviors in domains like health, safety, and addiction.

\subsection{Social and Organizational Factors}

\textbf{Attitude towards policies} is typically measured by surveying users' attitudes about following security rules. For example, Herath and Rao \cite{Herath2009Protection} used a set of Likert statements to capture policy attitudes (e.g. ``If I follow the organization's Information Security policies, I can make a difference in helping to secure [it].'') and reported high reliability in their scale. A recent instrument is the \textit{Attitudes toward Security Recommendations (ASR)} scale, a 9-item survey designed to measure employees’ attitudes toward organizational security recommendations along two distinct dimensions: perceived legitimacy and effectiveness (how right, useful, legitimate the recommendations seem) and perceived rigor (how strict, demanding, or burdensome they are). The scale has demonstrated good psychometric properties and shows that these two dimensions differently predict people’s intentions to follow security recommendations \cite{toro2024not}.  The \textit{Theory of Planned Behavior} \cite{ajzen1991theory} can be used to construct policy compliance scales, as it has been done extensively in the literature \cite{sommestad2019Theory}. Sommestad et al. \cite{sommestad2019Theory} reviewed the literature and proposed a 50-item questionnaire that measures eleven variables regarding intention to comply with information security policies. Finally, the already cited HAIS-Q \cite{parsons2014haisq} can be used to measure employees' attitude towards security policies. 

\textbf{Lack of resources} often refers to a lack of support and organizational learning, which can be estimated by using the \textit{Perceived Organizational Support} (POS) \cite{eisenberger1986perceived}. It asks employees if they feel the organization genuinely cares about their well-being and helps them do their job effectively. While not security-specific, a high POS generally implies that the organization provides resources (training, help) — indirectly fostering learning. There are also ``learning organization'' surveys like the 43-item \textit{Dimensions of the Learning Organization Questionnaire} (DLOQ) \cite{marsik2003DLOQ} and its shorter 21-item version \cite{yang2004construct} that measure continuous learning and support.

The concept of ``\textbf{trust}'' can refer to trust in technology or trust in people/organizations. In the interpersonal sense, \textit{Rotter's }\textit{Interpersonal Trust Scale} \cite{rotter1967new} is a classic, validated questionnaire comprising 25 items that assesses general trust in others. In an organizational context, the \textit{trust inventory} by Mayer et al.\cite{mayer1995integrative}, \textit{McAllister's trust scale} \cite{mcallister1995affect}, or \textit{Organizational Trust Inventory (OTI)} \cite{cummings1996organizational} can be used to measure employees' trust in management and coworkers -- these are validated in organizational psychology. For trust in technology (e.g., emails or software), scales have been adapted as well (e.g., Jian et al.'s \textit{Trust in Automation} (TiA) scale \cite{jian2000foundations} or \textit{McKnight et al. measures} \cite{McKnight2009TRUSTIT}).  

Adherence to \textbf{norms} is commonly evaluated via subjective items from models like the \textit{Theory of Planned Behavior} \cite{ajzen1991theory}. A typical approach is to ask respondents whether people important to them (or colleagues) expect them to behave securely, and whether others do so (descriptive norm). These 5-point Likert scale items have been used in many information security compliance studies and shown significant effects \cite{herath2009encouraging}. Another approach involves building measurement items based on the \textit{Focus Theory of Normative Conduct} by Cialdini \cite{CIALDINI1991201}, which categorizes norms into perceived common behavior (descriptive) or perceived social approval (injunctive). Additionally, the \textit{Consumer Susceptibility to Interpersonal Influence} (CSII) scale \cite{netemeyer1992consumer} is a validated 12-item measure with two subscales: Normative influence (8 items) and Informational influence (4 items). This questionnaire, although originally devised for marketing purposes, may be repurposed to assess how the behaviors of workplace peers influence an individual.

Evaluation of \textbf{security posture} is a higher-level assessment typically done via audits or composite indices rather than a single survey. One example is \textit{GoSafe} \cite{al2022gosafe}, an auditing and ranking system designed to assess an organization's overall security posture. GoSafe produces a score based on multiple checkpoints (policies, controls, incident history). It is a more technical measure and a proprietary approach. The \textit{Attitudes Towards Cybersecurity and cybercrime in Business} (ATC-IB) questionnaire \cite{hadlington2017human} can measure an organization's security posture with items such as ``I don't have the right skills to be able to protect the organization from cybercrime''. Moreover, Parsons et al. \cite{parsons2014haisq} note that the HAIS-Q can be used collectively to gauge an organization's human security posture. Finally, the \textit{Global Cyber Risk Perception Survey} can be administered to organizations' leaders to gain insights into the cybersecurity posture of their organizations \cite{MarshMicrosoft2019CyberRisk}.

\textbf{Social influence} can be measured by susceptibility to common persuasion tactics. The \textit{Susceptibility to Persuasion-II scale} (StP-II) \cite{modic2018stp2} is a psychometric tool that covers social influence mechanisms, designed to measure an individual's general susceptibility to persuasion, particularly in the context of scams and fraudulent offers. It essentially reveals how easily an individual can be persuaded by various strategies, and it has been specifically used in cybersecurity research to understand who is more likely to be scammed. A shorter alternative questionnaire for this factor is the \textit{Normative Influence subscale} \cite{netemeyer1992consumer}. A different solution is the \textit{Affective and Normative Commitment} scales, which assess the social influence factor, specifically focusing on organizational commitment and loyalty \cite{allen1990measurement}. It measures the affective attachment to the institution and the internalization of moral obligations regarding organizational loyalty. 

\textbf{Social proof}, a specific type of social influence where people follow the crowd, can be assessed with scenario-based questions. There is no standalone ``social proof scale'' by name, but the StP-II scale mentioned includes a sub-component for Social Influence susceptibility \cite{modic2018stp2}.

\section{Analytical Walkthroughs of the Framework}
\label{sec:operational_use}

While MORPHEUS serves as a robust conceptual framework, its true value can be better realized when it is applied to mitigate vulnerabilities in the complex reality of organizational security. To bridge the gap between theory and practice, we proposed a \textit{Human-Centric Risk Assessment Process}. This logic guides analysts in identifying and mitigating vulnerabilities through three core phases, which can be adapted to specific operational workflows (e.g., prevention, forensics, or red teaming):

\begin{enumerate}
    \item \textbf{Factor Identification \& Filtering:} The process begins by retrieving the full set of human factors associated with a specific threat via the framework's mapping (Table \ref{tab:hfs_vertical_larger}). The analyst then applies a \textit{contextual filter}—based on environmental constraints (e.g., time pressure), user roles (e.g., experts vs. novices), or incident history—to prioritize the subset of factors most likely to be active drivers, thereby preventing analysis paralysis. For instance, if an incident is linked to a time-sensitive workflow, the analyst would proactively filter for proximal state factors (e.g., \textit{Cognitive fatigue}, \textit{Stress}) rather than static distal traits.
    \item \textbf{Quantification \& Assessment:} The selected subset is assessed using the validated instruments provided in the Measurement Layer (Table \ref{tab:measurement_tools}) tailored to the organization's operational constraints. For example, if rapid, large-scale assessment is needed to minimize disruption, brief self-report scales (e.g., the 3-item Cognitive Reflection Test) are selected; if passive monitoring is preferred, behavioral proxies (e.g., log analysis) are utilized.
    \item \textbf{Systemic Analysis:} The interactions between these identified factors are analyzed using the network mechanisms (Section \ref{sec:interplay}) to identify and disrupt compounding vulnerability pathways and reinforcing behavioral patterns. By cross-referencing the locally measured factors against the 12 interaction mechanisms, practitioners can hypothesize how an external constraint (e.g., \textit{Lack of resources}) probabilistically exacerbates an internal state (e.g., \textit{Frustration}), guiding systemic rather than localized interventions.
\end{enumerate}

To demonstrate this process, we present \textbf{two in-depth} synthetic \textbf{operational scenarios} as vignettes illustrating the step-by-step application of the framework. To preserve the breadth of the framework's applicability without cluttering the main text, \textbf{six additional scenarios} (covering threat mapping, monitoring, red teaming, forensics, access control, and insider threats) are reported in \hyperref[sec:appendix_5]{Appendix 5}.

Since MORPHEUS represents a comprehensive synthesis of validated constructs rather than a singular hypothesis, the goal of this section is not to provide a complete empirical validation of the framework, but to demonstrate its descriptive power and operational utility in depth. We employ analytical vignettes---synthesized from common patterns identified in our literature review (e.g., \cite{Jalali2020Employees, nifakos2021influence, yeng2022investigation})---to illustrate how MORPHEUS enables practitioners to unpack complex risks that unidimensional models may overlook. Broader methodological and operational limitations are discussed in Section~\ref{sec:discussion}.

\subsection{Scenario A: Human-Centric Risk Diagnosis}
\textbf{Challenge:} Organizations often face ``mystery recurrences''—incidents that keep happening despite robust technical controls. In these cases, MORPHEUS shifts the diagnostic lens from a generic ``user error'' label to specific environmental deficits, mapping how external pressures degrade internal cognitive processing.

\begin{tcolorbox}[colback=gray!10, breakable, enhanced, colframe=gray!50, title=\textbf{Vignette: Spear-Phishing in a Clinical Setting}]
\small
\textbf{Context:} A high-pressure hospital environment faces repeated spear-phishing compromises despite advanced email gateways. Traditional post-incident reviews simply state that employees ``failed to check the sender address,'' leading to repetitive and ineffective awareness campaigns.

\textbf{Phase 1: Factor Identification \& Filtering:} Guided by the MORPHEUS mapping (Table \ref{tab:hfs_vertical_larger}), the security team notes over 20 factors linked to phishing. However, considering the context of \textit{emergency wards} and \textit{shift work}, they apply a contextual filter. They deprioritize stable internal modulators (e.g., Personality traits) and focus on External Modulators (\textbf{Lack of resources/time}, \textbf{Norms}) and their proximal Direct Factors: \textbf{Stress} and \textbf{Shame} (Affective), and \textbf{Vigilance} and \textbf{Cognitive fatigue} (Cognitive).

\textbf{Phase 2: Quantification \& Assessment:} To move beyond assumptions, analysts deploy validated tools from the Measurement Layer (Table \ref{tab:measurement_tools}). They utilize the \textit{NASA-TLX} to measure cognitive load and the \textit{Experiential Shame Scale (ESS)} to evaluate cultural friction in incident reporting. The data confirms acute decision fatigue during night shifts and a pervasive fear of reprimand.

\textbf{Phase 3: Systemic Analysis:} Applying the framework's interaction network reveals two compounding causal loops. First, the \textbf{Cognitive-Affective Bottleneck (Mechanism \hyperref[int_pat:cognitive_emotional]{10})}: acute \textit{Stress} (Affective) actively consumes executive functions, degrading \textit{Vigilance} (Cognitive) regardless of the user's actual knowledge. Second, the \textbf{Silence Loop (Mechanism \hyperref[int_pat:social_silence]{6})}: rigid authority \textit{Norms} combined with \textit{Shame} discourage junior nurses from verifying suspicious emails supposedly sent by senior doctors.

\textbf{Outcome \& Intervention:} The diagnosis shifts the root cause from ``user ignorance'' to ``environmental failure.'' Instead of assigning more training videos (which would only exacerbate cognitive fatigue), the hospital implements systemic mitigations. They alter the UI to aggressively flag external senders exclusively during high-stress night shifts (reducing cognitive burden) and establish a blameless, one-click reporting channel to deliberately break the Silence Loop.
\end{tcolorbox}

\subsection{Scenario B: Guiding Targeted and Ethical Interventions}
\textbf{Challenge:} One-size-fits-all training is often inefficient and perceived as punitive, causing organizational friction. By distinguishing between direct drivers and modulators, MORPHEUS enables ethical, targeted interventions tailored to specific psychological profiles.

\begin{tcolorbox}[colback=gray!10, breakable, enhanced, colframe=gray!50, title=\textbf{Vignette: Adaptive Controls for University Staff}]

\textbf{Context:} A university needs to reduce high click-through rates on phishing emails among administrative staff, but mandatory multi-hour training modules are disrupting essential workflows and generating pushback.

\textbf{Phase 1: Factor Identification \& Filtering:} The objective is to mitigate automated, unreflective clicking. The team focuses strictly on proximal Direct Factors rather than systemic organizational issues. They select \textbf{Impulsive behavior} (Behavioral) and \textbf{Cognitive reflectiveness} (Cognitive) as the primary targets for investigation.

\textbf{Phase 2: Quantification \& Assessment:} Instead of treating the workforce as a monolith, the security team baselines the population using the \textit{Cognitive Reflection Test (CRT)} and the \textit{Barratt Impulsiveness Scale (BIS-11)} from the framework's inventory. This psychological profiling identifies a specific high-risk cluster of users exhibiting high impulsivity and chronically low reflectiveness.

\textbf{Phase 3: Systemic Analysis:} The framework identifies this specific cluster as highly susceptible to the \textbf{Habitual Autopilot Loop (Mechanism \hyperref[int_pat:habitual]{11})}. Within the \textbf{Dual Cognitive Process (Mechanism \hyperref[int_pat:dual_cognitive]{2})}, these users process emails via fast, automatic heuristic thinking (System 1), entirely bypassing the deliberate, analytical reasoning (System 2) required to spot deceptive cues, regardless of how much theoretical knowledge they possess.

\textbf{Outcome \& Intervention:} Recognizing that generic training does not disrupt automatic habits at the moment of action, the university deploys an adaptive, targeted intervention. Only for the high-risk cluster, the system introduces ``cognitive friction'': micro-delays or forced confirmation pop-ups (e.g., ``Are you sure this link is safe?'') when clicking external links. This artificially disrupts System 1, forcing System 2 engagement exactly when needed, while ethically sparing conscientious, reflective users from unnecessary workflow disruption.
\end{tcolorbox}

\section{Discussion}
\label{sec:discussion}

This section positions MORPHEUS within the existing literature and discusses the main methodological and operational limitations that should be considered when interpreting and applying the framework.

\subsection{Positioning MORPHEUS within the Literature}
While the operational scenarios in Section~\ref{sec:operational_use} demonstrate the practical utility of MORPHEUS, it is crucial to articulate how this framework advances the existing theoretical landscape discussed in Section~\ref{sec:relwork}.

Previous research has produced highly valuable models, but these often remain bounded by either scope or theoretical abstraction. For instance, threat-specific frameworks---such as Choong's cognitive-behavioral model for passwords \cite{choong2014cognitive} or Green et al.'s Drivers of Insider Threats \cite{green2023understanding}---provide deep but isolated insights. MORPHEUS overcomes this fragmentation by providing a universal architectural scaffolding. By anchoring the taxonomy in established psychological constructs (the CAB model and Attribution Theory), MORPHEUS demonstrates that the underlying psychological drivers are shared across seemingly disparate vectors, from social engineering to system misconfigurations.

Furthermore, comprehensive organizational models like the NIST Cybersecurity Framework \cite{shen2014nist} or the Cybersecurity Culture model \cite{mwim2023conceptual} traditionally treat human factors as flat, static checklists. Even recent interdisciplinary efforts, such as the Human-Centric Cybersecurity Framework \cite{khadka2025human}, offer excellent conceptual pillars but lack granular causal mapping. MORPHEUS shifts this paradigm from a static taxonomy to a dynamic system. By charting 302 empirical interactions and distilling them into 12 recurring mechanisms, our framework explicitly models how distal situational modulators (e.g., lack of organizational resources) dynamically degrade proximal cognitive and affective states (e.g., inducing fatigue and stress). This systemic view is vital: it shifts the security narrative from merely blaming ``user error'' to identifying the environmental and psychological conditions that precipitate it.

Finally, the most significant differentiator of MORPHEUS is its explicit focus on operational measurability. Theoretical models often struggle to transition into practice because analysts lack the validated tools to assess abstract constructs. By integrating an inventory of 99 psychometric instruments, MORPHEUS bridges the gap between conceptual human-centric security and evidence-based, targeted interventions.

\subsection{Methodological and Operational Limitations}
It is essential to recognize that the proposed scenarios are theoretical vignettes. While grounded in empirical evidence, the framework has not yet undergone longitudinal ecological validation. Future work must address several critical challenges and methodological constraints in operationalizing MORPHEUS:

\begin{itemize}
    \item \textbf{Holistic Empirical Validation:} Methodologically, MORPHEUS is a theoretical synthesis built upon the aggregation of existing empirical studies. While its individual components and interactions are validated in isolation, the framework as a unified, holistic system has yet to be empirically tested in a longitudinal, real-world enterprise setting to quantify its overarching efficacy.
    \item \textbf{Heterogeneity of Interaction Data:} The 302 interactions mapped in the framework are derived from highly heterogeneous studies involving different methodologies, demographics, and threat contexts. Consequently, these relationships should be interpreted as strong statistical associations and probabilistic pathways rather than universal, deterministic causal laws.
    \item \textbf{Ethical Profiling:} Collecting psychometric data (e.g., Impulsivity in Scenario B) raises significant privacy concerns. Practitioners must ensure data is aggregated/anonymized to prevent discrimination and comply with regulations like GDPR.
    \item \textbf{Interdisciplinary Skill Gap:} Cybersecurity professionals may lack the expertise to interpret psychological scales correctly. Successful implementation likely requires collaboration with HR or organizational psychologists to avoid misdiagnosis.
    \item \textbf{Temporal Dynamics:} Factors like \textit{Stress} or \textit{Vigilance} are highly volatile. Interventions based on static measurements may lose effectiveness if not updated dynamically in response to real-time context (e.g., time of day, workload spikes).
    \item \textbf{Measurement Tools Selection:} Measurement tools are often measured with varying dimensions or adapted items~\cite{Borgert2024, chen2025deterrence}. However, because we did not conduct a quality comparison of the same constructs across different sources, readers should carefully scrutinize the choice of a specific measurement tool.
\end{itemize}

\section{Conclusions and future works}
\label{sec:conclusions}

This paper introduced MORPHEUS, a holistic framework that systematizes 50 human factors across cognitive, affective, behavioral, personality, demographic, and social dimensions to address the human-centric vulnerabilities underlying the majority of security breaches. By synthesizing psychological theories—specifically, the CAB model and Attribution Theory—with cybersecurity practice, we mapped these factors to specific critical threats, identified 302 interactions, and consolidated 99 validated measurement tools. This extensive taxonomy, achieved through a novel hybrid methodology that combines AI-assisted analysis with expert validation, provides the operational granularity often lacking in previous high-level models.

Future research will focus on extending the framework's ecological validity. A priority is the systematic mapping of external triggers (e.g., deceptive UI patterns, social engineering cues, environmental stressors) to specific human factors. Understanding this causality is crucial: while MORPHEUS currently models the user's \textit{internal} susceptibility, identifying how \textit{external} stimuli activate these vulnerabilities will enable the design of context-aware defenses that can neutralize triggers before they precipitate a failure. Additionally, we plan to pursue empirical validation of the identified feedback loops in longitudinal settings and develop adaptive security interfaces that can respond to real-time biometric indicators.

\section*{Declaration of generative AI}
During the preparation of this work, the author(s) used \textit{Grammarly Pro} in order to fix grammatical errors and improve text quality. 
The authors acknowledge the use of Large Language Models (LLMs) as a support tool during the systematic literature screening process, as rigorously detailed and validated in Section~\ref{sec:hybrid_protocol}. 

\section*{Data Availability}
The datasets, protocols, and validation logs supporting the research are available in the Figshare repository: \url{https://doi.org/10.6084/m9.figshare.30919370}.

\begin{acks}
This work has been supported by the Italian Ministry of University and Research (MUR) and by the European Union-NextGenerationEU, under grant PRIN 2022 PNRR "DAMOCLES: Detection And Mitigation Of Cyber attacks that exploit human vuLnerabilitiES" (Grant P2022FXP5B) CUP: H53D23008140001. 
    
This work is partially supported by the co-funding of the European Union - Next Generation EU: NRRP Initiative, Mission 4, Component 2, Investment 1.3 - Partnerships extended to universities, research centres, companies and research D.D. MUR n. 341 del 5.03.2022 – Next Generation EU (PE0000014 – "Security and Rights In the CyberSpace – SERICS" - CUP: H93C22000620001). 
\end{acks}

\bibliographystyle{ACM-Reference-Format}
\bibliography{acmart}

\clearpage
\section*{APPENDIX 1: Description of the MORPHEUS human factors}
\label{sec:appendix_1}

In the following, each human factor is described to clarify its meaning in the context of this study. Of course, human factors whose meaning is trivial (i.e., age and gender) are not described.  

\subsection*{A1.1 Demographic factors}

\textit{Education} refers to the level of formal schooling acquired by an individual (e.g., high school, undergraduate, postgraduate) and the specific disciplinary field of study (e.g., STEM vs. non-STEM).

\subsection*{A1.2 Personality traits}

\textit{Agreeableness} denotes a propensity for cooperative, empathetic, and trusting conduct in social contexts \cite{zhang2006thinking}. Individuals who rate high on this dimension are attached to interpersonal harmony and positive relations and are usually kind, considerate, and willing to help others \cite{graziano1997self}. Conversely, individuals exhibiting limited agreeableness tend to remain skeptical, critical, and competitive, which can influence relational interpretation, intent inference, and interpersonal responsiveness.  

\textit{Conscientiousness} is concerned with organisation, self-discipline, and goal-directedness. Highly conscientious individuals tend to be reliable, structured, and task-oriented, with plans and a tendency to stick to regular practices \cite{zhang2006thinking}. These individuals possess high impulse control and a strong commitment to task completion over the long term. Low conscientiousness, conversely, is associated with spontaneity, reduced organization, and greater openness to flexible or adaptive decision-making.  

\textit{Extraversion} denotes the degree to which individuals seek and sustain social engagement. Individuals scoring high in extraversion are characterised by vigour, extroverted demeanour, and assertiveness, actively pursuing external stimuli and interacting with large cohorts \cite{zhang2006thinking}. In comparison, introverts tend to prefer solitude or smaller, more intimate environments. This dimension influences the trends of social participation and the climate where the interaction takes place.  

\textit{Greed} denotes a dispositional personality tendency to pursue material or symbolic rewards, such as money, prizes, or advantages, often prioritizing immediate gratification over cautious or secure behavior. Within cybersecurity contexts, higher levels of greed increase susceptibility to attacks that exploit promises of financial gain or exclusive benefits, such as prize-based phishing or fraudulent reward schemes.

\textit{Openness} to experience describes the extent to which an individual engages with imaginative, intellectually exploratory, emotionally expressive, art- and beauty-sensitive, and idea-oriented stimuli. High openness is associated with the enjoyment of abstract conceptualization, novel experiences, and creative endeavors, whereas low openness indicates a preference for routine, familiarity, and concrete thought. McCrae and Costa \cite{costa1999five} delineate six facets within openness: fantasy, aesthetics, feelings, actions, ideas, and values.
This dimension also encompasses the construct of \textit{sensation seeking}, which reflects a dispositional tendency to pursue novel, intense, or stimulating experiences. 

\textit{Narcissism} denotes a personality trait characterized by an inflated view of the self, a heightened need for admiration, recognition, and status, and a sense of entitlement. It involves a focus on self-enhancement and maintaining a grandiose self-image.

\textit{Neuroticism} represents a dispositional tendency toward negative affective states such as anxiety, stress, mood instability, and self-doubt \cite{barlow2014origins}. 
Conversely, low neuroticism—often referred to as \textit{emotional stability}—reflects resilience, composure, and the ability to maintain equilibrium in the face of adverse or stressful conditions. 

\subsection*{A1.3 Cognitive factors}

\textit{Bias}, in general, denotes systematic departures from rational judgment that influence perception, memory, and decision-making processes \cite{kumar2023overconfidence}. Cognitive biases lead individuals to draw conclusions based on subjective or heuristic reasoning rather than objective evaluation. These biases influence information processing and can lead to persistent errors in judgment. 

\textit{Cognitive fatigue} constitutes a state of diminished mental capability that manifests following prolonged engagement with cognitively demanding tasks. It is usually manifested through diminished attention, reduced processing speed, and overall decline in cognitive effectiveness \cite{audiffren2023effort}. Such performance losses may lead to a transition between deliberate and automatic operations, thereby impairing performance later. 

\textit{Cognitive reflectiveness} denotes the propensity to override intuitive, heuristic-based judgments — commonly referred to as System 1 thinking, which operates automatically and rapidly — and to engage instead in deliberate, analytical reasoning (System 2 thinking), which is slower and more effortful \cite{kahneman2011thinking}. Individuals with higher cognitive reflectiveness are less influenced by affect-driven or superficial cues and show improved discrimination between legitimate and fraudulent content, particularly in phishing contexts.

\textit{Cyber risk beliefs} denote individuals’ general beliefs about the likelihood and potential severity of cyber threats. Rather than reflecting procedural knowledge or momentary awareness, CRBs capture a stable cognitive stance toward cyber risk that shapes motivation and protective intentions. 

\textit{Cybersecurity self-monitoring} (metacognitive).
Cybersecurity self-monitoring refers to an ongoing metacognitive process in which individuals track their own behavior, attention, and adherence to secure practices (e.g., verifying sender identity, checking URLs, flagging anomalies). Unlike \textit{vigilance}, which concerns sustained attention to external cues, self-monitoring emphasizes reflexive oversight of one’s own actions and standards. 

\textit{Decision fatigue} is operationally distinct yet conceptually proximate; it refers to diminished decision quality after a sequence of choices \cite{Pignatiello2020Decision}. When resources are scarce, people struggle to effectively compare and contrast the available alternatives, and the risk of making an impulsive judgment or avoiding the decision increases.

\textit{Distraction} refers to an individual’s tendency to divert attention from a primary task. Such attentional shifts can arise from environmental stimuli, cognitive overload, or competing demands, thereby reducing the ability to maintain sustained concentration. Highly distractible individuals may struggle to concentrate, particularly in situations that require focused attention.

\textit{Lack of awareness} denotes an individual’s limited capacity to perceive salient elements of the environment or to observe aspects of their own behavior \cite{dupont1997dirty}. The causal mechanisms frequently involve a lack of exposure to the environment, inadequate training, or an inability to attend to relevant cues \cite{berthet2021measurement}. Awareness must be distinguished from knowledge, which refers to the explicit understanding, rather than perception.

\textit{Lack of knowledge}, conversely, signifies an absence of requisite information or expertise. Unlike misperception, the interpretation of a stimulus as other than it manifestly is, lack of knowledge reflects an inability to access or apply necessary information during decision-making.

\textit{Misperception} occurs when individuals interpret information incorrectly, resulting in discrepancies between perception and reality. This phenomenon can stem from cognitive biases, inadequate information processing, or external influences that distort understanding \cite{jervis1968hypotheses}. Misperception differs from lack of knowledge in that it involves an active distortion of information rather than a simple absence of comprehension.

\textit{Overconfidence} represents a cognitive bias characterized by an exaggerated belief in one’s abilities, judgments, or performance outcomes. It is usually expressed in the form of overestimates of the abilities, overrating in comparison with others, or great accuracy in the perceived accuracy \cite{kumar2023overconfidence}. By attenuating the inclination to solicit external validation or reassess one’s assumptions, overconfidence can foster inflexible decision-making.

\textit{Risk attitude}, by contrast, refers to an overarching dispositional orientation toward uncertainty, ranging from pronounced risk aversion to a propensity toward high-stakes or uncertain situations \cite{herman2018risk}. Whereas risk-taking denotes actions, risk attitude represents a broader personality trait that guides perceptions and strategies across multiple life domains. 

\textit{Security self-efficacy} is an individual’s belief in their capability to recognize, avoid, and handle security threats such as phishing. Higher security self-efficacy supports systematic checking (sender identity, URL inspection, attachment types), fosters adherence to protective routines, and predicts better avoidance behavior \cite{Arachchilage2014Security}.

\textit{Uncertainty} denotes a state in which ambiguous outcomes or missing information render decision-making more complex. Individuals confronted with uncertainty may exhibit hesitation, rely on heuristics, or experience increased cognitive load, each of which influences how information is processed and responded to \cite{leder2024background}.

\textit{Vigilance} denotes an individual’s capacity to maintain sustained attention and remain alert to relevant stimuli over extended periods. High vigilance is associated with strong attentional control and the capability to detect inconsistencies or patterns. Conversely, low vigilance may lead to lapses in focus and diminish the ability to recognize important cues \cite{Warm2008Vigilance}.

\subsection*{A1.4 Affective factors}

\textit{Anxiousness} refers to a future-oriented emotional state marked by heightened alertness, uncertainty, and physiological arousal. Anxiety is typically described as the worrying and expectation of negative future states, but it extends beyond the immediate dangers of the situation, influencing anticipatory appraisals and subsequent behavior \cite{beckers2023understanding}. 

\textit{Digital anxiety} denotes apprehension, tension, or worry specifically associated with the use of digital systems, technologies, and online services. Elevated digital anxiety increases cognitive load and narrows attention to surface cues, fostering reliance on heuristics (e.g., urgency, familiarity) and hastier, less accurate judgments; as a result, it can impair the detection of deceptive content (e.g., phishing) and reduce willingness to engage with protective actions (e.g., verifying URLs, reporting).
Unlike \textit{anxiousness}, which represents a broad, future-oriented emotional state related to general uncertainty or anticipation of negative outcomes, \textit{digital anxiety} is context-specific and arises directly from the interaction with technology and online environments.

\textit{Fear} constitutes a fundamental defensive emotion elicited in response to the apprehension of immediate or proximal danger. The related physiological indicators are autonomic arousal, behavioural avoidance, and quick decision-making, aiming to reduce possible damage \cite{beckers2023understanding}. In contrast to anxiety, which is characterised by future orientation and broad, diffuse anticipation, fear is typically tied to discrete and imminent threats.

\textit{Frustration} arises when an individual’s efforts to attain a goal are repeatedly obstructed \cite{jeronimus2017frustration}. Associated with irritability and tension, particularly when affected persons perceive little control over impeding factors, frustration differs from stress in that it is specifically linked to blocked goal attainment rather than generalised pressure or demands.

\textit{Shame}, a self-conscious affect, is linked to perceptions of personal failure, inadequacy, and social disapproval \cite{budiarto2021shame}. Distinct from guilt, which is anchored to specific behaviours, shame carries a more encompassing self-evaluation. Intensity levels of shame exert influence over social engagement, disclosure behaviour, and self-perception.

\textit{Stress}, in turn, emerges from external or internal pressures that challenge an individual’s capacity to maintain equilibrium \cite{staal2004stress}. Although moderate stress can enhance alertness and problem-solving, excessive or chronic stress impairs cognitive processing, amplifies emotional reactivity, and leads to fatigue. Unlike anxiety, which stems from anticipation, stress is a reaction to specific present-day demands or contingencies.

\subsection*{A1.5 Behavioral factors}

\textit{Complacency} involves a sense of self-satisfaction that causes the reduction of awareness of risks or deficiencies  \cite{merritt2019automation}. Unlike laziness, which arises through lack of motivation, complacency reflects an unfounded belief in the sufficiency of prior effort.  

\textit{Compulsive behavior} is a repetitive action that is perceived as being compulsory by the individual, usually in relation to an intrusive thought or an urge to have control over a situation\cite{croft2022risking}. In contrast to impulsivity, which is spontaneous and unplanned, compulsive behavior is purposeful and regularly performed to mitigate discomfort. People with compulsive behaviors can feel that they do not need too many routines, but cannot stop them.  

\textit{Impulsive behaviour} stems from a predisposition to act quickly without adequate deliberation of consequences and frequently involves difficulties in delayed gratification and resisting immediate urges \cite{bakhshani2014impulsivity}. Impulsive behavior differs from risk-taking: it is not necessarily driven by the anticipation of rewards, but rather stems from temporary difficulties in self-regulation.

\textit{Internet addiction} denotes a pathological behavioral pattern characterized by compulsive, excessive, and uncontrolled use of the Internet. Distinguished from high-frequency use (\textit{Internet usage}), it involves specific psychological symptoms, such as salience, tolerance, withdrawal, and the persistence of online engagement despite negative consequences on daily life or psychosocial functioning \cite{young2009internet}.

\textit{Internet usage} refers to the amount of time an individual spends engaging with online environments and digital services (e.g., email, social media, browsing, streaming). Higher Internet usage increases exposure to potential cyber threats and reinforces habitual message processing, thereby expanding the opportunity for deceptive contact unless counterbalanced by strong awareness and self-regulatory controls. Extreme levels of internet usage can degenerate into internet addiction behaviors.

\textit{Laziness} is characterized by an aversion to effortful activity despite having the capacity to perform it, a pattern associated with diminished productivity and influenced by factors such as perceived lack of control, task disinterest, or external circumstances that curtail engagement \cite{madsen2018conception}.  

\textit{Recurrence} designates the re-emergence of a previously exhibited behavioral pattern following a period of change or remission, a pattern frequently observable in individuals who relapse to unhealthy habits or emotional reactions \cite{de2019empirical}. The term differs with mere repetition in that it refers to the renewal of actions that have been repressed or altered.  

\textit{Risk-taking} is defined as a behavioral orientation that involves engaging in actions susceptible to adverse consequences. It is often motivated by personal differences in reward sensitivity, whereby some people perceive gains as greater than losses \cite{stanford1996impulsiveness}. Unlike \textit{risk attitude}, which reflects a dispositional preference toward uncertainty and the cognitive evaluation of risky options, \textit{risk-taking} denotes the actual enactment of such preferences in specific contexts, influenced by situational and psychological moderators such as impulsivity and perceived control \cite{sitkin1992reconceptualizing, weber2002dospert}. This construct also diverges from impulsivity insofar as risk-taking can be deliberate rather than spontaneous.

\subsection*{A1.6 Organizational and social factors}

\textit{Attitude toward policies} captures an individual’s evaluative stance toward formal regulations and organizational measures. This position is shaped by political beliefs, life experiences, and views of policy success \cite{vesely2023policy}. Individuals who regard a policy as advantageous may actively support it, whereas those who deem it restrictive or ineffective may resist or challenge it \cite{vesely2023policy}. Policy attitudes influence the manner in which individuals engage with institutional rules and the extent to which they comply with or contest regulations.  

\textit{Lack of communication} occurs when information is neither shared nor interpreted effectively, thereby generating misunderstandings, inefficiency, and diminished coordination within collaborative contexts \cite{dupont1997dirty}. Communicative difficulties may arise due to unclear messages, unwillingness to share important details, or knowledge gaps.

\textit{Lack of resources} refers to conditions in which individuals or organizations do not possess the requisite tools, support structures, training opportunities, or informational assets necessary for effective and secure action \cite{dupont1997dirty}. This deficit can manifest as limited technical assistance, insufficient time, inadequate materials, or the absence of an institutional culture that promotes continuous learning and knowledge sharing. In contrast, environments that provide organizational support and foster collective learning enhance members’ ability to respond effectively to security challenges by developing adaptive strategies based on shared experience and expertise. 

\textit{Lack of trust} denotes diminished confidence toward individuals, institutions, or systems. Collaboration and cooperation cannot occur without trust; without it, communication and effective decision-making are hindered \cite{getha2019collaborating}. These shortcomings can result from past negative experiences, inconsistent leadership, or system inefficiencies. A lack of trust in organizational systems or institutions can foster skepticism, rejection of directives, and withdrawal from collective endeavors.  

\textit{Norms} denote the collective expectations operative within a group or society, providing guidelines that delineate acceptable from unacceptable conduct \cite{dupont1997dirty}. This is because such expectations create social stability and order. In organizational contexts, norms govern workplace interactions, decision-making processes, and adherence to policies. They are informal or formal rules that maintain uniformity among people who are in the same social or professional setting.  

\textit{Online exposure} denotes the extent to which an individual’s personal information, credentials, or digital traces are accessible or stored across interconnected online systems and platforms. High exposure—whether through public sharing or centralized storage—heightens the likelihood of data exploitation, profiling, or targeted attacks (e.g., credential theft or account compromise), even when overall Internet usage remains limited.

\textit{Organizational tenure} indexes accumulated exposure to training, norms, and incident learning. Increased tenure can strengthen protective routines and pattern recognition for anomalies.

\textit{Security posture} refers to the overall security status of an individual or organization, based on the configuration, maintenance, and exposure of its digital assets and behaviors \cite{al2022gosafe}. A robust security posture involves proactive risk management, sustained vigilance, and the implementation of best practices. Conversely, when security controls are perceived as inconvenient or effortful—due to complex configurations, frequent manual approvals, or excessive alerts—users may exhibit resistance, circumvention, or non-compliance, leading to weakened configuration hygiene and increased vulnerability.

\textit{Social influence} denotes the modification of behavior in response to external pressures from social groups, authority figures, or cultural expectations \cite{cialdini2004social}. It encompasses conformity, compliance, and obedience, shaping individuals’ security practices through both explicit directives and implicit social norms.
Within organizational contexts, social influence also manifests through \textit{organizational commitment and loyalty}, whereby individuals internalize collective values, identify with the institution, and align their behavior with its goals and security expectations. 

\textit{Social proof} refers to the propensity for individuals to emulate the behavior observed in others \cite{cialdini2009influence}. People believe that most people possess a better understanding of the right lines of action. When uncertainty prevails, group behavior is often adopted to avert error. This inclination encourages conformity and strengthens patterns that have been largely accepted within a specific social environment. 

\textit{Type of organization} influences exposure and control maturity through policies, training cadence, and incident learning culture, indirectly shaping users’ secure practices. Workers placed in institutions that have strict cybersecurity measures and frequent training sessions are likely to display safer practices compared to other workers.

\subsection*{A1.7 Mapping between the human factors and the cyberthreats}

\begin{center}
    \renewcommand{\arraystretch}{1.0}
    \setlength{\tabcolsep}{3pt} 
   \begin{longtable}{|p{2.5cm}|p{1.7cm}|p{1.7cm}|p{1.7cm}|p{1.7cm}|p{1.7cm}|p{1.7cm}|}
        \hline
        \textbf{Human Fac.} & \textbf{Phishing} & \textbf{Spear} & \textbf{SMishing} & \textbf{Malware} & \textbf{Password} & \textbf{Misconfig.} \\
        \hline
        \endfirsthead

        \hline
        \textbf{Human Factor} & \textbf{Phishing} & \textbf{Spear} & \textbf{SMishing} & \textbf{Malware} & \textbf{Password} & \textbf{Misconfig.} \\
        \hline
        \endhead
        
        \hline
        \multicolumn{7}{|c|}{\textit{(Table continues on the next page)}} \\
        \hline
        \endfoot
        \hline
        \endlastfoot
        \rowcolor[gray]{0.9} 
        \multicolumn{7}{|c|}{\textbf{Demographics (\textit{modulator}/\textit{external} factors)}} \\
        \hline
        Age & \cite{greitzer2021experimental, sheng2010falls, Huseynov2024Using, ge2021personal, Redmiles2018Examining} & \cite{lin2019susceptibility, eftimie2022spear} & \cite{Rahman2023Users, Huseynov2024Using, tabassum2024drives} & \cite{levesque2018technological} & \cite{whitty2015individual} &  \\
        \hline
        Education & \cite{abroshan2021covid, Butavicius2022People, Huseynov2024Using, moody2017phish} & \cite{moody2017phish} & \cite{Huseynov2024Using} &  &  & \cite{mushi2018human} \\
        \hline
        Gender & \cite{Abdelhamid2020concerns, iuga2016baiting, yeboah2014phishing, sheng2010falls, Huseynov2024Using, greitzer2021experimental, abroshan2021phishing, ge2021personal} & \cite{lin2019susceptibility, alhaddad2023study} & \cite{Timko2025Understanding, Huseynov2024Using} &  &  &  \\
        \hline

        \rowcolor[gray]{0.9} 
        \multicolumn{7}{|c|}{\textbf{Personality (\textit{modulator/internal} factors)}} \\
        \hline
        Agreeableness & \cite{Workman2008Wisecrackers, LopezAguilar2025Phishing, Wright2010Influence, Abdelhamid2020concerns} & \cite{eftimie2022spear, Abdelhamid2020concerns} &  & \cite{Huseynov2024Using} &  &  \\
        \hline
        Conscientiousness & \cite{Frauenstein2020Susceptibility, Vishwanath2015Examining, Huseynov2024Using, LopezAguilar2025Phishing, marin2023influence, ge2021personal} & \cite{eftimie2022spear, moody2017phish} & \cite{Huseynov2024Using} &  &  &  \\
        \hline
        Extraversion & \cite{Huseynov2024Using, LopezAguilar2025Phishing, marin2023influence} & \cite{eftimie2022spear} & \cite{Huseynov2024Using} &  &  &  \\
        \hline
        Greed &  &  & \cite{Rahman2023Users} & \cite{Mott2024Ransomware} &  &  \\
        \hline
        Openness & \cite{ge2021personal, moody2017phish, Huseynov2024Using, Buckley2023Indicators} & \cite{eftimie2022spear, benenson2017unpacking} & \cite{yeng2022investigation} & \cite{Mott2024Ransomware}  &  & \\
        \hline
        Narcissism & \cite{Curtis2018DarkTriad, Hart2025Phishing} &  &  &  &  &  \\
        \hline
        Neuroticism & \cite{Vishwanath2015Examining, LopezAguilar2025Phishing, greitzer2021experimental, ge2021personal} & \cite{eftimie2022spear} & \cite{Huseynov2024Using} &  &  & \\
        \hline

        \rowcolor[gray]{0.9} 
        \multicolumn{7}{|c|}{\textbf{Cognitive (\textit{direct/internal} factors)}} \\
        \hline
        Bias & \cite{Frauenstein2020Susceptibility, Vishwanath2015Examining, iuga2016baiting, Chou2021Mindless, Waqas2023Enhancing} & \cite{Williams2018Exploring} &  &  &  & \cite{chen2024, rahman2024towards} \\ 
        \hline
        Cognitive fatigue & \cite{vishwanath2011people, Jalali2020Employees, Musuva2019Cognitive, nifakos2021influence} &  &  &  & \cite{das2014tangled, Komanduri2011, allendoerfer2005human} & \cite{renaud2021shame, manfredi2021, rahman2024towards} \\
        \hline
        Cognitive reflecti-veness & \cite{Gallo2024HumanFactor, Buckley2023Indicators, Butavicius2022People, Vishwanath2015Examining, Vishwanath2018Suspicion, ge2021personal, greitzer2021experimental, dawn2024who, Waqas2023Enhancing, Wang2017Coping} & \cite{Wang2012Phishing, iuga2016baiting} & \cite{Kamar2023Moderating} &  &  &  \\
        \hline
        Cyber risk beliefs & \cite{greitzer2021experimental, Martens2019Investigating, jampen2020don, Musuva2019Cognitive, moody2017phish} & \cite{Kwak2020Spear, aleroud2020examination} & \cite{Timko2025Understanding} & \cite{Abraham2010Overview, Choi2024Enhancing, Yilmaz2023Personality, Onarlioglu2012Insights} & \cite{allendoerfer2005human} &  \\
        \hline
        Cyb-sec. self-monit. &  & \cite{Kwak2020Spear} &  &  &  &  \\
        \hline
        Decision fatigue & \cite{Jalali2020Employees} &  &  &  &  &   \\
        \hline
        Distraction & \cite{zhuo2023sok} &  &  &  &  &  \\
        \hline
        Lack of awareness & \cite{Jaeger2021Eyes, desolda2021human, Ion2015Noone, tornblad2021characteristics, nifakos2021influence, jampen2020don} & \cite{Wang2012Phishing} & \cite{Timko2025Understanding, Huseynov2024Using} & \cite{Abraham2010Overview, Choi2024Enhancing} & \cite{fagan2017investigation, Stanton2005Analysis, das2014tangled, Ion2015Noone, allendoerfer2005human} & \cite{mahajan2002understanding, Rahman2023Misconfigurations, chen2024, gupta2023vulnerability} \\
        \hline
        Lack of knowledge & \cite{dawn2024who, ge2021personal, Arachchilage2014Security, desolda2021human, Huseynov2024Using, Greco2025Enhancing, Ribeiro2024Factors, Wash2020Experts, Ion2015Noone, zhuo2023sok, nifakos2021influence} & \cite{desolda2021human, Wang2012Phishing, allendoerfer2005human} & \cite{Timko2025Understanding, Huseynov2024Using} & \cite{levesque2018technological,Mott2024Ransomware, ovelgonne2017understanding} & \cite{fagan2017investigation,ngandu2025strengthening, inglesant2010true, hadlington2017human, Ion2015Noone, Ion2015Noone} & \cite{ mahajan2002understanding, manfredi2021, Rahman2023Misconfigurations} \\
        \hline
        Misperception &  & \cite{benenson2017unpacking, Distler2023Influence} &  &  & \cite{allendoerfer2005human} & \cite{rahman2024towards} \\
        \hline
        Overconfidence & \cite{wang2016overconfidence, Canfield2019Metacognition, jampen2020don} &  & \cite{yeng2022investigation}  & \cite{levesque2018technological} &  &  \\
        \hline
        Risk attitude & \cite{moody2017phish} &  &  &  &  &  \\
        \hline
        Secur. self-efficacy & \cite{Martens2019Investigating, Arachchilage2014Security, Buckley2023Indicators, Wright2010Influence, Ribeiro2024Factors, Jansen2018Persuading, House2020Phishing, marin2023influence, boer1996protection, chen2024motivates} & \cite{Kwak2020Spear} & \cite{Verkijika2019SelfEfficacy, Lee2023Thwarting} & \cite{Abraham2010Overview} &  \cite{fagan2017investigation} & \cite{manfredi2021} \\
        \hline
        Uncertainty & \cite{benenson2017unpacking, Williams2018Exploring}  & \cite{Williams2018Exploring, Distler2023Influence} &  &  & &  \\
        \hline
        Vigilance & \cite{greitzer2021experimental, jampen2020don} & \cite{Wang2012Phishing} &  &  &  &  \\
        \hline

        \rowcolor[gray]{0.9} 
        \multicolumn{7}{|c|}{\textbf{Affective (direct/internal factors)}} \\
        \hline
        Anxiousness & \cite{Abdelhamid2020concerns, abroshan2021covid} & \cite{alhaddad2023study} &  &  &  &  \\
        \hline
        Digital anxiety & \cite{moody2017phish} & \cite{alhaddad2023study} &  &  &  &  \\
        \hline
        Fear & \cite{abroshan2021covid, Jansen2018Persuading, House2020Phishing} & \cite{Williams2018Exploring, benenson2017unpacking, Distler2023Influence, Martens2019Investigating} & \cite{Rahman2023Users, Verkijika2019SelfEfficacy} & \cite{Abraham2010Overview} & &  \\
        \hline
        Frustration &  &  &  &  & \cite{inglesant2010true, allendoerfer2005human} & \cite{hasan2025} \\
        \hline
        Shame &  & \cite{Distler2023Influence} &  &  &  & \cite{renaud2021shame}  \\
        \hline
        Stress & \cite{abroshan2021covid, zhuo2023sok, chen2024motivates, nifakos2021influence} &  &  &  &  &  \\
        \hline

        \rowcolor[gray]{0.9} 
        \multicolumn{7}{|c|}{\textbf{Behavioral (direct/internal factors)}} \\
        \hline
        Complacency & \cite{desolda2021human} &  &  &  & \cite{das2014tangled}  &  \\
        \hline
        Compulsive behav. & \cite{abroshan2021covid, Vishwanath2015Examining} &  &  &  &  &  \\
        \hline
        Impulsive behavior & \cite{greitzer2021experimental, dawn2024who, Butavicius2015HumanFirewall, tornblad2021characteristics, jampen2020don} & \cite{benenson2017unpacking, Butavicius2015HumanFirewall} &  & \cite{Neupane2016Neural} & \cite{whitty2015individual, hadlington2017human} &  \\
        \hline
        Internet addiction & &  &  &  & \cite{hadlington2017human} &  \\
        \hline
        Internet usage & \cite{iuga2016baiting, ge2021personal,alhaddad2023study, moody2017phish, Vishwanath2015Habitual, Greco2025Enhancing, Huseynov2024Using, Ribeiro2024Factors, Redmiles2018Examining, Reyns2015Routine, Ngo2020Victimization} & \cite{moody2017phish} & \cite{Huseynov2024Using} & \cite{levesque2018technological, Reyns2015Routine, Ngo2020Victimization} &  &  \\
        \hline
        Laziness &  &  &  &  & \cite{whitty2015individual} &  \\
        \hline
        Recurrence & \cite{greitzer2021experimental, vishwanath2011people, tornblad2021characteristics, Williams2018Exploring, marin2023influence} & \cite{Williams2018Exploring} &  &  &  &  \\
        \hline
        Risk-taking & \cite{Abdelhamid2020concerns, abroshan2021covid, abroshan2021phishing} &  &  &  &  & \\
        \hline
        
        \rowcolor[gray]{0.9} 
        \multicolumn{7}{|c|}{\textbf{Social and Organizational (modulator/external factors)}} \\
        \hline
        
        Attitude toward policies & \cite{Lee2022Phishing, Petric2022Impact} & \cite{Martens2019Investigating} &  &  & \cite{Stanton2005Analysis, Komanduri2011, ngandu2025strengthening, inglesant2010true, hadlington2017human, karlsson2021effect, allendoerfer2005human} &  \\
        \hline
        Lack of commun. & \cite{chen2024motivates} & \cite{Distler2023Influence, Williams2018Exploring} &  & \cite{Mott2024Ransomware, Yamagishi2025Collaborative} & \cite{ ngandu2025strengthening} & \cite{renaud2021shame} \\
        \hline
        Lack of resources & \cite{desolda2021human, chen2024motivates, sheng2010falls, Williams2018Exploring, nifakos2021influence, Butavicius2022People, Greco2025Enhancing, Petelka2019Warning, iuga2016baiting, zhuo2023sok, Chen2024GroupDiscussion} & \cite{desolda2021human, burns2019spear, Williams2018Exploring, rastenis2025credulitySpear} & \cite{tabassum2024drives, yeng2022investigation, Kamar2023Moderating} & \cite{Mott2024Ransomware, Abraham2010Overview, Vaclavik2025Lessons, Neupane2016Neural, Choi2024Enhancing, Yamagishi2025Collaborative} & \cite{ngandu2025strengthening, inglesant2010true} & \cite{manfredi2021, Rahman2023Misconfigurations, manfredi2022empirical, rahman2024towards, gupta2023vulnerability} \\
        \hline
        Lack of trust & \cite{greitzer2021experimental, moody2017phish, jampen2020don, tornblad2021characteristics} & \cite{aleroud2020examination, Williams2018Exploring, moody2017phish, benenson2017unpacking} & \cite{yeng2022investigation} &  & \cite{ fagan2017investigation, allendoerfer2005human} & \cite{renaud2021shame} \\
        \hline
        Norms & \cite{desolda2021human, Williams2018Exploring, Petric2022Impact, marin2023influence} & \cite{Distler2023Influence, Williams2018Exploring} &  &  & \cite{karlsson2021effect, allendoerfer2005human}  &  \\
        \hline
        Online exposure & \cite{Lee2022Phishing, Reyns2015Routine, Ngo2020Victimization} & \cite{Williams2018Exploring, nifakos2021influence} & \cite{Reyns2015Routine, Ngo2020Victimization} &  & \cite{fagan2017investigation} &  \\
        \hline
        Organizational ten-ure & \cite{Gallo2024HumanFactor, Taib2019SocialEngineering, greitzer2021experimental} & \cite{bullee2017spear} & \cite{Huseynov2024Using} &  &  &  \\    \hline
        Security posture &  &  &  & \cite{Mott2024Ransomware, Vaclavik2025Lessons} &  & \cite{Rahman2023Misconfigurations, mushi2018human} \\
        \hline
        Social influence & \cite{vishwanath2011people, Chou2021Mindless, Wright2014Influence, Taib2019SocialEngineering, Gallo2024HumanFactor, DeBona2020RealWorld, Butavicius2015HumanFirewall, Workman2008Wisecrackers} & \cite{lin2019susceptibility, Distler2023Influence, Butavicius2015HumanFirewall} & \cite{Rahman2023Users} &  & \cite{allendoerfer2005human} & \\
        \hline
        Social proof &  & \cite{Williams2018Exploring} &  &  & \cite{Wang2023SecurityLocal, allendoerfer2005human} & \\
        \hline
        Type of organiz. &  & \cite{rastenis2025credulitySpear}  &  & \cite{yuryna2020empirical} & \cite{karlsson2021effect, allendoerfer2005human}  &  \\
        \hline
    \caption{Influencing human factors on cyberthreats' susceptibility.}
    \label{tab:hfs}
    \end{longtable}
\end{center}

\clearpage
\section*{APPENDIX 2: Interactions among the human factors in MORPHEUS}
\label{sec:appendix_2}

\subsection*{A2.1 Detailed analysis of the interactions}
In the following subsections, we report a detailed analysis of the interactions among the 50 human factors, divided by dimension.

\subsection*{A2.1.1 Interactions with demographic factors}

\begin{figure}[htbp]
    \centering
    \includegraphics[width=\textwidth]{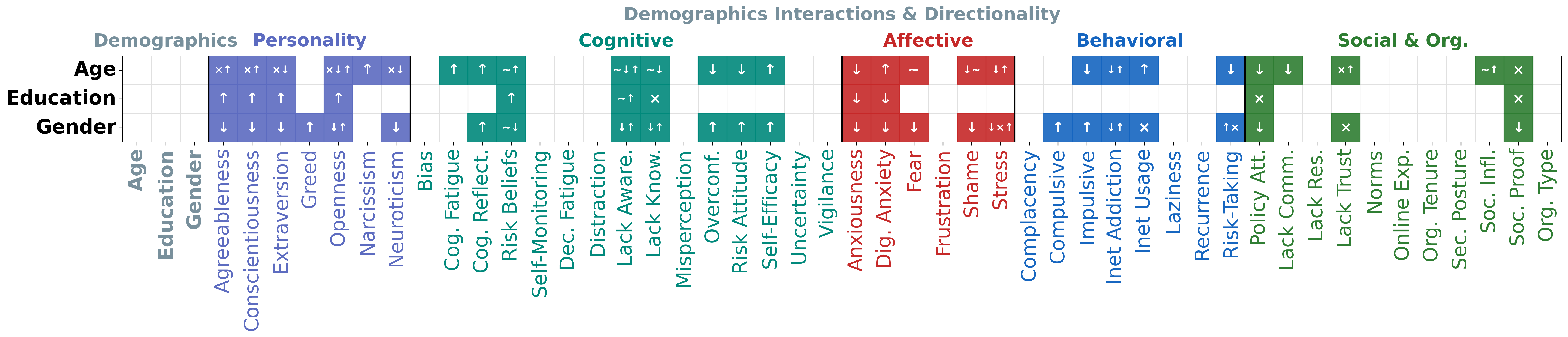}
    \caption{Summary of the interactions among the demographic human factors and the other ones. \textit{Legend}: $\uparrow$ indicates a positive association between the traits, $\downarrow$ a negative association, $\sim$ a mixed or complex association, and $\times$ a modulation effect between the two factors.}
    \Description{A table summarizing the interactions between demographic human factors (rows) and other human factor categories (columns: Personality, Cognitive, Affective, Behavioral, Social/Organizational). The legend indicates that $\uparrow$ means positive association, $\downarrow$ negative association, $\sim$ mixed/complex association, and $\times$ modulation effect.For Age:Personality: Positive association with Conscientiousness and Agreeableness; negative with Neuroticism, Openness, Extraversion, Narcissism, and Greed.Cognitive: Positive association with Security self-efficacy, Risk aversion, Vigilance, and Cognitive fatigue; negative with Cognitive biases and Overconfidence.Affective: Positive association with Digital anxiety and Fear; negative with Stress, Anxiousness, and Frustration.Behavioral: Positive association with Complacency; negative with Impulsive behavior, Risk-taking, Internet addiction, and Internet usage.Social/Organizational: Positive association with Trust and Attitude toward policies; negative with Social influence and Social proof.For Gender (Female vs. Male):Personality: Females show higher Neuroticism, Agreeableness, and Conscientiousness. Males show higher Narcissism and Greed.Cognitive: Females show higher Cyber risk beliefs and lower Security self-efficacy. Males show higher Overconfidence and Risk attitude.Affective: Females show higher Digital anxiety, Anxiousness, Fear, Stress, and Shame.Behavioral: Females show higher Complacency. Males show higher Impulsive behavior, Risk-taking, and Internet usage.Social/Organizational: Females show higher Social influence and Attitude toward policies. Males show higher Trust in tech.For Education:Personality: Positive association with Openness, Extraversion, and Narcissism; negative with Neuroticism.Cognitive: Positive association with Cognitive reflectiveness, Vigilance, and Security self-efficacy; negative with Cognitive biases, Cyber risk beliefs, and Lack of knowledge. Complex association with Overconfidence.Affective: Negative association with Digital anxiety, Anxiousness, and Fear.Behavioral: Positive association with Internet usage; negative with Complacency and Impulsive behavior.Social/Organizational: Positive association with Attitude toward policies; negative with Social influence and Trust.}
    \label{fig:matrix_demographics}
\end{figure}

\subsubsection*{Age interactions}

Age was found to have several interactions with other human factors described in MORPHEUS. 

Firstly, several studies have investigated the development of personality traits with age, yielding interesting findings. As one ages from adolescence to old age, traits such as conscientiousness and agreeableness tend to increase, while extraversion, openness, neuroticism, and narcissism decrease~\cite{Soto2012Development, Allemand2008Personality, Leite2023Dark}. Moreover, a study on adolescents and young adults specifically observed a rise in openness and conscientiousness traits from 16 to 20 years old~\cite{Vecchione2012Gender}, suggesting that openness does not decrease linearly across the lifespan.

Results from the literature highlight an unclear relationship between users' age and their cybersecurity awareness and knowledge: some results highlight better knowledge and awareness among older people~\cite{Sari2023Demographic, sheng2010falls, Zhou2020Risk, reeves2020whose}, while others yield mixed results~\cite{bell2022exploring} or contradict this, e.g., with older women exhibiting the highest susceptibility~\cite{lin2019susceptibility}.  Typically, older individuals self-report lower ICT skills and literacy, which may be due to them being under-confident in their cybersecurity knowledge~\cite{bell2022exploring}. The age of a user can also heighten their perception of the severity of cyberthreats and their perceived vulnerability~\cite{fatokun2019Impact}. 
Age also has a direct effect on people's mental resources. Cognitive fatigue, for example, is frequently reported in older adults~\cite{Bouche2025Mental}. Specifically, age was found to be an important factor in moderating the effects of cognitive issues (e.g., memory issues) according to personality traits. In older adults, lower extraversion, openness, and agreeableness were associated with more cognitive problems, together with higher neuroticism and conscientiousness, while this effect was reduced in middle-aged adults~\cite{aschwanden2025longitudinal}. Nonetheless, older people have a greater tendency to perform accurate elaboration of information, e.g., reflecting more on phishing emails~\cite{ge2021personal}, and generally be less impulsive~\cite{iuga2016baiting}.

Older individuals may have more internet experience (e.g., if they have used work-related online technologies for several years)~\cite{Zhou2020Risk}, and are generally more averse to risk~\cite{Nicholson2005Personality}. Older employees are also more likely to seek social support compared to younger individuals~\cite{Zhou2020Risk}.

Age can also have an impact on how individuals are emotionally affected. For example, in the context of the COVID-19 pandemic, younger individuals reported being more stressed compared to older people~\cite{nwachukwu2020covid}.

Finally, there is evidence that the impact of social proof may be higher on older users, i.e., they are more likely to consider their peers' security behavior compared to younger individuals~\cite{fatokun2019Impact}. 

\subsubsection*{Education interactions}

Education is another factor that can impact the cybersecurity resilience of users, particularly their awareness of cyberthreats. Surprisingly, a higher educational level does not correspond to higher awareness; in fact, there is evidence that the opposite is true, with undergraduate students showing better cybersecurity awareness than the post-graduate students~\cite{fatokun2019Impact}. Sari and colleagues found that more educated employees are less aware of their passwords and the organization's data~\cite{Sari2023Demographic}, even if they did not find a substantial association between educational level and security behavior overall. 
There is also evidence that contradicts this, showing that a higher education level actually increases concern about one's information privacy~\cite{Lee2019IPC}.

There is evidence that university education can even shape people's personalities, indicating that university education can help develop traits of conscientiousness, agreeableness, and extraversion, especially in students from disadvantaged backgrounds~\cite{Kassenboehmer2018University}. 
Finally, more educated people generally experience less anxiety while using technology~\cite{Yoon2016Computer} and in general~\cite{Huseynov2024Using}, compared to individuals with a lower education.

\subsubsection*{Gender interactions}

Gender can be a decisive component that influences cybersecurity behavior in several ways and interacts with other human factors of different dimensions. 
Regarding \textit{personality traits}, research indicates significant and clear differences between males and females. It is well established that women exhibit higher values of neuroticism~\cite{Weisberg2011Gender, Leslie2025Gender, Vecchione2012Gender, Murphy2021International, Costa2001GenderPersonality, AlosFerrer2016CognitiveReflection} and agreeableness~\cite{Weisberg2011Gender, Leslie2025Gender, Vecchione2012Gender, Otterbring2022Pandemic, Costa2001GenderPersonality} compared to men. Evidence also suggests that women are, in general, moderately more conscientious~\cite{Vecchione2012Gender, Otterbring2022Pandemic} and more extroverted than men~\cite{Weisberg2011Gender, AlosFerrer2016CognitiveReflection}. Findings on openness are instead contradictory~\cite{Vecchione2012Gender, Costa2001GenderPersonality}.  

Overconfidence in one's skills is significantly influenced by gender, with men generally being more confident than women in perceiving their computer abilities~\cite{aldarwish2019framework, anwar2017gender, bell2022exploring}. This overconfidence can also render security self-assessment methods less precise, as disparities in results between men and women might not be due to actual differences, but rather to their individual perception of themselves~\cite{anwar2017gender}. 

Although there are contradicting findings in the literature regarding the effect of gender alone on cyber-susceptibility, some contributions report secondary effects of gender on other factors. For example, gender can greatly influence \textit{security self-efficacy} (and \textit{overconfidence}) of users, with male individuals generally being more confident of their skills than females~\cite{aldarwish2019framework, anwar2017gender, bell2022exploring, Verkijika2019SelfEfficacy, Sun2016Mediating}. 
However, when considering the effects of gender alone on cybersecurity awareness, there are contrasting results, with studies reporting better knowledge and awareness in males~\cite{Gratian2018Correlating, sheng2010falls, fatokun2019Impact} and others reporting the opposite~\cite{Sari2023Demographic, reeves2020whose, daengsi2022cybersecurity}. 
Regarding people's attitude towards risk and risk-taking behavior, generally, women are more careful and take fewer risks than men~\cite{kennison2020risks, Pavlicek2021Personality, Nicholson2005Personality, gustafsod1998gender, weber2002dospert}, generally evaluating cyber risk threats as more severe~\cite{fatokun2019Impact, kennison2020risks, McGill2018Gender}. Being male was also found to be associated with more compulsive behaviors such as online gaming, online gambling, and online sex~\cite{Kircaburun2018Dark}. Nonetheless, male individuals have been found to be more reflective than females and generally engage in a deeper elaboration of information~\cite{ge2021personal, AlosFerrer2016CognitiveReflection}.
In a study by Lee and colleagues, women aged 40 or younger exhibited more privacy concerns regarding their data than men, while men aged 50 or above showed greater concern than women in that age group~\cite{Lee2019IPC}. Additionally, males have been found to be less compliant with corporate policies~\cite{Ifinedo2014Information}. 

Gender has a clear influence on anxiousness. Women are, in general, more anxious than men, with females showing higher Trait Anxiety~\cite{Leslie2025Gender, Weisberg2011Gender} and digital anxiety~\cite{Yoon2016Computer}. Moreover, men tend to experience less stress induced by a data breach compared to women~\cite{sears2024individual}. There is also evidence that social anxiety is more prominent in younger female users with higher internet use~\cite{tiraboschi2023Adolescent}.

\subsection*{A2.1.2 Interactions with personality traits}
\label{sec:personality_interactions}

\begin{figure}[htbp]
    \centering
    \includegraphics[width=\textwidth]{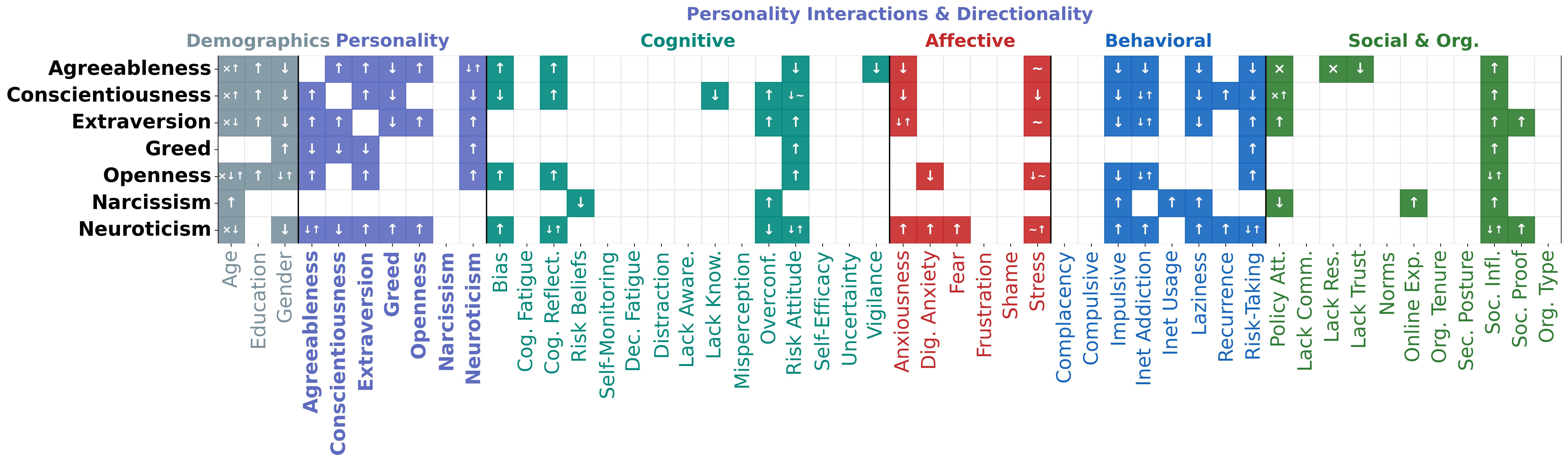}
    \caption{Summary of the interactions among the personality traits human factors and the other ones. \textit{Legend}: $\uparrow$ indicates a positive association between the traits, $\downarrow$ a negative association, $\sim$ a mixed or complex association, and $\times$ a modulation effect between the two factors.}
    \label{fig:matrix_personality}
    \Description{A table summarizing the pairwise interactions between personality human factors (rows) and other human factor categories (columns: Demographics, Personality, Cognitive, Affective, Behavioral, Social/Organizational).}
\end{figure}

Personality traits were found to have several interactions with other human factors. Moreover, some studies have explored how the Big 5 personality traits (agreeableness, conscientiousness, extraversion, neuroticism, openness) interact with one another.
DeYoung found that some personality traits are consistently related to each other and co-occur, forming two higher-order traits~\cite{DeYoung2015Cybernetic}. One combines emotional stability (i.e., the reverse of neuroticism), agreeableness, and conscientiousness---which can be interpreted as a broad \textit{Stability} factor, while the second combines extraversion and openness/intellect---which can be interpreted as a \textit{Plasticity} (or exploration) factor.
Another, more recent study on college students~\cite{Bhattacharjee2025Effect} revealed a strong positive correlation between openness and extraversion (as per the Plasticity factor), as well as between openness and agreeableness and neuroticism---which contradicts the previous findings~\cite{DeYoung2015Cybernetic}. Additionally, this study revealed a moderate positive correlation between extraversion, agreeableness, and neuroticism, as well as between conscientiousness and extraversion.

\subsubsection*{Agreeableness Interactions}

Agreeableness also affects individuals' risk-taking behavior and attitude, since agreeable individuals tend to take fewer risks than people who score low in this trait~\cite{Nicholson2005Personality}. Moreover, there is evidence that behaviors associated with agreeableness are more strongly affected by cybersecurity training courses~\cite{eftimie2022spear}.

If, on the one hand, agreeableness improves cautious behavior and compliance, reinforcing the effect of attitude toward policies on actual security behavior~\cite{Shropshire2015Personality}, on the other hand, agreeable individuals tend to be more vulnerable to social influence techniques~\cite{Cusack2018Personality, Shropshire2015Personality, Palm2025Influence}, as they tend to trust others more as per their nature~\cite{Rossier2004NEO16PF, Brandt2022Trust, Shropshire2015Personality}. This can also decrease their overall vigilance levels toward possible threats~\cite{Rossier2004NEO16PF}.

\subsubsection*{Conscientiousness Interactions}

Highly conscientious people are generally more reluctant to take risks~\cite{Fawad2020Personality, Nicholson2005Personality} and more compliant with cybersecurity policies~\cite{marin2023influence}. Moreover, conscientiousness strengthens the influence that employees' positive attitude has on actual security compliance~\cite{Shropshire2015Personality}.
Nevertheless, conscientiousness can lead to the creation of strong email habits that are not secure, such as diligently checking each communication and link in emails~\cite{Vishwanath2015Examining}, thus making users more susceptible. Moreover, a study by Kennison and Chan-Tin investigating cybersecurity risk found that conscientiousness predicted riskier self-reported cybersecurity behaviors among female participants, but not among men~\cite{kennison2020risks}. However, the same study also highlighted that higher levels of conscientiousness correspond to higher levels of self-reported knowledge of security passwords.

\subsubsection*{Extraversion Interactions}

Extraversion is a personality trait that can lead individuals to generally have riskier attitudes and seek more risks~\cite{Fawad2020Personality, Pavlicek2021Personality, Nicholson2005Personality}. 
Nonetheless, extraversion has been found to be a protective factor that positively influences the helpful behaviors of employees in protecting the organization and their colleagues~\cite{marin2023influence}. 
A study by Murday et al. investigated the role of personality traits, specifically extraversion, in affecting people's social baselines and their use of social resources (i.e., the help from another person) to accomplish a manual task. Participants with higher extraversion resorted to the social resource, seeking social proof, more often than others~\cite{Murday2021Extraversion}.

\subsubsection*{Greed Interactions}

Men are, in general, more greedy than women~\cite{krekels2015Dispositional}. 
A higher dispositional greed can lead to more risky behaviors and a riskier attitude in general, which are driven by the desire to potentially obtain profit and rewards~\cite{Li2019Neural}. 

On a personality level, the trait of greed is correlated positively with neuroticism~\cite{krekels2015Dispositional} and negatively with the traits of extraversion, agreeableness, and conscientiousness~\cite{krekels2015Dispositional, Zeelenberg2024Disentangling}.

\subsubsection*{Openness Interactions}

The openness to experience personality trait has been found to be higher in individuals with a higher education~\cite{Ludeke2014Truth}. However, these findings are based on self-reported data, which can be biased since more educated people tend to exaggerate their level of openness~\cite{Ludeke2014Truth}.  

The openness of an individual clearly influences their risk attitude, as more open people are generally less risk-averse~\cite{Pavlicek2021Personality, Nicholson2005Personality} and more likely to take risks~\cite{Nicholson2005Personality}.

\subsubsection*{Narcissism Interactions}

Narcissism is one of the traits of the so-called \textit{Dark Triad} (together with Machiavellianism and psychopathy) and is associated with behaviors that can be detrimental to cybersecurity. Narcissism, for example, can also lead to a superior sense of self and feeling overconfident about their skills and knowledge~\cite{Paulhus2002DarkTriad, Campbell2004Narcissism}, which can also lead to underestimating threats and consequences~\cite{Jones2011Impulsivity}.

Narcissism is also associated with procrastination, lazy behaviors~\cite{Meng2024Procrastinators}, and more impulsivity~\cite{Jones2011Impulsivity}. 
A narcissistic individual is also more prone to use social media and share their private information online to self-promote~\cite{Leite2023Dark, Kircaburun2018Dark}.

Highly narcissistic people are also more susceptible to social influence~\cite{Hart2025Phishing} and can have more negative attitudes towards organizational policies, with a higher likelihood of representing a potential insider threat~\cite{Maasberg2020DarkTriad}.

\subsubsection*{Neuroticism Interactions}

Neuroticism is profoundly tied to emotional instability. For example, highly neurotic individuals are generally more likely to experience anxiousness~\cite{Regzedmaa2024Systematic, Thorp1993Personality, Budimir2021CybersecurityEmotions, Dong2022Anxious, Leslie2025Gender, Rossier2004NEO16PF} and emotions such as anger and \textit{fear}~\cite{Dong2022Anxious}. 
On the contrary, low-neurotic individuals generally exhibit more calmness (less anxiety)~\cite{Dong2022Anxious}, are more \textit{reflective}, and can better resist impulsive behaviors~\cite{Waqas2023Enhancing}.

Neuroticism can also affect the risk attitude of individuals, as highly neurotic people are more likely to be more prone to risk~\cite{Pavlicek2021Personality, kennison2020risks} and take more risks~\cite{Zhang2025Emotional}. Moreover, highly neurotic individuals can develop the recurrence of bad email habits, such as carelessly reacting to emails, thus rendering them more susceptible to phishing attacks~\cite{Vishwanath2015Examining}. However, there is evidence that high levels of neuroticism lead to more risk aversion and less risky behaviors~\cite{Nicholson2005Personality}.

\subsection*{A2.1.3 Interactions with cognitive factors}
\label{sec:cognitive_interactions}

\begin{figure}[htbp]
    \centering
    \includegraphics[width=\textwidth]{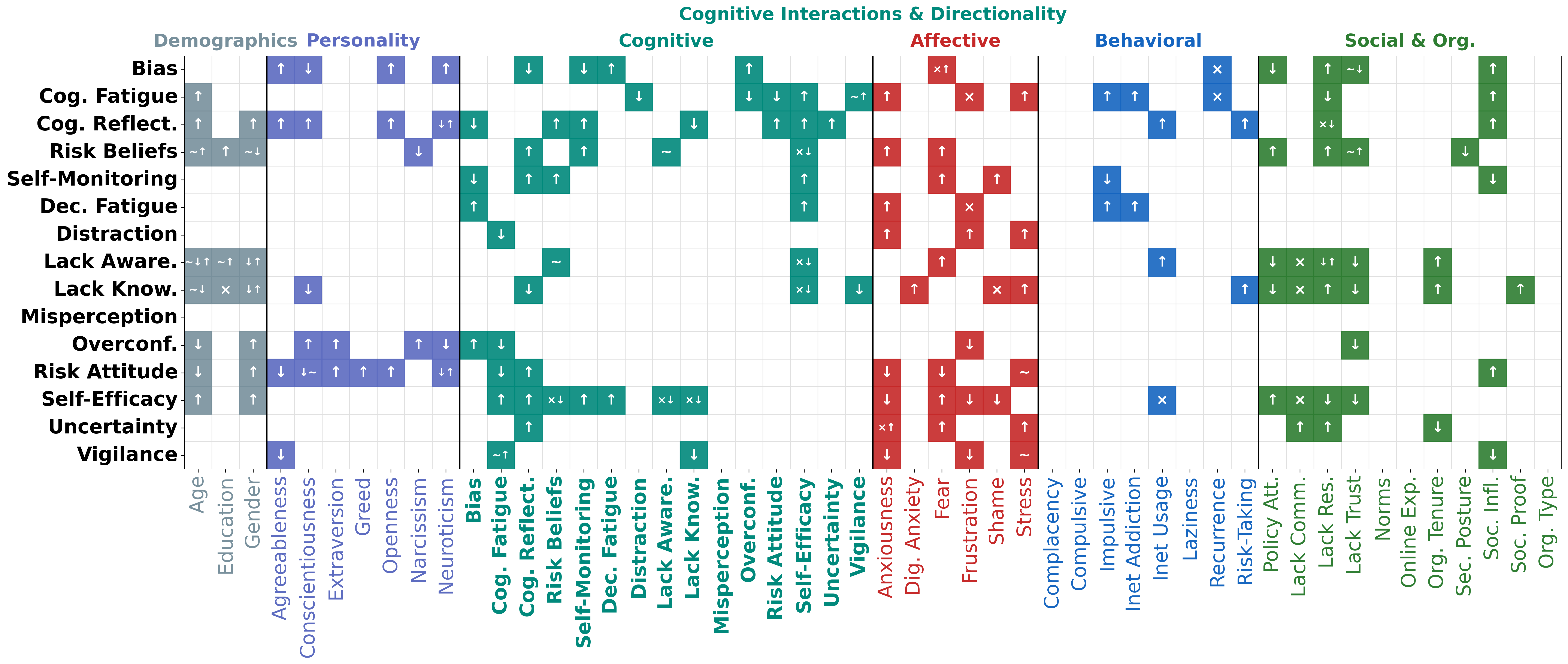}
    \caption{Summary of the interactions among the cognitive human factors and the other ones. \textit{Legend}: $\uparrow$ indicates a positive association between the traits, $\downarrow$ a negative association, $\sim$ a mixed or complex association, and $\times$ a modulation effect between the two factors.}
    \label{fig:matrix_cognitive}
    \Description{A table summarizing the pairwise interactions between cognitive human factors (rows) and other human factor categories (columns: Demographics, Personality, Cognitive, Affective, Behavioral, Social/Organizational).}
\end{figure}

\subsubsection*{Bias and Cognitive reflectiveness interactions} 

Cognitive reflectiveness and Biases can be considered two intricately linked factors that affect susceptibility to cyberthreats. The Heuristic-Systematic Model of persuasion (HSM)~\cite{Chaiken1980HSM} describes how people process persuasive information through either systematic or heuristic processing. The former involves a deep and careful analysis of the message's arguments, whereas heuristic processing relies on mental shortcuts and simple cues (i.e., relying on biases). 

Personality traits affect preferences towards using one or the other. Cognitive reflectiveness has been shown to be positively correlated with openness and conscientiousness~\cite{ge2021personal}, while being negatively associated with low self-regulation levels (i.e., high neuroticism)~\cite{Waqas2023Enhancing}. Specifically, conscientious people are less likely to rely on heuristic processes in the context of phishing in social networks~\cite{Frauenstein2020Susceptibility}. Other traits, including agreeableness, neuroticism, and openness, were instead found to be positively associated with both processing systems~\cite{Frauenstein2020Susceptibility}.

Biased, heuristic-based processing is particularly relied upon during states of decision fatigue~\cite{Pignatiello2020Decision, stanton2016security} or when time resources are limited~\cite {williams2017individual}, which can lead to decisions yielding undesirable outcomes. On the other hand, cognitive reflectiveness lowers commitment to biases (such as conjunction fallacy and base-rate neglect biases)~\cite{AlosFerrer2016CognitiveReflection, Chou2021Mindless} and can affect the user's cyber risk beliefs, heightening the perception of threats~\cite{Musuva2019Cognitive}. However, the cumulative effect of heuristic processing and recurrence of bad security habits may overwhelm the advantages coming from systematic processing~\cite{Vishwanath2015Examining}.

Technical knowledge and internet use have indeed been shown to increase cognitive reflection when elaborating emails~\cite{ge2021personal, Wash2020Experts}, as well as cybersecurity training~\cite{iuga2016baiting}. Cognitive reflectiveness can be fostered by a perceived uncertainty when receiving suspicious emails in order to address missing information~\cite{Vishwanath2015Examining}. In economic contexts, irrational risk aversion and low-risk-taking behaviors can be limited by more cognitive reflection~\cite{Li2023Risk}.

Therefore, cognitive reflectiveness can enhance protection against cyber-attacks. However, it should be combined with security awareness messages or training~\cite{Kamar2023Moderating}. Technical-based aids such as email alerts can indeed increase cognitive reflectiveness~\cite{Williams2018Exploring}.

\subsubsection*{Cognitive and decision fatigue interactions} 

In the context of phishing attacks, cognitive fatigue deriving from a high email load, combined with the recurrence of bad email habits, can increase user susceptibility~\cite{vishwanath2011people}. Another relevant factor is vigilance (or sustained attention), which increases mental workload and cognitive fatigue over time~\cite{Warm2008Vigilance}. However, focusing attention on specific elements, such as social influence or technical cues in phishing emails, can decrease the overall cognitive effort required for email processing~\cite{Wang2012Phishing}. 
While too much cognitive fatigue can hinder decision-making, too little can have negative effects alike, leading to users self-interrupting or being distracted from their tasks if they have too many mental resources available~\cite{Katidioti2016Interrupt}. 

On the emotional side, anxiousness can contribute to decision fatigue, as socially anxious individuals are generally more hesitant when making decisions~\cite{Hengen2021Stress}. In general, anxiousness can occupy many cognitive resources, thereby increasing cognitive fatigue and hindering emotional and cognitive processing~\cite{Botvinick2001Conflict}. 
Similarly, security fatigue---i.e., a state of mental and emotional exhaustion arising from repeated exposure to security demands---can lead to reduced productivity and increased anxiety~\cite{Mizrak2025DigitalDetox}. Security fatigue can also lead to security-related decisions being driven by immediate motivations, relying on simplified heuristics, and acting impulsively~\cite{stanton2016security}.

Cognitive fatigue can also affect individuals' risk attitude, as mentally fatigued individuals tend to be more risk-averse when facing risk options in gambling scenarios~\cite{Jia2022MentalFatigue}.
Finally, exposing users to training too frequently---especially when combined with other routine communications like health and safety information---can increase their workload and, subsequently, fatigue~\cite{Reeves2021Encouraging, Williams2018Exploring}.

\subsubsection*{Cybersecurity self-monitoring}

People with higher levels of cybersecurity self-monitoring and self-regulation have been shown to be more likely to exhibit cognitive reflectiveness and resist impulsive reactions in the context of phishing susceptibility~\cite{Waqas2023Enhancing}. Moreover, increased self-monitoring can enhance resistance to social influence techniques and reduce reliance on biased heuristic processing~\cite{Hutton1992SelfAwareness}. 

Cybersecurity self-monitoring is significantly influenced by security self-efficacy~\cite{Kwak2020Spear}; this means that the more users perceive they can handle threats, the more they will be able to control their cybersecurity behaviors. Additionally, the cyber risk beliefs of a user can increase their cybersecurity self-monitoring~\cite{Kwak2020Spear, Martens2019Investigating}. In the context of spear phishing, for example, expected negative outcomes such as feelings of fear or shame can improve behaviors and encourage users to report malicious emails more effectively~\cite{Kwak2020Spear}.

\subsubsection*{Cyber risk beliefs and Security self-efficacy interactions}

Cyber risk beliefs and security self-efficacy are crucial factors that, in combination, shape individuals' motivation to engage in cybersecurity behaviors (cf. Protection Motivation Theory~\cite{boer1996protection}) and can lead to a more secure use of technology~\cite{Kwak2020Spear, Lee2023Thwarting, Verkijika2019SelfEfficacy}. The cyber risk beliefs of a user can negatively affect their security self-efficacy~\cite{Wang2017Coping}; this implies that, as perceived threats increase, the user will question their ability to cope effectively with them, while, if they are perceived as less severe, they may feel more assured in their ability to handle them. 

Both security self-efficacy and cyber risk beliefs increase cognitive reflectiveness in users when dealing with suspicious emails~\cite{vishwanath2011people, Kwak2020Spear} since more severely perceived threats increase the amount of cognitive resources users are willing to expend to stay safe. 
Security self-efficacy can reduce susceptibility to cyber threats, but the actual user's knowledge and awareness significantly affect its impact~\cite{vishwanath2011people}. 
Moreover, security self-efficacy can be increased by workload (cognitive and decision fatigue), as more fatigued users perceive they can better comply with security regulations~\cite{yeng2022investigation}, which may increase their susceptibility to cyberthreats. 

At an organizational level, an employee's security self-efficacy and cyber risk beliefs significantly impact their attitude towards security policies and compliance therewith~\cite{DelsoVicente2025ComplianceReview, Li2022Investigation, Martens2019Investigating}. Specifically, self-efficacy more effectively predicts a positive attitude toward policies for users with higher internet use experience~\cite{Martens2019Investigating}. 
A very strong organizational security posture and large availability of security resources (e.g., specialist support) can lead employees to feel secure and perceive a low level of risk~\cite{Williams2018Exploring}. To improve their employees' self-efficacy in phishing contexts, organizations can employ cybersecurity training~\cite{Chen2024GroupDiscussion}.

\subsubsection*{Lack of awareness and knowledge interactions}

Users' cybersecurity awareness and knowledge are crucial factors in determining susceptibility to cyber threats. User awareness of cyber threats affects their cyber risk beliefs in a complex manner, as it positively influences their perceived threat severity but negatively influences their perceived vulnerability~\cite{Martens2019Investigating}. This means that users with a high level of awareness may acknowledge the severity of a threat but not feel significantly threatened by it. Conversely, users are more aware of cyber threats that are perceived as genuinely dangerous to their personal security~\cite{Reeves2021Encouraging}.
On the other hand, high levels of knowledge and awareness are correlated with a high level of self-efficacy~\cite{Martens2019Investigating, Arachchilage2014Security}. 

There is proof that users who tend to use the Internet for personal reasons in the workplace may exhibit lower cybersecurity awareness~\cite{hadlington2017cyberloafing}. Moreover, cybersecurity knowledge can also improve people's behavior, as higher self-reported levels of security knowledge can lead to a decrease in taken risks~\cite{kennison2020risks} and be more vigilant about specific security indicators, such as phishing cues in emails~\cite{Wang2012Phishing}. A higher level of knowledge can even benefit users on an emotional level, as it can decrease their levels of digital anxiety~\cite{Yoon2016Computer}.

Technological support, such as security alerts for phishing emails, can increase users' awareness of specific threats~\cite{Williams2018Exploring}. Moreover, employees tend to believe that educational support can benefit cybersecurity knowledge and awareness~\cite{Renaud2021CybersecurityEmotions}. However, security training that is too frequent can lead to employees negatively reacting to educational resources and ultimately decreasing their security awareness levels~\cite{reeves2020whose}.
Even if employees' knowledge is a critical factor in achieving better security overall, a review by Dawson and Thomson highlights that, in cybersecurity workforce contexts, users' knowledge is not sufficient on its own but requires social intelligence (i.e., communication with peers) in combination~\cite{Dawson2018Future}. 
An interesting finding suggests that exposure to the workplace (as little as one year) can negatively impact the cybersecurity behavior of employees and also affect their security knowledge~\cite{Hong2023Influence}. This effect might be exacerbated for the long-tenure employees, such as managers~\cite{reeves2020whose}.

\subsubsection*{Overconfidence Interactions}

From a personality perspective, overconfidence can be influenced by specific traits. Specifically, in the financial investments domain, there is evidence that people are more confident in their skills and rely less on unpredictable factors such as luck when they have higher values of conscientiousness~\cite{Fawad2020Personality}. Extraversion is another factor that influences overconfidence, with extroverted individuals generally being more optimistic about their knowledge, skills, and experience when making decisions~\cite{Fawad2020Personality, Schaefer2004Overconfidence}. Finally, neuroticism negatively influences overconfidence, as neurotic people generally possess lower self-esteem~\cite{Roberts1999Neuroticism} and are less confident in their skills, thereby relying more on advice coming from others~\cite{Fawad2020Personality}. 

In phishing email detection contexts, a user's overconfidence can decrease if they are mentally fatigued~\cite{wang2016overconfidence}. Moreover, users who heavily rely on heuristic decision rules (i.e., biases) are more overconfident in their decisions~\cite{wang2016overconfidence}.

\subsubsection*{Uncertainty interactions}

Uncertainty can lead to feelings of a lack of security in employees, who should instead be supported with actionable and concrete cybersecurity-related information (i.e., organizational resources)~\cite{Renaud2021CybersecurityEmotions}, especially for employees with low organizational tenure, who may be more uncertain about work norms~\cite{Williams2018Exploring}. 

Intolerance of uncertainty, i.e., the negative emotional attitude towards uncertain situations and events, can contribute to increased social anxiety~\cite{Spiroiu2025SocialAnxiety}. Conversely, feelings of fear and anxiety can cause higher levels of uncertainty in users when dealing with phishing emails~\cite{abroshan2021covid}.

\subsection*{A2.1.4 Interactions with affective factors}
\label{sec:emotional_interactions}

\begin{figure}[htbp]
    \centering
    \includegraphics[width=\textwidth]{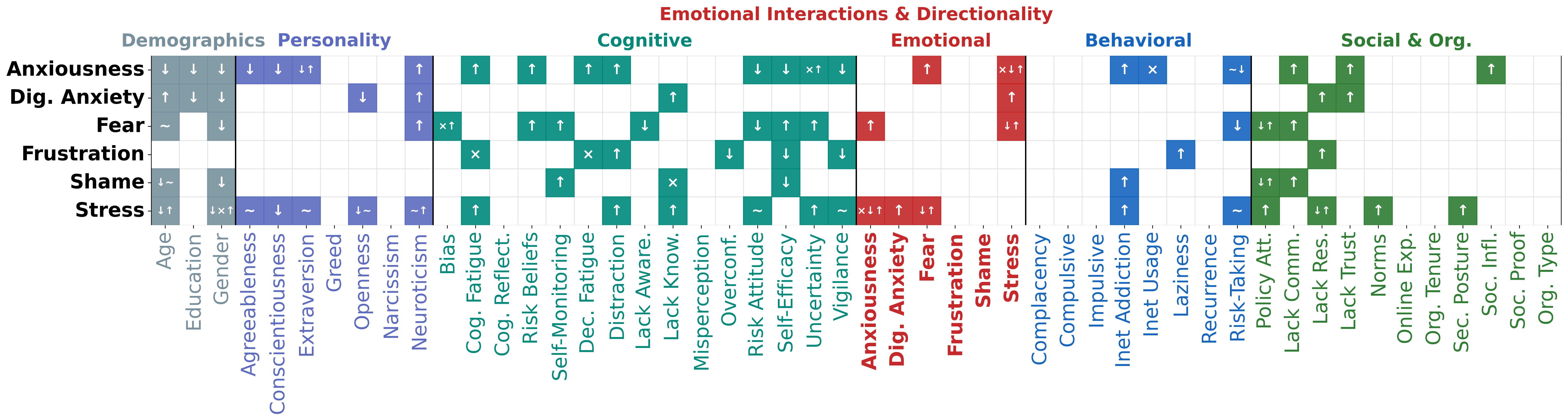}
    \caption{Summary of the interactions among the affective human factors and the other ones. \textit{Legend}: $\uparrow$ indicates a positive association between the traits, $\downarrow$ a negative association, $\sim$ a mixed or complex association, and $\times$ a modulation effect between the two factors.}
    \label{fig:matrix_emotional}
    \Description{A table summarizing the pairwise interactions between affective human factors (rows) and other human factor categories (columns: Demographics, Personality, Cognitive, Affective, Behavioral, Social/Organizational).}
\end{figure}

\subsubsection*{Anxiousness and Digital anxiety interactions} 

Anxiousness can vary based on different short- and long-term factors of an individual; for instance, personality traits have been shown to potentially influence anxiousness. Apart from neuroticism, which increases emotional intensity, including anxiety~\cite{Budimir2021CybersecurityEmotions, Rossier2004NEO16PF, Leslie2025Gender, Thorp1993Personality, Regzedmaa2024Systematic, Dong2022Anxious}, extraversion, agreeableness, and conscientiousness are negatively correlated with anxiousness~\cite{Leslie2025Gender, Thorp1993Personality}. 
However, there is evidence that extraversion can lead to more anxiousness-related behavior, such as \textit{fight/flight} reactions, rather than proactive reactions~\cite{Budimir2021CybersecurityEmotions}.

Studies have highlighted that anxiousness can decrease an individual's attentional control (i.e., vigilance) while performing their tasks~\cite{Allsop2014FlyingPressure, Bishop2009TraitAnxiety}.

Cyber risk beliefs can increase anxiety, for example, if the user has a major concern about a cyber threat, while high self-efficacy can decrease it~\cite {Wang2017Coping}. Conversely, a state of anxiousness can alter the perception of risk, leading to wrong estimations of negative outcomes and risk-averse choices~\cite{Hengen2021Stress, Li2019Exploring}. 
From a neuroscience perspective, specific neural activations tied to anxiety modulate risky decision-making and can predict both decreases and increases in risk-taking behavior~\cite{Nash2021AnxietyRisk}.

Anxiousness can be increased by internet usage~\cite{tiraboschi2023Adolescent}. In fact, a specific form of anxiety is digital anxiety, which can generate stress in non-savvy users~\cite{Pfaffinger2021DigitalisationAnxiety} and can decrease their willingness to trust a computer system~\cite{Hwang2007Customer}. Digital anxiety is generally experienced by older users (who are not digital natives) and individuals with low educational backgrounds or limited technical knowledge~\cite{Yoon2016Computer}. Digital anxiety is more common in high-neuroticism and low-openness individuals~\cite{Maricutoiu2014Meta} and can be decreased through user training~\cite{Maricutoiu2014Meta}.

Finally, anxiousness stemming from social connections (i.e., social anxiety) can hinder communication with others and generally lead to isolation~\cite{Dong2024SocialAnxiety}. In the context of spear-phishing, where the victim receives a malicious email from someone they apparently know, anxiousness can manifest as Fear-of-Missing-Out (FOMO) and lead to increased susceptibility~\cite{Klutsch2024PhishingFOMO}.

\subsubsection*{Fear Interactions}

Certain demographic groups are more prone to experiencing fear. In general, female individuals are more likely to be fearful than men~\cite{Cook2022Fear}. Moreover, age affects the fear of cybercrime in a non-linear manner, with an increase from 18 years old until middle age and a decrease thereafter~\cite {Cook2022Fear}.

Fear can have positive effects on the cybersecurity of an individual, such as more risk aversion~\cite{Ohman2001FearsPhobias, Hengen2021Stress}, reduced risk-taking~\cite{Hengen2021Stress}, and increased reliance on security organizational policies~\cite{Aggarwal2024Eustress}. In this sense, fear can also be used by organizations as a strategy to foster compliance with information security policies through the use of \textit{fear appeals}. However, their use is more effective when they are tailored to individual cognitive biases of employees~\cite{DelsoVicente2025ComplianceReview}. Organizations also need to be aware of the side effects of fear appeals on long-term responses of employees, who can instead manifest a paralyzed attitude in response~\cite{Brennan2010Fear}. Moreover, the use of fear appeals can induce \textit{lie bias} in phishing detection, increasing users' tendency to misjudge genuine emails as threats~\cite{Wang2017Coping}.

On the other hand, a complete lack of fear in users can lead to low awareness and misperceptions, such as not feeling ``important enough for anyone to want to take [their] information''~\cite{stanton2016security}. Fear appeals can thus be useful to adjust employees' cyber risk beliefs and security self-efficacy, if properly designed~\cite{Jansen2019Design}.

\subsubsection*{Frustration interactions} 

Frustration is an affective state that can severely compromise correct user behavior. Specifically, frustration was found to be associated with users' cognitive and decision fatigue, with frustrated and fatigued users more likely to dismiss security warnings without due consideration~\cite{stanton2016security}. Frustration can also negatively affect vigilance, with frustrated users being less likely to remain vigilant for sustained periods of time~\cite{Warm2008Vigilance}.
Frustration also dictates whether an individual will self-interrupt during their tasks; in fact, some studies suggest that this can happen if the task is either too easy (in which case, the distraction happens out of boredom) or too hard (out of frustration)~\cite{Katidioti2016Interrupt}.

When people have the misperception that they are not worth being attacked, they can feel frustrated if being told to protect themselves too frequently~\cite{stanton2016security}.
In these cases, policies that are too frustrating or stringent can lead users to adopt lazy workarounds, such as writing passwords down on post-its~\cite{inglesant2010true}.  
On the other hand, when users do not receive the proper supporting resources and information to use technological solutions, they can experience frustration, which can in turn decrease their self-confidence and self-efficacy~\cite{Wang2019Technology}.

\subsubsection*{Shame interactions} 

Shame can be influenced by gender and age, with females being more prone to feelings of shame than males~\cite{malinakova2020Psychometric}, and younger individuals typically experiencing higher levels of shame~\cite{Gambin2018Relations, Orth2010Shame}---although individuals over 50 may report increased shame compared to middle-aged individuals~\cite{Orth2010Shame}.

Shame can lead to decreased communication when a person is too ashamed to ask for help, especially if they have a major lack of technical knowledge on the subject they need help with~\cite{Wang2019Technology}. 
Sometimes employees may have a more positive attitude towards policies in order to avoid feelings of shame caused by non-compliance~\cite{Harris2012Routes}. However, leveraging shame as a motivational factor does not generally improve the employee's attitude~\cite{Farshadkhah2021Onlooker}, and can even be counterproductive for the organization, resulting in self-protective or paralyzing behavior~\cite{Brennan2010Fear}. In fact, shame can lead to reduced levels of security self-efficacy~\cite{Baldwin2006Relationship}. Therefore, it is not generally recommended for organizations to use emotions such as guilt or shame as drivers for increased compliance.

\subsubsection*{Stress interactions} 

Stress can be chronically dictated by personality traits such as neuroticism, as neurotic individuals can experience stress at a higher intensity compared to other people~\cite{Budimir2021CybersecurityEmotions, Maier2019Technostress, Cuadrado2024Technostress}. Openness can also have a mitigating effect on stress~\cite{Cuadrado2024Technostress}. However, Pflünger and colleagues argue that personality traits should be considered holistically to account for the finely granulated relationships; they indeed reported mixed effects of personality traits on stress, depending on the specific profile of individuals~\cite{Pflugner2021PersonalityTechnostress}.

Stress is not inherently a negative emotion, as it can actually serve as a driver for good behavior by increasing engagement with cybersecurity practices. This is the case of \textit{challenge stress} or \textit{eustress}~\cite{Shirish2021Eustress}, which proved to be correlated with emotions such as excitement and, ultimately, compliance with information security policies~\cite{Chen2022InfoSecStress, Aggarwal2024Eustress}. Moreover, eustress can also decrease emotions like anxiety and fear~\cite{Chen2022InfoSecStress}.
In particular, stress has a complex relationship with anxiousness and affects the risk attitude of individuals based on their level of social anxiety: in low socially anxious people, stress fosters risk-taking, while in high socially anxious people, stress does not change their behavior and they remain risk-averse~\cite{Hengen2021Stress}.
In cybersecurity contexts, stress and anxiousness can jointly worsen cybersecurity behavior~\cite{Mizrak2025DigitalDetox}.

However, stress in its negative sense (e.g., \textit{technostress} – the stress caused by technology) can decrease productivity, especially in male individuals~\cite{Amin2024TechnostressEducators}. Technostress, in particular, is more prevalent among younger, male individuals with low digital literacy~\cite{RaguNathan2008Technostress}. Moreover, stress is strictly linked to uncertainty~\cite{Amin2024TechnostressEducators} and feelings of anxiousness~\cite{Amin2024TechnostressEducators, Cuadrado2024Technostress} and fear~\cite{Aggarwal2024Eustress}.

Stress is also strongly related to vigilance, as attentional tasks are very demanding for people, rapidly draining cognitive resources and increasing stress over time~\cite{Warm2008Vigilance}. In fact, stress and cognitive fatigue are intertwined factors, as the former can increase the latter~\cite{Botvinick2001Conflict, Cuadrado2024Technostress, Amin2024TechnostressEducators} and vice versa~\cite{Tarafdar2019Technostress}. Frequent distractions and the norm of ``being always available'' can also increase stress in the workplace~\cite{Tarafdar2019Technostress}.

On an organizational level, support and training resources can affect the stress of employees at two extremes: on the one hand, too frequent training can lead to user overload and possibly burnout; on the other hand, lack of support can increase stress, as employees may feel unable to deal with highly sophisticated technological solutions in the workplace~\cite{Amin2024TechnostressEducators, Reeves2021Encouraging}.

\subsection*{A2.1.5 Interactions with behavioral factors}
\label{sec:behavioral_interactions}

\begin{figure}[htbp]
    \centering
    \includegraphics[width=\textwidth]{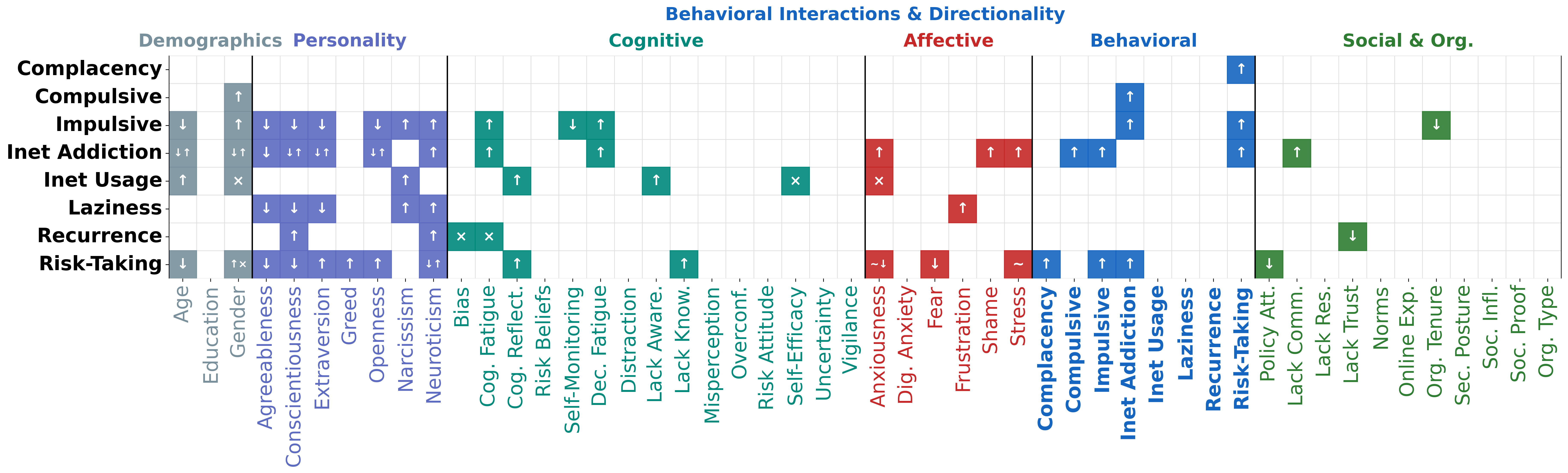}
    \caption{Summary of the interactions among the behavioral human factors and the other ones. \textit{Legend}: $\uparrow$ indicates a positive association between the traits, $\downarrow$ a negative association, $\sim$ a mixed or complex association, and $\times$ a modulation effect between the two factors.}
    \label{fig:matrix_behavioral}
    \Description{A table summarizing the pairwise interactions between behavioral human factors (rows) and other human factor categories (columns: Demographics, Personality, Cognitive, Affective, Behavioral, Social/Organizational).}
\end{figure}

\subsubsection*{Impulsive behavior Interactions}

Impulsive behaviors can stem from personality traits; in fact, impulsivity has been shown to be positively correlated with neuroticism, thereby leading to more impulsive behaviors~\cite{Zhang2025Emotional, Mao2018Selfcontrol, Schreiber2012Emotion, Waqas2023Enhancing}.
There is also evidence that gender may influence impulsive behaviors, as males tend to be more impulsive than females~\cite{Diotaiuti2022Internet}.
Impulsivity is also significantly associated with increased internet addiction, as shown by several studies~\cite{Diotaiuti2022Internet, Salehi2023Impulsivity, Marzilli2020Internet}. 

Organizational tenure can reduce impulsive behaviors (e.g., when processing emails), as employed individuals may be more inclined to think carefully due to exposure to workplace norms, as opposed to students and unemployed users~\cite{iuga2016baiting}.

\subsubsection*{Internet Addiction Interactions}

Internet addiction has been widely studied in the psychology and cybersecurity literature. 

Age has been found to be negatively associated with internet addiction, as younger people are the most affected~\cite{LozanoBlasco2022Internet, Sechi2021Addictive, Leite2023Dark}; however, this correlation might not be linear, as internet addiction was shown to increase in the 18--30 age range~\cite{Diotaiuti2022Internet}. Gender also dictates internet addiction, as males are generally more susceptible to it than female users~\cite{Diotaiuti2022Internet, Jojo2022Personality, Ghislieri2022Might}, even if there is contrasting evidence~\cite{LozanoBlasco2022Internet}.

Under a personality perspective, neurotic people are generally more prone to exhibit increased internet addiction~\cite{rachubinska2021analysis, Nwufo2024Personality, Jojo2022Personality, Kayis2016BigFive, Tian2021Associations, Sechi2021Addictive}. The other traits are instead generally correlated negatively with internet addiction~\cite{Kayis2016BigFive, Tian2021Associations, Jojo2022Personality}. However, there is some evidence showing that openness, extroversion, and conscientiousness can be heightening factors~\cite{Nwufo2024Personality}. 

Internet addiction has been shown to be correlated with higher levels of cognitive and decision fatigue~\cite{Ioannidis2019Cognitive}, shame~\cite{Craparo2014InternetAddiction}, anxiousness and stress~\cite{Sayed2022InternetAddiction, Kumar2025Netholicism}, but also with depression and compulsive behaviors such as substance abuse or online gambling~\cite{Kumar2025Netholicism}.
Other findings in the literature bear similar trends by showing positive correlations between social anxiety and internet addiction~\cite{Dong2024SocialAnxiety}. Internet addiction can also lead to social isolation and hinder communication skills~\cite{Kumar2025Netholicism}.

\subsubsection*{Risk-Taking Interactions}


Impulsivity and internet addiction are factors that were shown to be significant predictors of riskier cybersecurity behaviors~\cite{hadlington2017human, Zhang2025Emotional}. Moreover, more risk-taking behavior is linked to worse attitudes towards cybersecurity policies~\cite{hadlington2017human, hadlington2018employees}.
Risk-taking behaviors have also been linked to lower levels of self-reported security knowledge~\cite{kennison2020risks}. 

\subsubsection*{Laziness Interactions}

Laziness can be manifested in various forms, such as procrastination. Studies have investigated the correlation between procrastination and the Big 5 traits, finding it is negatively correlated with agreeableness, conscientiousness, and extraversion, while being positively correlated with neuroticism~\cite{Steel2007Procrastination, Meng2024Procrastinators}.

\subsection*{A2.1.6 Interactions with Social and Organizational Factors}
\label{sec:social_interactions}

\begin{figure}[htbp]
    \centering
    \includegraphics[width=\textwidth]{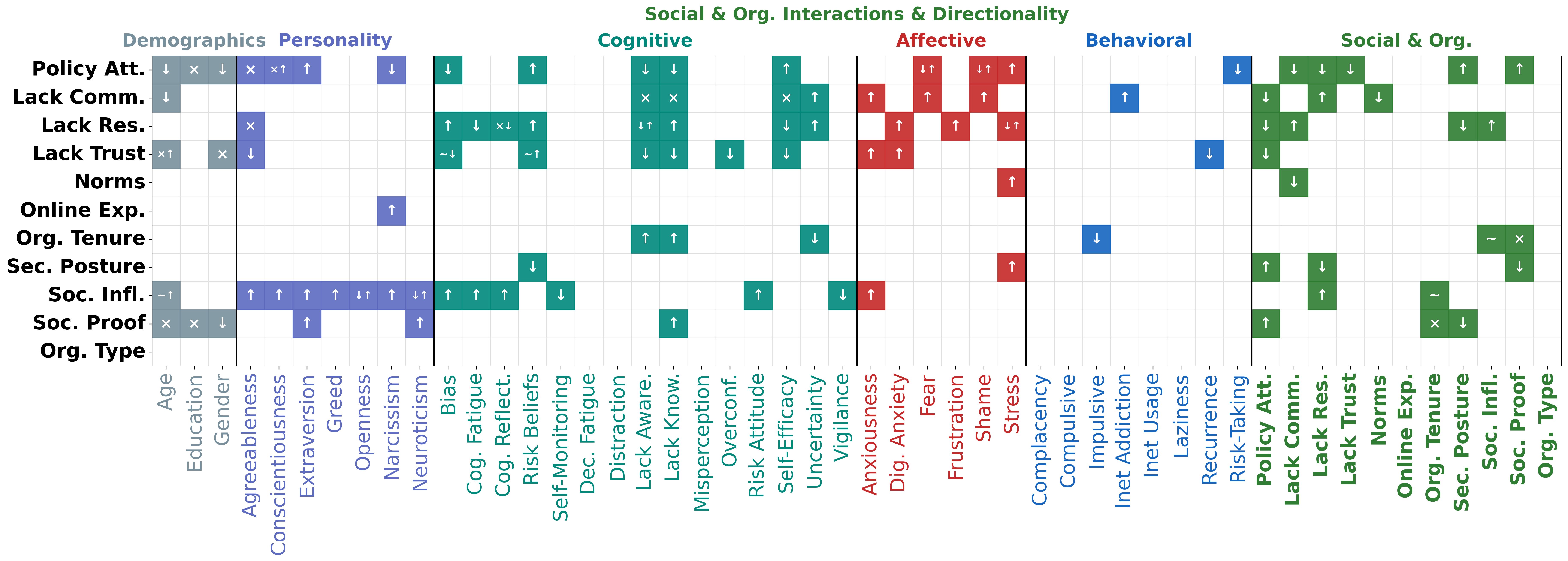}
    \caption{Summary of the interactions among the social and organizational human factors and the other ones. \textit{Legend}: $\uparrow$ indicates a positive association between the traits, $\downarrow$ a negative association, $\sim$ a mixed or complex association, and $\times$ a modulation effect between the two factors.}
    \label{fig:matrix_social}
    \Description{A table summarizing the pairwise interactions between social and organizational human factors (rows) and other human factor categories (columns: Demographics, Personality, Cognitive, Affective, Behavioral, Social/Organizational).}
\end{figure}

\subsubsection*{Attitude towards policies Interactions}

As reported in the previous sections, employees' attitudes towards their organization's policies can be influenced by various factors. Another individual factor that can affect behavior includes the employees' age, with younger users generally having a more positive attitude~\cite{nguyen2024investigation}. 

Employees' knowledge and awareness can improve their compliance with corporate policies~\cite{DelsoVicente2025ComplianceReview, nguyen2024investigation, parsons2017human, Safa2015Information}; a higher educational level among employees can further enhance the positive effect of their knowledge on policy compliance~\cite{nguyen2024investigation}. 

To increase compliance, organizations must also address the lack of resources and support for their employees, for example, by creating training programs tailored to their roles and responsibilities, or by including the top management as an active part in security initiatives~\cite{DelsoVicente2025ComplianceReview}.
Cognitive and cultural biases are also crucial factors in employee policy adherence and should thus be addressed individually by each organization~\cite{Tsohou2015CognitiveBiases}.
Finally, communication and relationships within the organization are critical in promoting more policy compliance~\cite{Ifinedo2014Information}. 

\subsubsection*{Lack of communication Interactions}

Lack of communication with peers and technical support can increase susceptibility to cyberthreats. In fact, the combination of security self-efficacy, risk awareness, and social support has emerged as a powerful defensive synergy in the context of smartphone use~\cite{Zhou2020Risk}.

Communication is generally driven by the desire to warn or protect others from immediate threats that are either observed in others or experienced by oneself~\cite{Das2014SocialInfluence}. Therefore, communication can be encouraged by observing others' behaviors (i.e., social norms), and can, in turn, improve social norms.

Support-seeking intentions can increase as organizations provide more training, leading to more frequent communications among peers~\cite{Chen2024GroupDiscussion}. 
Emotions can hinder communication in the context of phishing reporting, as users might avoid reporting an email for which they fell victim to avoid embarrassment or for fear of getting colleagues into trouble (in case their account was compromised as part of a \textit{lateral phishing} attack)~\cite{Distler2023Influence}. Uncertainty about how to properly communicate a phishing attack may also prevent reporting~\cite{Distler2023Influence}.

\subsubsection*{Lack of trust Interactions}

Trust can be influenced by cognitive processes that occur naturally. The Truth-Default Theory poses that people tend to believe others by default, unless they have a reason to believe otherwise~\cite{Levine2014TDT}. Previous experiences, in fact, create biases that influence trust in communications in online settings: a source that is considered trustworthy due to past communications will thus continue to be likely considered as such~\cite{williams2017individual}. This can imply that familiarity with the sender and entities involved in a phishing email can increase trust and victims' overconfidence in trusting the attacker~\cite{wang2016overconfidence}.  

The level of trust in phishing emails can vary based on the recipient's age (older users tend to be less trusting of emails) and the strength of their email habits and computer knowledge~\cite{Butavicius2022People, aleroud2020examination}. Moreover, the general perceptions and beliefs that people hold regarding the potential risks of online communications and technology in general---such as privacy concerns---can make users more suspicious~\cite{williams2017individual, aleroud2020examination}.

Affective aspects also affect trust, as anxious individuals tend to be less trusting of others~\cite{Rossier2004NEO16PF}. 
Trust in security systems is strictly tied to employees' perceived effectiveness of security measures and can ultimately increase their willingness to comply with policies~\cite{DelsoVicente2025ComplianceReview}. 

In social engineering contexts, trust is also affected by demographic characteristics of the attacker: for example, older individuals can appear to be more trustworthy and less likely to present a risk, while females are generally perceived as more trustworthy than males and can lead to more risk-taking in the victims~\cite{Abuelezz2025SocialEngineering}. However, these results can vary based on cultural factors that shape subjective perceptions.

\subsubsection*{Security posture Interactions}

Security posture refers to the cybersecurity readiness/level of either an organization as a whole, or just considering its members (e.g., employees). The security posture of an organization can highly depend on whether top management provides proper support for security initiatives~\cite{Hu2012InfoSecCompliance}. 
Moreover, promoting good user behavior with solid organizational policies has a positive effect on employees' attitude towards the organization and can increase security compliance~\cite{Safa2015Information}.

\subsubsection*{Social influence and Social Proof Interactions}

Social influence is a complex aspect that can be influenced by several factors. Social influence techniques often leverage aspects such as time pressure (i.e., a lack of time resources), greed, and risk attitude of victims, who may be motivated by financial gains or a desire for challenge~\cite{williams2017individual}. 
Social influence techniques in social engineering attacks can lead target users to rely on heuristic processing to deal with, e.g., urgency and authority cues, and save cognitive effort~\cite{Vishwanath2015Examining, Chou2021Mindless}; in fact, they aim at diverting the victim's attentional resources away from suspicious cues and leveraging biased processing strategies~\cite{vishwanath2011people}. However, when social influence techniques trigger suspicion or use very convincing arguments, they can activate the more reflective cognitive system of users~\cite{Chou2021Mindless, Musuva2019Cognitive, vishwanath2011people}. 

Resistance to social influence can vary depending on the age of users, but the evidence is contradictory~\cite{Steinberg2007Age, Taib2019SocialEngineering}; in fact, users of different ages may be susceptible to different social influence strategies~\cite{lin2019susceptibility}. Organizational tenure can also have varying effects on social influence susceptibility, with more senior employees being less susceptible to authority cues, but not to urgency cues, in contrast to middle-level employees~\cite{Gallo2024HumanFactor}. 

An individual's personality traits can also increase their susceptibility to social influence. As already reported in Section~\ref{sec:personality_interactions}, agreeable people are generally more vulnerable to social engineering due to their trusting nature~\cite{Cusack2018Personality, Shropshire2015Personality, Palm2025Influence}. Another trait that generally increases susceptibility to persuasion strategies and social influence is extraversion~\cite{Palm2025Influence, Cusack2018Personality}, while people with higher values of neuroticism and openness are less affected by persuasion strategies~\cite{Palm2025Influence}. However, there is evidence that higher traits of neuroticism, openness, and conscientiousness lead instead to increased susceptibility to social influence~\cite{Oyibo2019Relationship}. 

Regarding social proof, a study by Coopamootoo and colleagues found that, when looking for security advice, women tend to rely more on social connections, while men prefer online content~\cite{coopamootoo2023gender}. In general, social proof is a powerful factor in influencing employees' behavior, especially those with low organizational tenure~\cite{Williams2018Exploring}. In fact, social pressure from peers and superiors impacts an employee's behavior and positively (or negatively) influences their attitude and compliance with policies~\cite{Ifinedo2014Information, Cheng2013Understanding, Safa2015Information}. The social context and the impact of peers can also influence the beliefs and biases that people use for judgments, for example, when evaluating the quality of a web interface~\cite{Soper2020Informational}.
Social proof can even impact people's knowledge and security posture~\cite{Hong2023Influence}; interestingly, this is especially true for individuals with higher education levels~\cite{Hong2023Influence}. 
\section*{A2.2 Tables of interactions between human factors}

\textit{Legend:} In the following tables, the symbols indicate the nature of the relationship between the two factors: 
$\uparrow$ denotes a positive association (an increase in one factor correlates with an increase in the other); 
$\downarrow$ denotes a negative association; 
$\sim$ indicates a mixed, complex, or non-linear association; 
and $\times$ represents a modulation effect. 

\textit{Note}: The absence of an entry in the tables indicates that no interaction was found in the reviewed literature.

\subsection*{A2.2.1 Demographic factors interactions}

\begin{center}
\centering
\small
\begin{longtable}{p{0.2\textwidth}||c|c|c|}

\caption{Interactions for the \textbf{Demographics} factors. \textit{Legend}: $\uparrow$ indicates a positive association between the traits, $\downarrow$ a negative association, $\sim$ a mixed or complex association, and $\times$ a modulation effect between the two factors.}
\label{tab:demographics_interactions} \\
    & \textbf{Age} & \textbf{Education} & \textbf{Gender (1=M)} \\
    \toprule
    \endfirsthead

    & \textbf{Age} & \textbf{Education} & \textbf{Gender (1=M)} \\
    \toprule
    \endhead

    \multicolumn{4}{c}{\textit{(Table continues on the next page)}} \\
    
    \endfoot
    \endlastfoot



\textbf{Agreeableness} &
$\times$~\cite{aschwanden2025longitudinal}, $\uparrow$~\cite{Soto2012Development, Allemand2008Personality} &
$\uparrow$~\cite{Kassenboehmer2018University} &
$\downarrow$~\cite{Costa2001GenderPersonality,Otterbring2022Pandemic,Leslie2025Gender,Vecchione2012Gender,Weisberg2011Gender} \\
\hline

\textbf{Conscientiousness} &
$\times$~\cite{aschwanden2025longitudinal}, $\uparrow$~\cite{Soto2012Development, Allemand2008Personality, Vecchione2012Gender} &
$\uparrow$~\cite{Kassenboehmer2018University} &
$\downarrow$~\cite{Otterbring2022Pandemic,Vecchione2012Gender} \\
\hline

\textbf{Extraversion} &
$\times$~\cite{aschwanden2025longitudinal}, $\downarrow$~\cite{Soto2012Development, Allemand2008Personality} &
$\uparrow$~\cite{Kassenboehmer2018University} & 
$\downarrow$~\cite{Weisberg2011Gender, AlosFerrer2016CognitiveReflection} \\
\hline

\textbf{Greed} &  &  & $\uparrow$~\cite{krekels2015Dispositional} \\
\hline

\textbf{Openness} &
$\times$~\cite{aschwanden2025longitudinal}, $\downarrow$~\cite{Soto2012Development, Allemand2008Personality}, $\uparrow$~\cite{Vecchione2012Gender} &
$\uparrow$~\cite{Ludeke2014Truth} &
$\downarrow$~\cite{Vecchione2012Gender}, $\uparrow$~\cite{Costa2001GenderPersonality} \\
\hline

\textbf{Narcissism} &
$\uparrow$~\cite{Leite2023Dark} &
  &
  \\
\hline

\textbf{Neuroticism} &
$\times$~\cite{aschwanden2025longitudinal}, $\downarrow$~\cite{Soto2012Development,Allemand2008Personality} & 
  & 
$\downarrow$~\cite{Costa2001GenderPersonality, Leslie2025Gender,Weisberg2011Gender, Murphy2021International,Vecchione2012Gender, AlosFerrer2016CognitiveReflection}
  \\
\hline
\hline


\textbf{Cognitive fatigue} &
 $\uparrow$~\cite{Bouche2025Mental} &
  &
  \\
\hline

\end{longtable}
\end{center}

\subsection*{A2.2.2 Personality factors interactions}

\begin{center}
    \setlength{\tabcolsep}{3pt} 
\small
   \begin{longtable}{p{0.15\textwidth}||p{0.1\textwidth}|p{0.1\textwidth}|p{0.12\textwidth}|p{0.08\textwidth}|p{0.1\textwidth}|p{0.1\textwidth}|p{0.12\textwidth}|}
   
\caption{Interactions for the \textbf{Personality} factors. \textit{Legend}: $\uparrow$ indicates a positive association between the traits, $\downarrow$ a negative association, $\sim$ a mixed or complex association, and $\times$ a modulation effect between the two factors.}
\label{tab:personality_interactions} \\
& \textbf{Agree.ness} & \textbf{Consc.ness} & \textbf{Extraversion} & \textbf{Greed} & \textbf{Openness} & \textbf{Narcissism} & \textbf{Neuroticism} \\
\toprule
\endfirsthead

& \textbf{Agree.ness} & \textbf{Consc.ness} & \textbf{Extraversion} & \textbf{Greed} & \textbf{Openness} & \textbf{Narcissism} & \textbf{Neuroticism} \\
\toprule
\endhead

\multicolumn{8}{c}{\textit{(Table continues on the next page)}} \\

\endfoot
\endlastfoot

\textbf{Agree.ness} & \cellcolor{gray!30}  & \cellcolor{gray!30} & \cellcolor{gray!30} & \cellcolor{gray!30} & \cellcolor{gray!30} & \cellcolor{gray!30} & \cellcolor{gray!30} \cellcolor{gray!30}\\
\hline
\textbf{Consc.ness} & $\uparrow$~\cite{DeYoung2015Cybernetic} & \cellcolor{gray!30} & \cellcolor{gray!30}  & \cellcolor{gray!30} & \cellcolor{gray!30} & \cellcolor{gray!30} & \cellcolor{gray!30} \\
\hline
\textbf{Extraversion} & $\uparrow$~\cite{Bhattacharjee2025Effect} & $\uparrow$~\cite{Bhattacharjee2025Effect} & \cellcolor{gray!30} & \cellcolor{gray!30} & \cellcolor{gray!30} & \cellcolor{gray!30} & \cellcolor{gray!30} \\
\hline
\textbf{Greed} & $\downarrow$~\cite{krekels2015Dispositional} & $\downarrow$~\cite{Zeelenberg2024Disentangling} & $\downarrow$~\cite{krekels2015Dispositional} & \cellcolor{gray!30} & \cellcolor{gray!30} & \cellcolor{gray!30} & \cellcolor{gray!30} \\
\hline
\textbf{Openness} & $\uparrow$~\cite{Bhattacharjee2025Effect} &  & $\uparrow$~\cite{DeYoung2015Cybernetic, Bhattacharjee2025Effect} &  & \cellcolor{gray!30} & \cellcolor{gray!30} & \cellcolor{gray!30} \\
\hline
\textbf{Narcissism} &  &  &  &  &  & \cellcolor{gray!30} & \cellcolor{gray!30} \\
\hline
\textbf{Neuroticism} & $\downarrow$~\cite{DeYoung2015Cybernetic}, $\uparrow$~\cite{Bhattacharjee2025Effect}& $\downarrow$~\cite{DeYoung2015Cybernetic} & $\uparrow$~\cite{Bhattacharjee2025Effect} & $\uparrow$~\cite{krekels2015Dispositional} & $\uparrow$~\cite{Bhattacharjee2025Effect} &  & \cellcolor{gray!30} \\
\hline

\end{longtable}
\end{center}

\begin{landscape}
\subsection*{A2.2.3 Cognitive factors interactions}

\setlength{\tabcolsep}{2pt}
\small

\begin{longtable}{p{0.1\textwidth}||*{15}{p{0.069\textwidth}|}}

\caption{Interactions for the \textbf{Cognitive} factors. \textit{Legend}: $\uparrow$ indicates a positive association between the traits, $\downarrow$ a negative association, $\sim$ a mixed or complex association, and $\times$ a modulation effect between the two factors.}
\label{tab:cognitive_interactions}
\\

 & \textbf{Bias} & \textbf{Cog. fatigue} & \textbf{Cog. reflectiv.} & \textbf{Cyb. r. beliefs} & \textbf{Cyb. self-m.} & \textbf{Dec. fatigue} & \textbf{Distract.} & \textbf{Lack of awar.} & \textbf{Lack of know.} & \textbf{Misper.} & \textbf{Overco.} & \textbf{Risk attitude} & \textbf{Sec. self-eff.} & \textbf{Uncert.} & \textbf{Vigilan.} \\
\toprule
\endfirsthead

 & \textbf{Bias} & \textbf{Cog. fatigue} & \textbf{Cog. reflectiv.} & \textbf{Cyb. r. beliefs} & \textbf{Cyb. self-m.} & \textbf{Dec. fatigue} & \textbf{Distract.} & \textbf{Lack of awar.} & \textbf{Lack of know.} & \textbf{Misper.} & \textbf{Overco.} & \textbf{Risk attitude} & \textbf{Sec. self-eff.} & \textbf{Uncert.} & \textbf{Vigilan.} \\
\toprule
\endhead

\multicolumn{16}{c}{\textit{(Table continues on the next page)}}\\
\endfoot
\endlastfoot

\textbf{Bias} & \cellcolor{gray!30} & \cellcolor{gray!30} & \cellcolor{gray!30} & \cellcolor{gray!30} & \cellcolor{gray!30} & \cellcolor{gray!30} & \cellcolor{gray!30} & \cellcolor{gray!30} & \cellcolor{gray!30} & \cellcolor{gray!30} & \cellcolor{gray!30} & \cellcolor{gray!30} & \cellcolor{gray!30} & \cellcolor{gray!30} & \cellcolor{gray!30} \\
\hline

\textbf{Cog. fatigue} &  & \cellcolor{gray!30} & \cellcolor{gray!30} & \cellcolor{gray!30} & \cellcolor{gray!30} & \cellcolor{gray!30} & \cellcolor{gray!30} & \cellcolor{gray!30} & \cellcolor{gray!30} & \cellcolor{gray!30} & \cellcolor{gray!30} & \cellcolor{gray!30} & \cellcolor{gray!30} & \cellcolor{gray!30} & \cellcolor{gray!30}  \\
\hline

\textbf{Cog. reflect.} & $\downarrow$~\cite{AlosFerrer2016CognitiveReflection, Chou2021Mindless} &  & \cellcolor{gray!30} & \cellcolor{gray!30} & \cellcolor{gray!30} & \cellcolor{gray!30} & \cellcolor{gray!30} & \cellcolor{gray!30} & \cellcolor{gray!30} & \cellcolor{gray!30} & \cellcolor{gray!30} & \cellcolor{gray!30} & \cellcolor{gray!30} & \cellcolor{gray!30} & \cellcolor{gray!30} \\
\hline

\textbf{Cyb. r. bel.} &  &  & $\uparrow$~\cite{Kwak2020Spear, Musuva2019Cognitive} &  \cellcolor{gray!30} & \cellcolor{gray!30} & \cellcolor{gray!30} & \cellcolor{gray!30} & \cellcolor{gray!30} & \cellcolor{gray!30} & \cellcolor{gray!30} & \cellcolor{gray!30} & \cellcolor{gray!30} & \cellcolor{gray!30} & \cellcolor{gray!30} & \cellcolor{gray!30} \\
\hline

\textbf{Cyb. self-m.} & $\downarrow$~\cite{Hutton1992SelfAwareness} &  & $\uparrow$~\cite{Waqas2023Enhancing} & $\uparrow$~\cite{Kwak2020Spear, Martens2019Investigating} &  \cellcolor{gray!30} & \cellcolor{gray!30} & \cellcolor{gray!30} & \cellcolor{gray!30} & \cellcolor{gray!30} & \cellcolor{gray!30} & \cellcolor{gray!30} & \cellcolor{gray!30} & \cellcolor{gray!30} & \cellcolor{gray!30} & \cellcolor{gray!30}  \\
\hline
\end{longtable}

\end{landscape}
\subsection*{A2.2.4 Affective factors interactions}

\begin{center}
\setlength{\tabcolsep}{3pt}
\small
\begin{longtable}{p{0.15\textwidth}||p{0.12\textwidth}|p{0.12\textwidth}|p{0.12\textwidth}|p{0.12\textwidth}|p{0.12\textwidth}|p{0.12\textwidth}|}
\caption{Interactions for the \textbf{Affective} factors. \textit{Legend}: $\uparrow$ indicates a positive association between the traits, $\downarrow$ a negative association, $\sim$ a mixed or complex association, and $\times$ a modulation effect between the two factors.} 
\label{tab:emotional_interactions}
\\
& \textbf{Anxiousness} & \textbf{Digital anx.} & \textbf{Fear} & \textbf{Frustration} & \textbf{Shame} & \textbf{Stress} \\
\toprule

\endfirsthead
 & \textbf{Anxiousness} & \textbf{Digital anx.} & \textbf{Fear} & \textbf{Frustration} & \textbf{Shame} & \textbf{Stress} \\
\toprule
\endhead

\multicolumn{7}{c}{\textit{(Table continues on the next page)}}\\
\endfoot
\endlastfoot

\textbf{Anxiousness} & \cellcolor{gray!30} & \cellcolor{gray!30} & \cellcolor{gray!30} & \cellcolor{gray!30} & \cellcolor{gray!30} & \cellcolor{gray!30} \\
\hline

\textbf{Digital anx.} &  & \cellcolor{gray!30} & \cellcolor{gray!30} & \cellcolor{gray!30} & \cellcolor{gray!30} & \cellcolor{gray!30} \\
\hline

\textbf{Fear} & $\uparrow$~\cite{Spiroiu2025SocialAnxiety} &  & \cellcolor{gray!30} & \cellcolor{gray!30} & \cellcolor{gray!30} & \cellcolor{gray!30} \\
\hline

\textbf{Frustration} &  &  &  & \cellcolor{gray!30} & \cellcolor{gray!30} & \cellcolor{gray!30} \\
\hline

\textbf{Shame} &  &  &  &  & \cellcolor{gray!30} & \cellcolor{gray!30} \\
\hline

\textbf{Stress} & $\times$~\cite{Hengen2021Stress, Mizrak2025DigitalDetox}, $\downarrow$~\cite{Chen2022InfoSecStress}, $\uparrow$~\cite{Amin2024TechnostressEducators, Cuadrado2024Technostress} & $\uparrow$~\cite{Pfaffinger2021DigitalisationAnxiety} & $\downarrow$~\cite{Chen2022InfoSecStress}, $\uparrow$~\cite{Aggarwal2024Eustress} &  &  & \cellcolor{gray!30} \\
\hline
\hline

\textbf{Age} &
$\downarrow$~\cite{tiraboschi2023Adolescent} &
$\uparrow$~\cite{Yoon2016Computer} &
$\sim$~\cite{Cook2022Fear} &
 &
$\downarrow$~\cite{Gambin2018Relations}, $\sim$~\cite{Orth2010Shame} &
$\downarrow$~\cite{nwachukwu2020covid}, $\uparrow$~\cite{RaguNathan2008Technostress} \\
\hline

\textbf{Education} & $\downarrow$\cite{Huseynov2024Using} &
$\downarrow$~\cite{Yoon2016Computer} &
&
 &
 &
 \\
\hline

\end{longtable}
\end{center}

\subsection*{A2.2.5 Behavioral factors interactions}

\begin{center}
\setlength{\tabcolsep}{3pt}
\small
\begin{longtable}{p{0.13\textwidth}||p{0.07\textwidth}|p{0.1\textwidth}|p{0.1\textwidth}|p{0.13\textwidth}|p{0.1\textwidth}|p{0.08\textwidth}|p{0.06\textwidth}|p{0.1\textwidth}|}
\caption{Interactions for the \textbf{Behavioural} factors. \textit{Legend}: $\uparrow$ indicates a positive association between the traits, $\downarrow$ a negative association, $\sim$ a mixed or complex association, and $\times$ a modulation effect between the two factors.} 
\label{tab:behavioral_interactions}
\\
& \textbf{Compla.} & \textbf{Compulsive behavior} & \textbf{Impulsive behavior} & \textbf{Internet addiction} & \textbf{Internet usage} & \textbf{Laziness} & \textbf{Recurr.} & \textbf{Risk-taking}
\\
\toprule

\endfirsthead
& \textbf{Compla.} & \textbf{Compulsive behavior} & \textbf{Impulsive behavior} & \textbf{Internet addiction} & \textbf{Internet usage} & \textbf{Laziness} & \textbf{Recurr.} & \textbf{Risk-taking} \\
\toprule
\endhead

\multicolumn{9}{c}{\textit{(Table continues on the next page)}}\\
\endfoot
\endlastfoot

\textbf{Complacency} & \cellcolor{gray!30} & \cellcolor{gray!30} & \cellcolor{gray!30} & \cellcolor{gray!30} & \cellcolor{gray!30} & \cellcolor{gray!30} & \cellcolor{gray!30} & \cellcolor{gray!30} \\
\hline

\textbf{Compul. behav.} &  & \cellcolor{gray!30} & \cellcolor{gray!30} & \cellcolor{gray!30} & \cellcolor{gray!30} & \cellcolor{gray!30} & \cellcolor{gray!30} & \cellcolor{gray!30}\\
\hline

\textbf{Impul. behav.} &  &  & \cellcolor{gray!30} & \cellcolor{gray!30} & \cellcolor{gray!30} & \cellcolor{gray!30} & \cellcolor{gray!30} & \cellcolor{gray!30} \\
\hline

\textbf{Internet add.} &  & $\uparrow$~\cite{Kumar2025Netholicism} & $\uparrow$~\cite{Salehi2023Impulsivity, Diotaiuti2022Internet, Marzilli2020Internet} & \cellcolor{gray!30} & \cellcolor{gray!30} & \cellcolor{gray!30} & \cellcolor{gray!30} & \cellcolor{gray!30} \\
\hline

\textbf{Internet usage} &  &  &  &  & \cellcolor{gray!30} & \cellcolor{gray!30} & \cellcolor{gray!30} & \cellcolor{gray!30} \\
\hline

\textbf{Laziness} &  &  &  &  &  & \cellcolor{gray!30} & \cellcolor{gray!30} & \cellcolor{gray!30} \\
\hline

\textbf{Recurrence} &  &  &  &  &  &  & \cellcolor{gray!30} & \cellcolor{gray!30} \\
\hline

\textbf{Risk-taking} & $\uparrow$~\cite{Zhang2025Emotional} & & $\uparrow$~\cite{hadlington2017human, Zhang2025Emotional} & $\uparrow$~\cite{hadlington2017human} & & &  & \cellcolor{gray!30} \\
\hline

\end{longtable}
\end{center}

\begin{landscape}
\subsection*{A2.2.6 Social and organizational factors interactions}

\setlength{\tabcolsep}{2pt}
\small

\begin{longtable}{p{0.13\textwidth}||p{0.12\textwidth}|p{0.12\textwidth}|p{0.12\textwidth}|p{0.12\textwidth}|p{0.06\textwidth}|p{0.08\textwidth}|p{0.08\textwidth}|p{0.08\textwidth}|p{0.12\textwidth}|p{0.07\textwidth}|p{0.06\textwidth}|}

\caption{Interactions for the \textbf{Social and organizational} factors. \textit{Legend}: $\uparrow$ indicates a positive association between the traits, $\downarrow$ a negative association, $\sim$ a mixed or complex association, and $\times$ a modulation effect between the two factors.}
\label{tab:social_interactions}
\\

 & \textbf{Att. tow. policies} & \textbf{Lack of commun.} & \textbf{Lack of resources} & \textbf{Lack of trust} & \textbf{Norms} & \textbf{Online exposure} & \textbf{Organiz. tenure} & \textbf{Security posture} & \textbf{Social influence} & \textbf{Social proof} & \textbf{Type of org.} \\
\toprule
\endfirsthead

 & \textbf{Att. tow. policies} & \textbf{Lack of commun.} & \textbf{Lack of resources} & \textbf{Lack of trust} & \textbf{Norms} & \textbf{Online exposure} & \textbf{Organiz. tenure} & \textbf{Security posture} & \textbf{Social influence} & \textbf{Social proof} & \textbf{Type of org.} \\
\toprule
\endhead

\multicolumn{12}{c}{\textit{(Table continues on the next page)}}\\
\endfoot
\endlastfoot

\textbf{Att. tow. pol.} & \cellcolor{gray!30} & \cellcolor{gray!30} & \cellcolor{gray!30} & \cellcolor{gray!30} & \cellcolor{gray!30} & \cellcolor{gray!30} & \cellcolor{gray!30} & \cellcolor{gray!30} & \cellcolor{gray!30} & \cellcolor{gray!30} & \cellcolor{gray!30} \\
\hline

\textbf{Lack of comm.} & $\downarrow$~\cite{Ifinedo2014Information} & \cellcolor{gray!30} & \cellcolor{gray!30} & \cellcolor{gray!30} & \cellcolor{gray!30} & \cellcolor{gray!30} & \cellcolor{gray!30} & \cellcolor{gray!30} & \cellcolor{gray!30} & \cellcolor{gray!30} & \cellcolor{gray!30} \\
\hline

\textbf{Lack of resour.} & $\downarrow$~\cite{DelsoVicente2025ComplianceReview} & $\uparrow$~\cite{Chen2024GroupDiscussion} & \cellcolor{gray!30} & \cellcolor{gray!30} & \cellcolor{gray!30} & \cellcolor{gray!30} & \cellcolor{gray!30} & \cellcolor{gray!30} & \cellcolor{gray!30} & \cellcolor{gray!30} & \cellcolor{gray!30} \\
\hline

\textbf{Lack of trust} & $\downarrow$~\cite{DelsoVicente2025ComplianceReview} &  &  & \cellcolor{gray!30} & \cellcolor{gray!30} & \cellcolor{gray!30} & \cellcolor{gray!30} & \cellcolor{gray!30} & \cellcolor{gray!30} & \cellcolor{gray!30} & \cellcolor{gray!30}  \\
\hline

\textbf{Norms} &  & $\downarrow$~\cite{Das2014SocialInfluence} &  &  & \cellcolor{gray!30} & \cellcolor{gray!30} & \cellcolor{gray!30} & \cellcolor{gray!30} & \cellcolor{gray!30} & \cellcolor{gray!30} & \cellcolor{gray!30}  \\
\hline

\textbf{Online expos.} &  &  &  &  &  & \cellcolor{gray!30} & \cellcolor{gray!30} & \cellcolor{gray!30} & \cellcolor{gray!30} & \cellcolor{gray!30} & \cellcolor{gray!30} \\
\hline

\textbf{Organiz. tenure} &  &  &  &  &  &  & \cellcolor{gray!30} & \cellcolor{gray!30} & \cellcolor{gray!30} & \cellcolor{gray!30} & \cellcolor{gray!30} \\
\hline

\textbf{Sec. posture} & $\uparrow$~\cite{Safa2015Information} &  & $\downarrow$~\cite{Hu2012InfoSecCompliance} &  &  &  &  & \cellcolor{gray!30} & \cellcolor{gray!30} & \cellcolor{gray!30} & \cellcolor{gray!30} \\
\hline

\end{longtable}

\end{landscape}

\clearpage
\section*{APPENDIX 3: Measurement instruments for assessing human factors}
\label{sec:appendix_3}

\setlength{\tabcolsep}{6pt}
\renewcommand{\arraystretch}{1.2}
\begingroup
\small
\begin{longtable}{|>{\raggedright\arraybackslash}p{5cm}|
                        >{\raggedright\arraybackslash}p{4.2cm}|
                        >{\raggedright\arraybackslash}p{2.2cm}|
                        >{\raggedright\arraybackslash}p{1.9cm}|}
\caption{Summary of the 99 measurement instruments for assessing human factors in cybersecurity, organized by dimension and data collection type. The table includes all measurement instruments reported in the text and supplementary material.}
\label{tab:measurement_tools}\\

\hline
\rowcolor[HTML]{EFEFEF}
\textbf{Name of Solution} & \textbf{Human Factor (HF)} & \textbf{Citation} & \textbf{Type} \\
\hline
\endfirsthead

\hline
\rowcolor[HTML]{EFEFEF}
\textbf{Name of Solution} & \textbf{Human Factor (HF)} & \textbf{Citation} & \textbf{Type} \\
\hline
\endhead

\hline
\multicolumn{4}{|r|}{\textit{(Table continues on the next page)}}\\
\hline
\endfoot

\hline
\endlastfoot

\multicolumn{4}{|c|}{\textbf{Demographic Factors}} \\ \hline
Self-report demographic questionnaire & Age, Gender, Education & \textit{n/a (standard demographic form)} & First-person \\ \hline

\multicolumn{4}{|c|}{\textbf{Personality Traits}} \\ \hline
Big Five Inventory (BFI) & Personality traits (Agreeableness, Conscientiousness, Extraversion, Openness, Neuroticism) & \cite{john1991big} & First-person \\ \hline
Big Five Inventory-2 (BFI-2) & Personality traits (Agreeableness, Conscientiousness, Extraversion, Openness, Neuroticism) & \cite{soto2016bfi2} & First-person \\ \hline
Dispositional Greed Scale (DGS) & Greed & \cite{zeelenberg2021dispositional} & First-person \\ \hline
Five-Factor Narcissism Inventory (FFNI) & Narcissism & \cite{glover2012five} & First-person \\ \hline
HEXACO-24 & Personality traits (Agreeableness, Conscientiousness, Extraversion, Openness, Neuroticism) & \cite{DEVRIES2013871} & First-person \\
\hline

\end{longtable}
\endgroup

\clearpage
\section*{APPENDIX 4: Prompts used in the AI-assisted Systematic Screening}
\label{sec:appendix_4}

\subsection*{A4.1 Prompts for human factors interactions}


\begin{lstlisting}[style=prompt, caption={Prompt template to find the interactions of the human factors of a specific dimensions with all the 50 human factors.}, label={lst:interactions_prompt}]
I'm conducting a systematic literature review on interactions between [DIMENSION] human factors and other human factors in the domains of cybersecurity and human-computer interaction (HCI).

Specifically, I am looking for peer-reviewed empirical or theoretical studies that investigate relationships (e.g., correlation, causality, co-occurrence, or influence) between the following [DIMENSION] human factors:
[LIST OF ALL THE HUMAN FACTORS OF THE DIMENSION]

and the following categories of human factors:
- Demographics: Age, Education, Gender
- Personality traits: Agreeableness, Conscientiousness, Extraversion, Greed, Narcissism, Neuroticism, Openness
- Cognitive factors: Bias, Cognitive fatigue, Cognitive reflectiveness, Cyber risk beliefs, Cybersecurity self-monitoring, Decision fatigue, Distraction, Lack of awareness, Lack of knowledge, Misperception, Overconfidence, Risk attitude, Security self-efficacy, Uncertainty, Vigilance
- Emotional factors: Anxiousness, Digital anxiety, Fear, Frustration, Shame, Stress
- Behavioral factors: Complacency, Compulsive behavior, Impulsive behavior, Internet addiction, Internet usage, Laziness, Recurrence, Risk-taking
- Social and organizational factors: Attitude toward policy, Lack of communication, Lack of resources, Lack of trust, Norms, Online exposure, Organizational tenure, Security posture, Social influence, Social proof, Type of organization

Please search for all possible pairwise interactions between each [DIMENSION] factor and each of the other factors above.

Return the results in a table format with the following columns:
- Human Factor 1 ([DIMENSION] factor)
- Human Factor 2
- Source(s) in APA style
- Type of relationship observed (e.g., correlation, causal influence, co-occurrence, moderation)
- Study context (e.g., cybersecurity, HCI, psychology)
- Exact quote(s) from the paper describing the interaction

 If no studies are found for a given pair, explicitly write "No evidence found".

 Focus only on peer-reviewed scientific publications. Use reliable academic sources such as IEEE, ACM, Springer, Elsevier, Wiley, Taylor & Francis, or peer-reviewed psychology/cybersecurity journals.
\end{lstlisting}


\subsection*{A4.2 Prompts for human factors measurement solutions}

\begin{lstlisting}[style=prompt, caption={Prompt template to find the measurement solutions for all the 50 human factors.}, label={lst:measurements_prompt}]
Conduct a systematic review of validated measurement instruments used to assess human factors that influence cybersecurity behavior and decision-making.

For each of the following 50 human factors, identify relevant psychometric scales, behavioral assessment tools, or standardized methods. Focus on instruments used in psychology, behavioral science, cybersecurity, or human-computer interaction (HCI).

For each human factor, list:
- Name of the instrument
- Brief description
- Peer-reviewed references demonstrating validation and use in academic research

Prioritize instruments with established validity and reliability. Organize results by human factor. 

Human factors include:
[LIST OF ALL THE HUMAN FACTORS]

Search should cover scholarly databases (e.g., Google Scholar, PubMed, IEEE Xplore, ACM DL, Scopus, PsycINFO).
\end{lstlisting}

\clearpage
\clearpage
\section*{Appendix 5: Additional Operational Scenarios}
\label{sec:appendix_5}

The following scenarios provide brief conceptual walkthroughs demonstrating the versatility of the MORPHEUS framework across different operational security contexts.

\subsection*{A5.1 Scenario C: Risk Prioritization via Threat Mapping}
\textbf{Challenge:} Security resources are finite. Organizations need to prioritize mitigations based on the specific human factors that trigger their most critical attack vectors.

\begin{tcolorbox}[colback=gray!10, breakable, enhanced, colframe=gray!50, title=\textbf{Vignette: Dashboard Design for Cloud Services}]
\small
\textbf{Context:} UX designers are building a new interface for cloud resource management used by sysadmins, a group prone to specific configuration errors \cite{Rahman2023Misconfigurations}.

\textbf{Phase 1: Factor Identification \& Filtering:} Using Table~\ref{tab:hfs_vertical_larger} for ``Misconfiguration,'' the team filters for factors relevant to \textit{expert users}. They prioritize \textbf{Overconfidence} and \textbf{Recurrence}, while deprioritizing \textbf{Lack of Knowledge}.

\textbf{Phase 2 \& 3: Assessment \& Systemic Analysis:} Recognizing that expert users are prone to Overconfidence \cite{wang2016overconfidence}, the design team maps the critical threat of accidental server deletion to these specific traits, identifying a susceptibility to the \textbf{Trust and Bias Overconfidence Trap (Mechanism \hyperref[int_pat:trust]{5})}.

\textbf{Outcome:} The team removes ``one-click'' deployments. They introduce a ``cognitive wedge''---a mandatory confirmation dialog requiring the user to type the resource name with actionable mitigations \cite{manfredi2021}---to break the habit (\textbf{Recurrence}) loop.
\end{tcolorbox}

\subsection*{A5.2 Scenario D: Monitoring with Validated Tools}
\textbf{Challenge:} ``Security culture'' is often a buzzword. MORPHEUS transforms it into a measurable metric to track progress over time.

\begin{tcolorbox}[colback=gray!10, breakable, enhanced, colframe=gray!50, title=\textbf{Vignette: Periodic Screening in Finance}]
\small
\textbf{Context:} A bank wants to assess the effectiveness of its security awareness program without causing employee burnout \cite{Aggarwal2024Eustress}.

\textbf{Phase 1 \& 2: Filtering \& Assessment:} The CISO selects the \textit{Big Five Inventory-2} and \textit{HAIS-Q} (Table \ref{tab:measurement_tools}) to monitor the baseline of the workforce, given the link between personality and financial risk \cite{Fawad2020Personality}. Hypothetical data reveal a specific department with high \textbf{Neuroticism} and low \textbf{Security self-efficacy}.

\textbf{Phase 3: Systemic Analysis:} This specific profile breaks the \textbf{Motivation Balance (Mechanism \hyperref[int_pat:motivation]{9})} and activates a fear-paralysis dynamic: high threat perception without efficacy leads to avoidance rather than action.

\textbf{Outcome:} The CISO halts planned phishing simulations for that group (which would be counterproductive \cite{Brennan2010Fear}) and pivots to positive reinforcement strategies to build self-efficacy.
\end{tcolorbox}

\subsection*{A5.3 Scenario E: Psychological Red Teaming}
\textbf{Challenge:} Traditional Red Teaming focuses on technical flaws. MORPHEUS provides a menu of psychological vulnerabilities to test the ``human surface.''

\begin{tcolorbox}[colback=gray!10, breakable, enhanced, colframe=gray!50, title=\textbf{Vignette: The ``Friday Afternoon'' Attack}]
\small
\textbf{Phase 1: Factor Identification:} Instead of random attacks, the Red Team filters for factors exacerbated by the \textit{end-of-week context}, specifically \textbf{Lack of resources (time)} \cite{chowdhury2020time, Butavicius2022People}.

\textbf{Phase 2: Execution (Assessment):} They send CEO fraud emails timed strictly to arrive at 5:00 PM on Friday, leveraging the desire to clear the inbox~\cite{vishwanath2011people}.

\textbf{Phase 3: Systemic Analysis \& Outcome:} Post-test interviews reveal the \textbf{Resource-Constraint Cascade (Mechanism~\hyperref[int_pat:resource]{7})}: \textbf{Fear} (Affective) of delaying a superior's request overrode the users' \textbf{Knowledge} (Cognitive) of verification procedures \cite{Gallo2024HumanFactor}. This confirms the vulnerability was rooted in the Organizational Culture exerting pressure on cognitive resources.
\end{tcolorbox}

\subsection*{A5.4 Scenario F: Blameless Post-Incident Forensics}
\textbf{Challenge:} Avoiding the ``blame and shame'' culture requires understanding the user's psychological state at the moment of error.

\begin{tcolorbox}[colback=gray!10, breakable, enhanced, colframe=gray!50, title=\textbf{Vignette: Ransomware Patient Zero}]
\small
\textbf{Phase 1 \& 2: Reconstruction \& Contextualization:} Forensic analysts use the MORPHEUS dimensions to identify that the incident occurred during a major software migration. They determine the user was experiencing high \textbf{Digital Anxiety} (Affective) due to the new tools \cite{wilson2023development} and was acting under \textbf{Social Influence}.

\textbf{Phase 3: Systemic Analysis \& Outcome:} The root cause is identified not as individual negligence, but as a systemic failure involving a \textbf{Lack of Resources} (support/training) \cite{ngandu2025strengthening}. This triggered the \textbf{Resource-Constraint Cascade (Mechanism~\hyperref[int_pat:resource]{7})}, exacerbating the user's anxiety and paralyzing their response \cite{Renaud2021CybersecurityEmotions}.
\end{tcolorbox}

\subsection*{A5.5 Scenario G: Breaking the Habitual Loop in Access Control}
\textbf{Challenge:} Employees persistently reuse passwords despite rigorous policy updates. Traditional enforcement leads to ``shadow IT'' workarounds.

\begin{tcolorbox}[colback=gray!10, breakable, enhanced, colframe=gray!50, title=\textbf{Vignette: The Credential Stuffing Incident}]
\small
\textbf{Phase 1 \& 2: Identification \& Diagnosis:} Following an attack, audits reveal high password reuse. The CISO filters for behavioral factors and identifies \textbf{Recurrence} (Habits) as the primary driver. Interviews reveal reuse is not due to a \textbf{Lack of Knowledge}, but to cognitive automaticity driven by high \textbf{Cognitive fatigue} and \textbf{Laziness} \cite{das2014tangled, whitty2015individual}.
\textbf{Phase 3: Systemic Analysis \& Outcome:} Engaging the \textbf{Habitual Autopilot Loop (Mechanism~\hyperref[int_pat:habitual]{11})} and \textbf{Dual Cognitive Process (Mechanism~\hyperref[int_pat:dual_cognitive]{2})}, the organization deploys a password manager with forced ``contextual friction'' (visually flagging reused passwords). This disrupts the habitual loop at the moment of creation, forcing a System 2 decision \cite{fagan2017investigation}.
\end{tcolorbox}

\subsection*{A5.6 Scenario H: Insider Threat Pre-emption}
\textbf{Challenge:} Mitigating risks from ``high-performing but deviant'' insiders, where technical skill masks deeper behavioral vulnerabilities.

\begin{tcolorbox}[colback=gray!10, breakable, enhanced, colframe=gray!50, title=\textbf{Vignette: The ``Privileged'' Data Leak}]
\small
\textbf{Phase 1 \& 2: Profiling \& Assessment:} An expert sysadmin bypasses safety protocols. Evaluations show high \textbf{Security self-efficacy} and \textbf{Knowledge}. However, targeted profiling identifies high \textbf{Narcissism} combined with elevated \textbf{Risk-taking}.

\textbf{Phase 3: Systemic Analysis:} This combination triggers the \textbf{``Dark Traits'' Risk Pathway (Mechanism~\hyperref[int_pat:dark]{8})}. The violation is driven by a sense of entitlement: the negative \textbf{Attitude toward policies} stems from a belief that rules apply only to ``lesser'' users \cite{Maasberg2020DarkTriad, Campbell2004Narcissism}.

\textbf{Outcome:} Recognizing that awareness training cannot correct personality-driven negligence, the strategy pivots from Education to Constraint, implementing granular ``Zero Trust'' privileges and Behavioral Monitoring \cite{green2023understanding}.
\end{tcolorbox}
\end{document}